\documentclass{aa}  

\usepackage{graphicx}
\usepackage{txfonts}
\newcommand{\teff}{$T_{\rm eff}$}

\newcommand{\msun}{${\rm M}_\odot$}
\newcommand{\lsun}{${\rm L}_\odot$}
\newcommand{\rsun}{${\rm R}_\odot$}

\newcommand{\msunyr}{${\rm M}_\odot {\rm yr}^{-1}$}

\def\deg{$^\circ$} 

\def\mas{mas}
 
\def\kms{km~s$^{-1}$}

\def\teff{$T_{\rm eff}$}
\def\logg{$\log g$}

\def\vsini{$v\sin i$}
\def\prot{${\rm P}_{rot}$}
\def\vrad{${\rm V}_{r}$}
\def\vmic{${\rm V}_{mic}$}
\def\ha{${\rm H}\alpha$}
\def\hb{${\rm H}\beta$}
\def\hei{${\rm HeI}$}
\def\ewha{$EW({\rm H}\alpha$)}
\def\brg{${\rm Br}\gamma$}
\def\pab{${\rm Pa}\beta$}
\def\heiir{HeI}
\def\caii{CaII}

\def\macc{$\dot{M}_{\rm acc}$}
\def\lacc{$L_{\rm acc}$}

\def\rstar{${\rm R}_\star$}
\def\mstar{${\rm M}_\star$}
\def\lstar{${\rm L}_\star$}
\def\prot{${\rm P}_{rot}$}

\def\phirot{$\Phi_{rot}$}

\def\rcor{$r_{cor}$}
\def\rsub{$r_{sub}$}

\begin{document}

   \title{Investigating the magnetospheric accretion process in the young pre-transitional disk system DoAr 44 (V2062~Oph)}

   \subtitle{A multiwavelength interferometric, spectropolarimetric, and photometric observing campaign\thanks{Based on observations obtained at the Canada-France-Hawaii Telescope (CFHT), at the European Organisation for Astronomical Research in the Southern Hemisphere under ESO programme 0103.C-0097, and at the Las Cumbres Observatory global telescope network (LCOGT).}}

 \titlerunning{DoAr 44}
 \authorrunning{Bouvier, Alecian, Alencar, et al.}

   \author{J. Bouvier  \inst{1}
          \and
          E. Alecian \inst{1}
          \and
          S.H.P. Alencar \inst{2}
          \and
          A. Sousa \inst{1}
          \and
          J.-F. Donati \inst{3}
          \and
          K. Perraut \inst{1}
          \and
          A. Bayo \inst{4,5}
          \and
          L.M. Rebull \inst{6}
          \and
          C. Dougados \inst{1}
          \and
          G. Duvert \inst{1}
          \and
          J.-P. Berger \inst{1}
          \and
          M. Benisty \inst{1,7}
          \and
          K. Pouilly \inst{1}
          \and
          C. Folsom \inst{3}
          \and
          C. Moutou \inst{3}
          \and
          the SPIRou consortium
         }

   \institute{Univ. Grenoble Alpes, CNRS, IPAG, 38000 Grenoble, France 
         \and
         Departamento de Fisica -- ICEx -- UFMG, Av. Antonio Carlos 6627, 30270-901 Belo Horizonte, MG, Brazil
         \and
         Univ. de Toulouse, CNRS, IRAP, 14 avenue Belin, 31400 Toulouse, France
         \and
         Instituto de F\'{\i}sica y Astronom\'{\i}a, Facultad de Ciencias, Universidad de Valpara\'{\i}so, Chile
         \and
         N\'ucleo Milenio de Formaci\'on Planetaria - NPF, Universidad de Valpara\'{\i}so, Chile
         \and
         Infrared Science Archive (IRSA), IPAC, 1200 E. California Blvd., California Institute of Technology, Pasadena, CA 91125, USA
         \and
         Unidad Mixta Internacional Franco-Chilena de Astronom\'{\i}a (CNRS, UMI 3386), Departamento de Astronom\'{\i}a, Universidad de
Chile, Camino El Observatorio 1515, Las Condes, Santiago, Chile
          }

   \date{Received 10 July 2020; accepted 10 September 2020}

  \abstract
{Young stars interact with their accretion disk through their strong magnetosphere.}  
{We aim to investigate the magnetospheric accretion/ejection process in the young stellar system DoAr 44 (V2062 Oph).}
{We monitored the system over several rotational cycles, combining high-resolution spectropolarimetry at both optical and near-IR wavelengths with long-baseline near-IR inteferometry and multicolor photometry.}
{We derive a rotational period of 2.96~d from the system's  light curve, which is dominated by stellar spots.  We fully characterize the central star's properties from the high signal-to-noise, high-resolution optical spectra we obtained during the campaign. DoAr 44 is a young 1.2\msun\ star, moderately accreting from its disk (\macc = 6.5 10$^{-9}$ \msunyr), and  seen at a low inclination ($i~\simeq 30\degr$). Several optical and near-IR line profiles probing the accretion funnel flows (\ha, \hb, \hei\ 1083~nm, \pab) and the accretion shock (\hei\ 587.6~nm) are modulated at the stellar rotation period. The most variable line profile is HeI 1083~nm, which exhibits modulated redshifted wings that are a signature of accretion funnel flows, as well as deep blueshifted absorptions indicative of transient outflows. The Zeeman-Doppler analysis suggests the star hosts a mainly dipolar magnetic field, inclined by about 20$\degr$ onto the spin axis, with an intensity reaching about 800~G at the photosphere, and up to 2$\pm$0.8~kG close to the accretion shock. The magnetic field appears strong enough to disrupt the inner disk close to the corotation radius, at a distance of about 4.6\rstar\ (0.043 au),  which is consistent with the 5\rstar\ (0.047 au) upper limit we derived for the size of the magnetosphere in our Paper I from long baseline interferometry. } 
{ DoAr 44 is a pre-transitional disk system, exhibiting a 25-30 au gap in its circumstellar disk, with the inner and outer disks being misaligned. On a scale of 0.1 au or less, our results indicate that the system is steadily accreting from its inner disk through its tilted dipolar magnetosphere. We conclude that in spite of a highly structured disk on the large scale, perhaps the signature of ongoing planetary formation, the magnetospheric accretion process proceeds unimpeded at the star-disk interaction level. 
 } 

   \keywords{Stars: pre-main sequence -- Stars: variables: T Tauri, Herbig Ae/Be -- Stars: magnetic field -- Protoplanetary disks -- Stars: individual: DoAr~44, V2062 Oph, Haro~1-16 
               }

   \maketitle

%

\section{Introduction}

T Tauri stars are low-mass pre-main sequence stars at an early stage of their evolution. For a few million years, they accrete material from their circumstellar disks, a remnant of the protostellar collapse and the birthplace of planetary systems. The so-called classical T Tauri stars (cTTSs) have specific properties, such as a rich emission-line spectrum, large photometric variability, and a strong continuum excess flux superimposed onto a late-type photospheric spectrum \citep{Herbig62}. The unique properties of this class of objects eventually vanish as the disk dissipates on a timescale of a few million years, and the young stellar object transitions to a weak-line T Tauri star \citep[wTTS,][]{Menard99}. Indeed, many of the specific properties of cTTSs are thought to derive from the accretion process \citep{Kenyon87, Bertout88, Basri89} and, more specifically, to arise from the interaction region between the inner disk and the star. 

The currently accepted paradigm that describes this interaction is the magnetospheric accretion scenario where the strong stellar magnetic field controls, at least partly, the accretion flow from the inner disk edge to the stellar surface \citep[][see reviews by \citealt{Bouvier07b, Hartmann16}]{Camenzind90, Konigl91}. With large-scale surface magnetic fields of a few hundred to a few kG \citep[e.g.,][]{Donati20a, Sokal20}, the stellar magnetosphere truncates the inner gaseous disk at a distance of a few  stellar radii (typically 3-8 \rstar) above the photosphere. From there, the accreted material is channeled along magnetic funnel flows, and hits the stellar surface almost at free-fall velocity to create a localized accretion shock where the kinetic energy of the flow is dissipated. In this scenario, the emission line spectrum of cTTSs is partly formed in the accretion funnel flows \citep{Hartmann94}, and the accretion shock is responsible for the continuum excess flux from the X-ray range to the optical wavelengths, which is referred to as veiling \citep{Calvet98}. 

The magnetospheric process extends over a scale less than 0.1~au from the star, which corresponds to $\leq$1~\mas\ angular separation on the sky at the distance of nearby star forming regions. Therefore, most constraints on the physics of magnetospheric accretion come from observational campaigns monitoring the spectral variability of selected cTTSs over successive rotation periods. The rotational modulation over a timescale of weeks of emission line profiles, veiling, and Zeeman signatures allows one to draw magnetic maps of the stellar surface and to reconstruct the structure of the magnetospheric accretion region and investigate its dynamics. Previous monitoring campaigns have been quite successful in interpreting the observed properties and variability of young stellar objects in the framework of the magnetospheric accretion model \citep[e.g.,][]{Bouvier07a, Alencar12, Alencar18, Donati19}. These provided new insights into the physics of the interaction region between the inner disk edge and the stellar surface, which impacts both early stellar evolution, and, potentially, planet formation and/or migration at the inner disk edge. 

We present here the results of a new campaign targeting the young star \object{DoAr~44} (also known as V2062~Oph, Haro 1-16, ROXs 44, HBC 268) located at a distance of 146$\pm$1~pc \citep{Gaia16, Gaia18} in the L1688 cloud of the Rho Ophiuchi molecular complex. This moderately bright source of spectral type K3 exhibits strong \ha\ emission with \ewha$\simeq$50\AA\ \citep{Bouvier92} and accretes at a substantial rate from its circumstellar disk, \macc =  6.0-9.3 10$^{-9}$ \msunyr\ \citep{Espaillat10, Manara14}. From the measurement of the Zeeman broadening of near-IR FeI lines, \cite{Lavail17} derived a mean surface magnetic field strength amounting to 1.8$\pm$0.4~kG, and possibly up to 3.6~kG. These authors further derived a projected rotational velocity of \vsini=17.5$\pm$1.0~\kms, and estimated a mass of 1.25~\msun\  for this source. 

The dedicated monitoring campaign of this source we present here combines, for the first time, high angular resolution ($\simeq$ 1~mas) long-baseline interferometry, optical and near-IR high-resolution spectropolarimetry, and multicolor photometry. DoAr~44 was selected for this campaign based on its brightness (V=12.6, K=7.6), which is amenable to both long-baseline interferometry and spectropolarimetry; the presence of a \brg\ line in emission in its near-IR spectrum, with EW(\brg)$\simeq$4~\AA\ measured on an archival ESO/XSHOOTER spectrum; and the recent report of its high surface magnetic field \citep{Lavail17}. In addition, below we derive a photometric period of 2.96~d from a publicly available ASAS-SN light curve (see Section 3.2), which we ascribe to the stellar rotation period. These properties made it a suitable target for a dedicated large-scale monitoring campaign, with the additional interest of probing the magnetospheric accretion process in a pre-transitional disk system. This system featured a large dust-depleted cavity in its circumstellar disk, extending up to a distance of 30 au from the central star \citep{Andrews11}. The goal of this campaign was to constrain the structure and dynamics of the magnetospheric accretion region between the inner edge of the compact inner disk and the central star by measuring the magnetic field strength at the stellar surface, reconstructing the geometry of the magnetospheric funnel flows and accretion shocks from the analysis of line profile variability and photometric variations, and, ultimately, attempting to resolve the interaction region over scales of a few stellar radii above the stellar surface using long-baseline interferometry. The VLTI/GRAVITY interferometric results are reported in an accompanying paper \citep[][hereafter Paper~I]{Bouvier20}. Section 2 describes the observations and data reduction, Section 3 features our results, and Section 4 offers a consistent interpretation of the observed variability of the system in the framework of the magnetospheric accretion model.

\section{Observations}

\begin{figure}
   \centering
   \includegraphics[width=0.9\hsize]{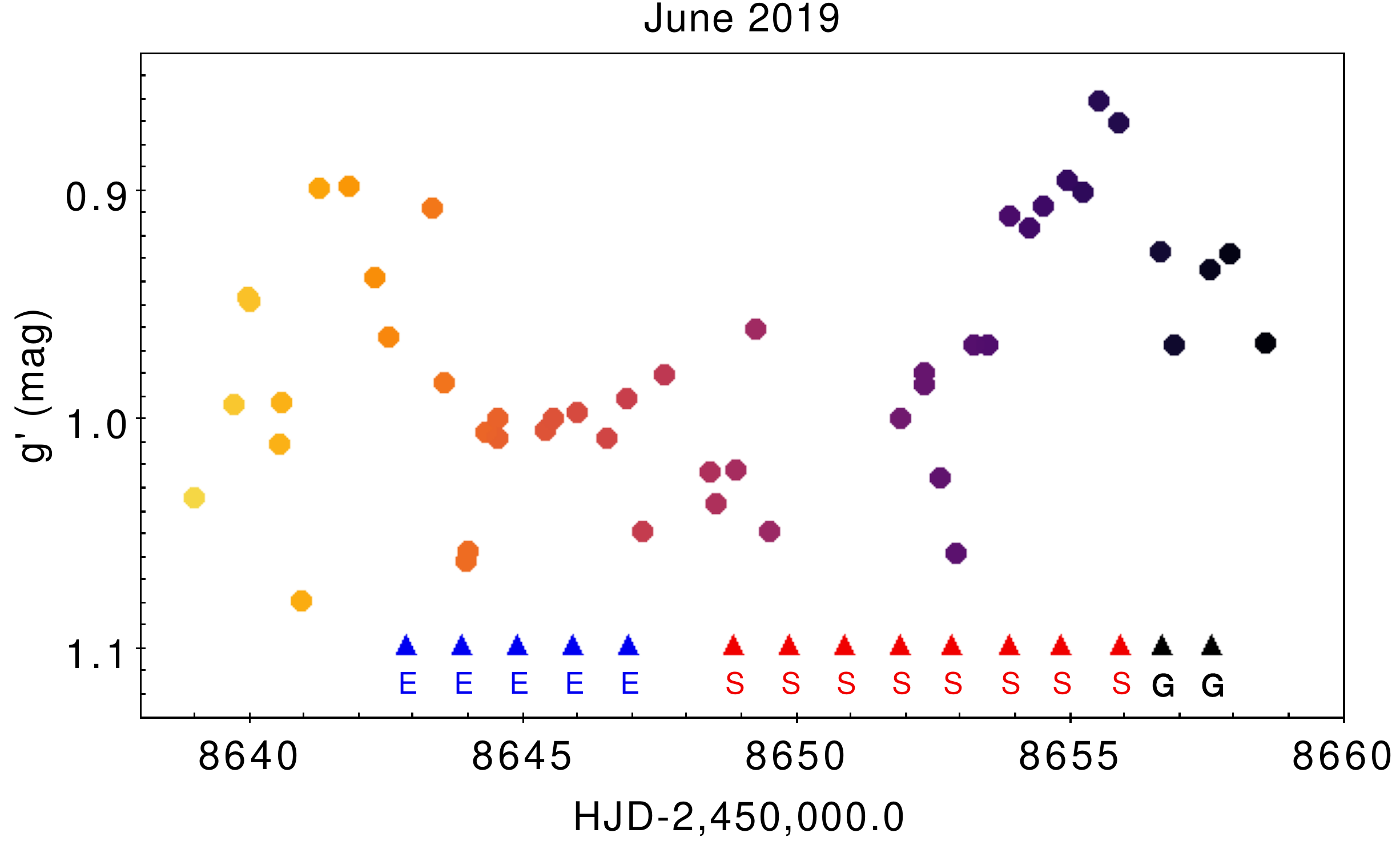}
   \caption{Visual summary of observations obtained in June 2019, as part of the DoAr~44 coordinated campaign. The differential g'-band light curve (see Section 3.2) is shown as colored circles, with the color scale reflecting the Julian Date, while the epochs of CFHT/ESPaDOnS, CFHT/SPIRou, and ESO VLTI/Gravity observations are indicated by triangles labeled with "E", "S", and "G", respectively. Note that the March 2019 ESPaDOnS run, which is part of the campaign reported here, is not shown in this figure. The average g-band magnitude of the system during the campaign was g=13.35 according to the publicly available Zwicky Transient Facility light curve \citep{Masci19}.}
              \label{sampling}%
    \end{figure}

In this section, we describe the acquisition and data-reduction processes of photometric and spectropolarimetric datasets obtained during the large-scale campaign performed on the cTTS DoAr~44, which also included long-baseline near-IR interferometry. A summary plot of the main part of the DoAr~44 observing campaign is provided in Fig.~\ref{sampling}. 

\subsection{LCOGT photomery}

\begin{figure}
   \centering
   \includegraphics[width=0.9\hsize]{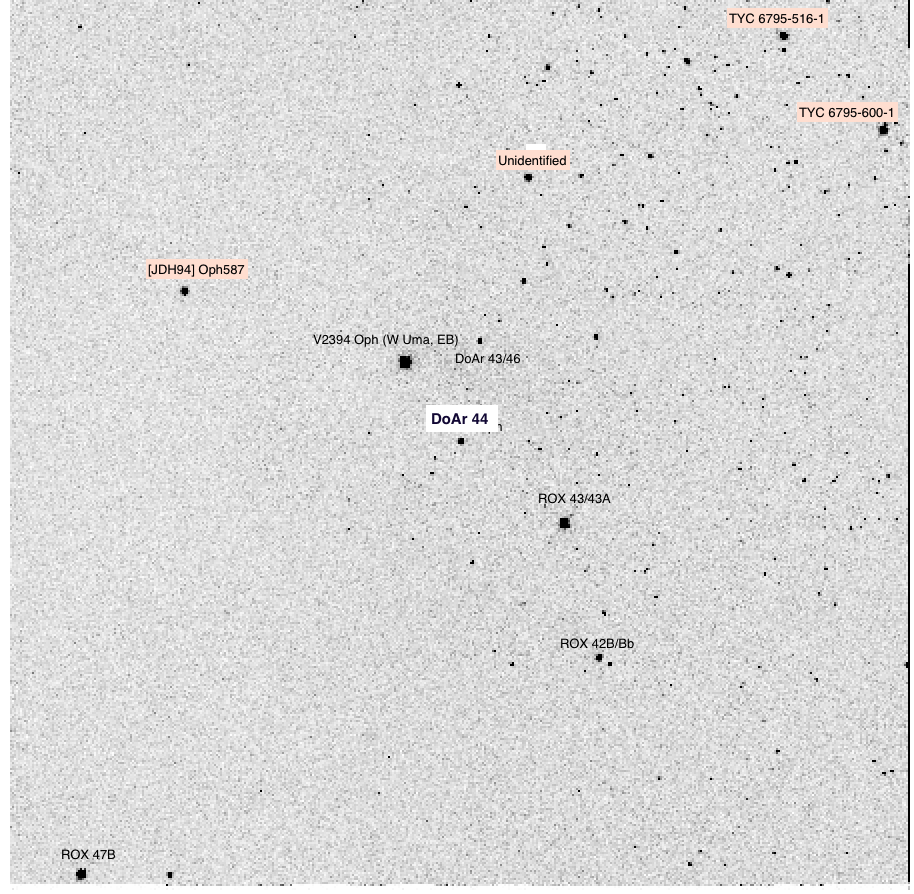}
   \caption{Field of view of LCOGT images, with the target, as well as comparison and check stars identified. North is up, east is left. The FOV is 27 arcmin on the side.}
              \label{fov}%
    \end{figure}

Photometric observations were obtained at the Las Cumbres Observatory Global Network \citep[LCOGT,][]{Brown13} with the 1m Sinistro telescopes from June 4 to June 24, 2019. We used  SDSS/PanSTARRS u'g'r'i' filters with exposure times of 60, 30, 10, and 5s, respectively. A total of 219 images centered on DoAr~44 were obtained over 20 days with a cadence of nearly three photometric sequences per 24 hours. Point-spread function (PSF) photometry was performed using IRAF/DAOPHOT on the pipe-line reduced images provided by LCOGT. Over the field of view (FOV) of 27 arcmin (see Figure~\ref{fov}), the PSF photometry includes the target DoAr~44, the nearby W Uma eclipsing binary \object{V2394 Oph}, the T Tauri stars \object{ROX 42B}, \object{ROX 43}, \object{ROX 47B}, and \object{DoAr 43}, as well as four potential comparison and check stars: \object{[JDH94]Oph 587}, \object{TYC 6795-600-1}, \object{TYC 6795-516-1}, and an anonymous star (RA, Dec: 247.8519d, -24.3306d), referred to as C5 in the following. We verified that Oph 587 and C5, located within 10 arcmin of DoAr~44, were stable at a few 0.01 mag level in all filters. While the two TYC stars are brighter than Oph 587 and C5, and also appear to be photometrically stable at similar levels, they are located in a corner of the image, and are therefore more prone to systematics (e.g. flat-field correction).  

\begin{table}
\caption{LCOGT u'g'r'i' differential photometry (first 10 of 229 measurements). The columns list: Julian date, differential magnitude (see text) between DoAr~44 and the comparison star, between the comparison star and the check star, filter, airmass, and Moon distance from the target. We note that the results are given for different filters on each row. In any given filter, the comparison-check star values listed in the third column are constant within errors (see text). The full table is available electronically at CDS, Strasbourg.}             
\label{phot}      
\centering                          
\begin{tabular}{l l l l l l l }        
\hline\hline                        
Julian date  &  Target  &  Check  &  Filt.  &  Airmass    & Moon \\
(2,450,000.+) & (mag) & (mag) && & dist. (\degr)\\
\hline                                   
8638.9847732  &  1.03  &  0.39  &  gp  &  1.14  &  163  \\
8638.9858978  &  0.09  &  0.28  &  ip  &  1.13    &  163  \\
8638.9867561  &  0.5  &  0.43  &  rp  &  1.13    &  163  \\
8638.9876614  &  0.95  &  0.02  &  up  &  1.13    &  163  \\
8639.7314887  &  0.99  &  0.43  &  gp  &  1.04   &  152  \\
8639.7326015  &  0.09  &  0.3  &  ip  &  1.04    &  152  \\
8639.7334362  &  0.46  &  0.43  &  rp  &  1.04   &  152  \\
8639.7343175  &  0.97  &  0  &  up  &  1.04    &  152  \\
8639.9927617  &  0.94  &  0.4  &  gp  &  1.1    &  149  \\
8639.9939326  &  0.06  &  0.3  &  ip  &  1.1    &  149  \\
...\\
\hline                                   
\end{tabular}
\end{table}

\begin{table}
\caption{Journal of ESPaDOnS observations of DoAr~44. All observations consist of sequences of four subexposures, each lasting 950 s. Columns respectively list, for each observation, the UT date, time, heliocentric Julian date (HJD), rotational phase, and signal-to-noise ratio (SNR) at 731 nm (per 2.6~\kms velocity bin).}             
\label{espadons_journal}      
\centering                          
\begin{tabular}{c c c c c}        
\hline\hline                 
            Date      &  UT  & HJD & \phirot & SNR \\
            (2019) & (hh:mm:ss) & (2,450,000.+) && (731nm)\\ 
\hline                        
Mar 15  &  14:31:32  &  8558.10852  &  0.48 & 137  \\ 
Mar 16  &  15:32:19  &  8559.15082  &  0.83 & 141  \\ 
Mar 18  &  15:11:40  &  8561.13665  &  1.50 & 133  \\ 
Mar 19  &  14:47:48  &  8562.12015  &  1.83 & 124  \\ 
Mar 20  &  14:13:53  &  8563.09667  &  2.16 & 131  \\ 
Mar 21  &  14:50:36  &  8564.12225  &  2.51 & 132  \\ 
Mar 22  &  13:32:39  &  8565.06818  &  2.83 & 143  \\ 
Jun 08  &  09:16:27  &  8642.89230  &  29.12 & 139  \\ 
Jun 09  &  09:21:13  &  8643.89558  &  29.46 & 135  \\ 
Jun 10  &  09:19:12  &  8644.89415  &  29.80 & 135  \\ 
Jun 11  &  10:15:19  &  8645.93308  &  30.15 & 134  \\ 
Jun 12  &  10:07:45  &  8646.92779  &  30.48 & 130  \\ 
\hline                                   
\end{tabular}
\end{table}

\begin{table}
\caption{Journal of SPIRou observations of DoAr~44. All observations consist of sequences of four subexposures, each lasting 300 s. Columns respectively list, for each observation, the UT date, time, HJD, and SNR at 2135 nm per 2.3~\kms\  velocity bin.}             
\label{spirou_journal}      
\centering                          
\begin{tabular}{c c c c c}        
\hline\hline                 
            Date      &  UT  & HJD & \phirot & SNR \\
            (2019) & (hh:mm:ss) & (2,450,000.+) && (2135nm)\\ 
\hline                        
Jun 14  &  08:43:03  &  8648.8680960  &  31.14 & 223  \\
Jun 15  &  08:54:21  &  8649.8759002  &  31.48 & 227  \\
Jun 16  &  09:17:09  &  8650.8916925  &  31.82 & 227  \\
Jun 17  &  09:39:18  &  8651.9070323  &  32.16 & 137  \\
Jun 18  &  08:28:37  &  8652.8579025  &  32.49 & 216  \\
Jun 19  &  09:27:16  &  8653.8985925  &  32.84 & 237  \\
Jun 20  &  08:07:01  &  8654.8428144  &  33.16 & 237  \\
Jun 21  &  10:17:10  &  8655.9331438  &  33.52 & 211  \\
\hline                                   
\end{tabular}
\end{table}

We therefore used Oph 587 as a reference star and C5 as a check star to derive the differential light curves for DoAr~44. Stars Oph 587 and C5 are within one magnitude of difference from the target's brightness in all filters. The rms error of the photometric measurement in each filter amounts to about 0.02 mag. DoAr~44 is found to be variable in all four filters: the rms dispersion of the u'g'r'i' differential light curves (DoAr~44 - Oph 587) is 0.17, 0.09, 0.07, and 0.09 mag, respectively, while the corresponding values for the comparison and check stars (Oph 587 - C5) are 0.03, 0.03, 0.04, and 0.04 mag.  The nearby bright Moon toward the middle of the photometric campaign (June 16, 2019; JD 2458650.5) strongly increased the sky background, especially in the u'-band, leading to poorer quality photometry around this date. DoAr~44's differential photometry is provided in Table~\ref{phot}. 

\subsection{CFHT spectropolarimetry}

Optical spectropolarimetric observations of DoAr~44 were completed during two runs, from March 15 to 22 and from June 8 to 12, 2019, with the ESPaDOnS spectropolarimeter at the Canada-France-Hawaii Telescope (CFHT), at a resolving power of 65,000 covering 370 to 1,000 nm \citep{Donati03}. During the first run, seven circularly polarized spectra were collected, and five additional spectra were obtained during the second run. Raw frames were reduced with the standard ESPaDOnS reduction package, and a least-squares deconvolution \citep[LSD,][]{Donati97} was applied to all spectra, using a line list appropriate to DoAr~44. The journal of observations is presented in Table~\ref{espadons_journal}. Near-IR spectropolarimetric observations of DoAr~44 were completed from June 14 to 21, 2019 with the SPIRou spectropolarimeter at the CFHT, covering from 0.95 to 2.50 microns, at a spectral resolution of~70,000 \citep{Donati20b} at a rate of one visit per night. Raw data were processed by the SPIRou pipe-line at CFHT and the reduced data provided by the Canadian Astronomy Data Center. The journal of observations is presented in Table~\ref{spirou_journal}.

\begin{table}
\caption{Stellar properties of the DoAr~44 system. The V-band and 2MASS J-band magnitudes, and the Gaia DR2 distance are taken from the literature (see text). All other values are derived in this study. }             
\label{starprop}      
\centering                          
\begin{tabular}{l l }        
\hline\hline                        
V, J & 12.6, 9.1\\
SpT & K2-K3\\
A$_V$ & 2.0$\pm$0.2 mag\\
\teff & 4600$\pm$120~K\\
\lstar/\lsun & 1.6$\pm$0.2\\
\rstar/\rsun & 2.0$\pm$0.15 \\
\mstar/\msun & 1.2$\pm$0.15 \\
EW(LiI) & 430 m\AA \\
\vsini & 17.0$\pm$1.1~\kms\\
\prot & 2.960$\pm$0.018~d\\
$i$ & 30$\pm$5~\degr \\
\macc & 7$\pm$3 10$^{-9}$ \msunyr\\
$d$ & 146$\pm$1 pc \\
\rcor & 0.043 au (4.63 \rstar)\\
\rsub & 0.044 au (4.74\rstar)\\
\hline                                   
\end{tabular}
\end{table}

\subsection{ESO/VLTI interferometry}

DoAr 44 was observed on June 22 and 23, 2019 in the K-band with the GRAVITY instrument \citep{Grav17} combining the four unit telescopes of the ESO/VLTI (ESO run 60.A-9256). At 2.2 $\mu$m, with a maximum baseline of 130~m, we thus reach an angular resolution of $\lambda$/2Bmax = 1.7~mas, corresponding to 0.25 au at 146~pc. We used the instrument in its high spectral-resolution mode (R $\sim$ 4000), thus partly resolving the \brg\ line profile. The observations, data reduction, and interferometric results are described in full in Paper~I.  

\section{Results}

\subsection{Stellar properties}

We derived estimates of the system's properties based on the newly acquired high-resolution spectra. We followed the same procedure as we used in \cite{Pouilly20} to which the reader is referred for details. We selected 16 spectral windows between 600 and 800 nm where the mean ESPaDOnS spectrum of DoAr~44 is fit with a synthetic spectrum generated from the ZEEMAN code \citep{Folsom12}  based on MARCS stellar atmosphere grids \citep{Gustafsson08} and VALD line lists \citep{Ryabchikova15}. Assuming \logg = 4.0,  we derived the following parameters: \vsini =  17.0 $\pm$ 1.1 km/s, \vrad = -5.70 $\pm$ 0.24 km/s, \teff = 4600 $\pm$ 120 K, and \vmic = 0.27 $\pm$ 0.27 \kms. The veiling, that is to say, the ratio of the continuum excess flux to the stellar photospheric flux,  is measured as constant over time and amounts to $r\simeq$0.18$\pm$0.05 over the wavelength range 610-670 nm. 

Using the optical photometry of \cite{Bouvier88} and 2MASS J-band, and adopting a K2-K3 spectral type with intrinsic colors listed in \cite{Pecaut13}, we derived color excesses in the (V-I$_c$) and (V-J) colors that correspond to a visual extinction A$_V$=2.0$\pm$0.2~mag. From the J-band magnitude, bolometric corrections from \cite{Pecaut13} for K2-K3 young stars, and a distance of 146$\pm$1~pc \citep{Gaia16, Gaia18}, we thus obtained \lstar=1.6$\pm$0.2~\lsun. Combined with \teff=4600~K derived above, this yields \rstar=2.0$\pm$0.15~\rsun. From its position in the HR diagram, we derived  \mstar=1.23$\pm$0.15~\msun\ using the CESAM evolutionary models \citep{Marques13}, and the star appears to have a modest radiative core, with M$_{rad}$/\mstar=0.22$\pm$0.20 and R$_{rad}$/\rstar=0.35$\pm$0.2, where M$_{rad}$ and R$_{rad}$ are the mass and radius of the radiative core, respectively. We note, however, that the latter values are strongly model dependent, as, using \cite{Siess00} models, we would derive  \mstar=1.4~\msun, M$_{rad}$/\mstar=0.73 and R$_{rad}$/\rstar=0.50. 

Finally, we used the above estimate of the stellar radius to derive the inclination of the stellar rotational axis on the line of sight (LoS). Using \vsini=17~\kms\ and \prot=2.96~d (see below), and propagating the uncertainties, yields $\sin i$=0.50$\pm$0.05, which translates to $i$=30$\pm$5\degr, meaning the system is seen at low inclination. 
Table~\ref{starprop} summarizes the properties of the stellar system with their uncertainties.  

\subsection{Photometric variability}

   \begin{figure}
   \centering
      \includegraphics[width=\hsize]{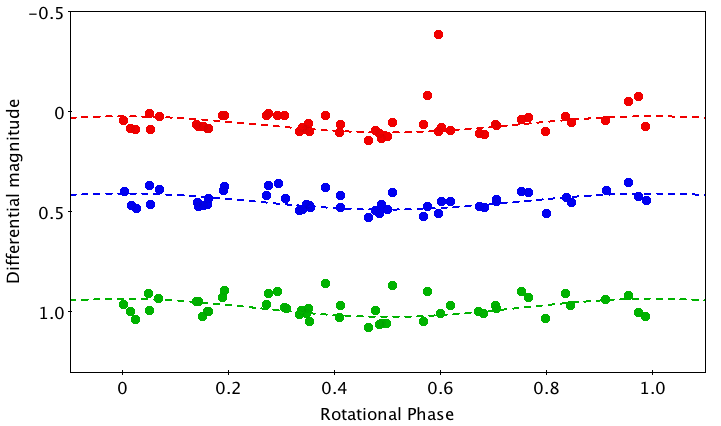}
   \caption{DoAr~44 photometric variations. The g'r'i' light curves (green, blue, and red, respectively) are shown folded in phase with a period of 2.960 days derived from periodogram analysis (see text). A sine curve was eye fit to the light curves and is shown for illustration with a peak-to-peak amplitude of 0.08~mag in the g'r'i' bands. } %
              \label{lc}%
    \end{figure}

The LCOGT observations yielded 20 day-long differential light curves for DoAr~44 in the u'g'r'i' filters. We searched for a periodicity in the photometric variations, excluding about five measurements per filter taken when the bright moon was under 15 degrees away from the target. We used both the CLEAN periodogram analysis \citep{Roberts87} and the string-length method \citep{Dworetsky83}. The former yields a period of 3.09~d in the r' filter only, while the latter returns a most probable period in all filters in the range from 2.97~d to 3.07~d. In order to confirm the period, we downloaded ASAS-SN photometry \citep{Shappee14}, which contains 288 single filter measurements over a single season extending from MJD 57780 to MJD 58030 (February-September 2017). Using the same period-search algorithms, we found a clear periodicity of 2.960 +/- 0.018~d in this much richer dataset, with a peak-to-peak amplitude of about 0.1 mag. The LCOGT g'r'i' light curves folded in phase with this period are shown in Fig.~\ref{lc} and suggest a low-level modulation by surface spots, of the order of a 0.08 mag peak to peak in the g'r'i' filters. While a period is found in the u'-band light curve as well, the phased light curve exhibits a scatter of the same order as the photometric amplitude. We therefore adopted the more accurate 2.960~d period derived from the  ASAS-SN dataset and ascribed it to the rotational period of the star. We thus define an ephemeris such that the origin of phase corresponds to the epoch of maximum brightness at the time of our observations, so \begin{equation} HJD (d) = 2,458,556.7 + 2.960\times E .\end{equation} Finally, we note that the uncertainty on the derived period, $\delta$P=0.018d, could induce a phase shift between the March and June ESPaDOnS runs up to $\delta\Phi$=0.16 rms.

   \begin{figure}
   \centering
   \includegraphics[width=0.32\hsize]{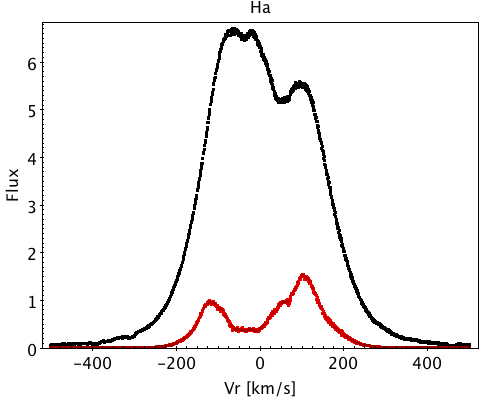}
      \includegraphics[width=0.32\hsize]{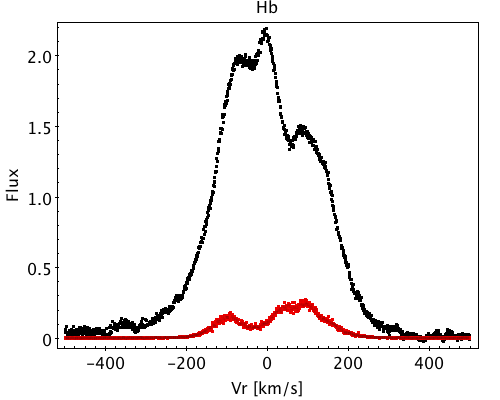}
      \includegraphics[width=0.32\hsize]{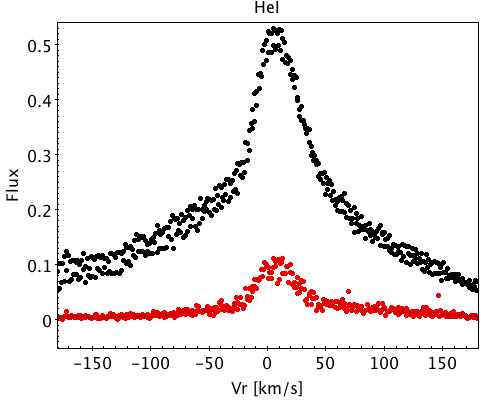}
   \caption{ {\it From left to right:} mean (black) and variance (red) residual profiles for \ha, \hb, and \hei\ 587.6 nm. \ha\ and \hb\ appear on overlapping spectral orders, which are both shown here. For \hei, the variance has been multiplied by 10. The increased scatter over the blue wing of the \hei\ line profile is due to spectral order overlap.  }
              \label{meanvar}%
    \end{figure}

   \begin{figure}
   \centering
   \includegraphics[width=0.48\hsize]{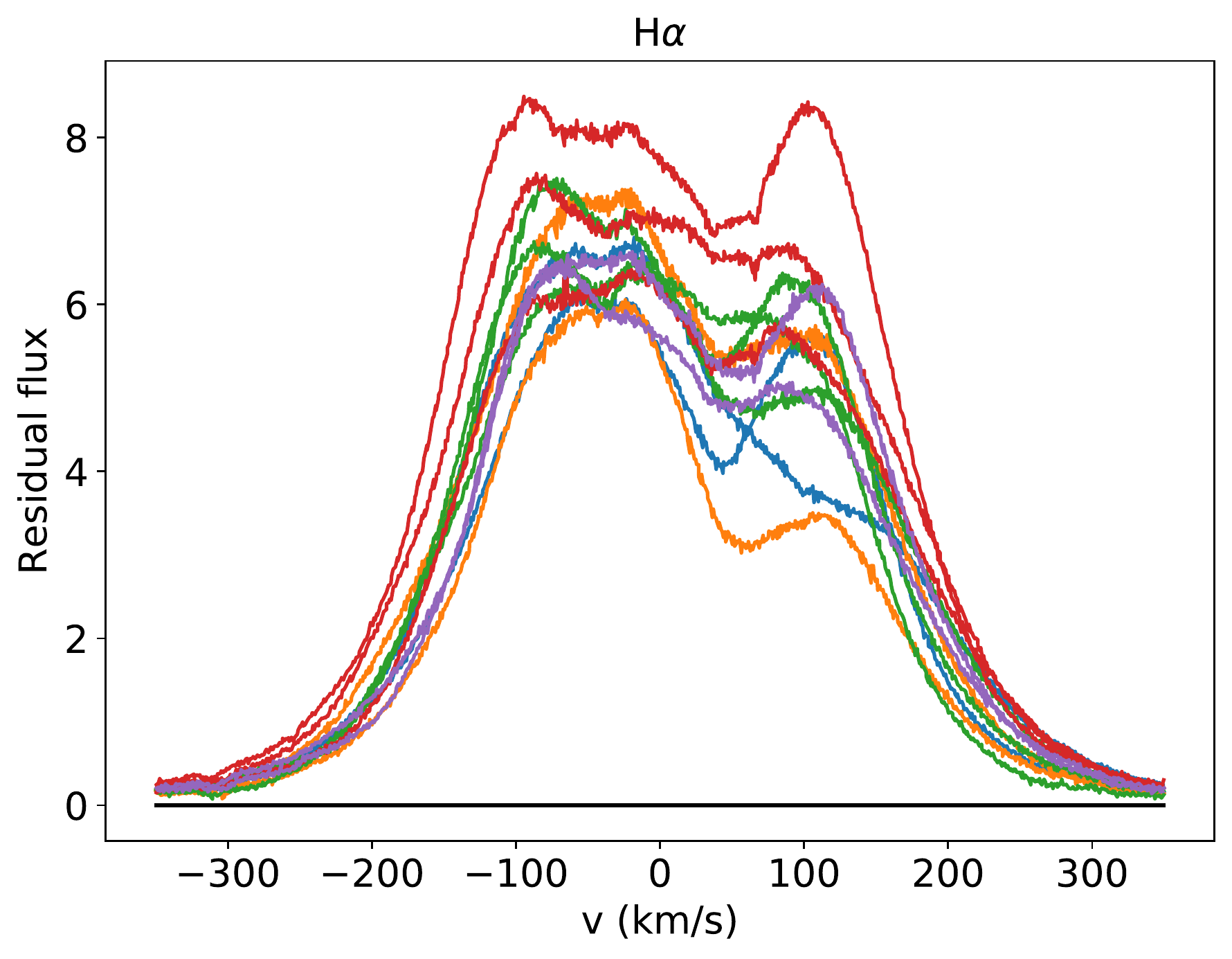}
      \includegraphics[width=0.48\hsize]{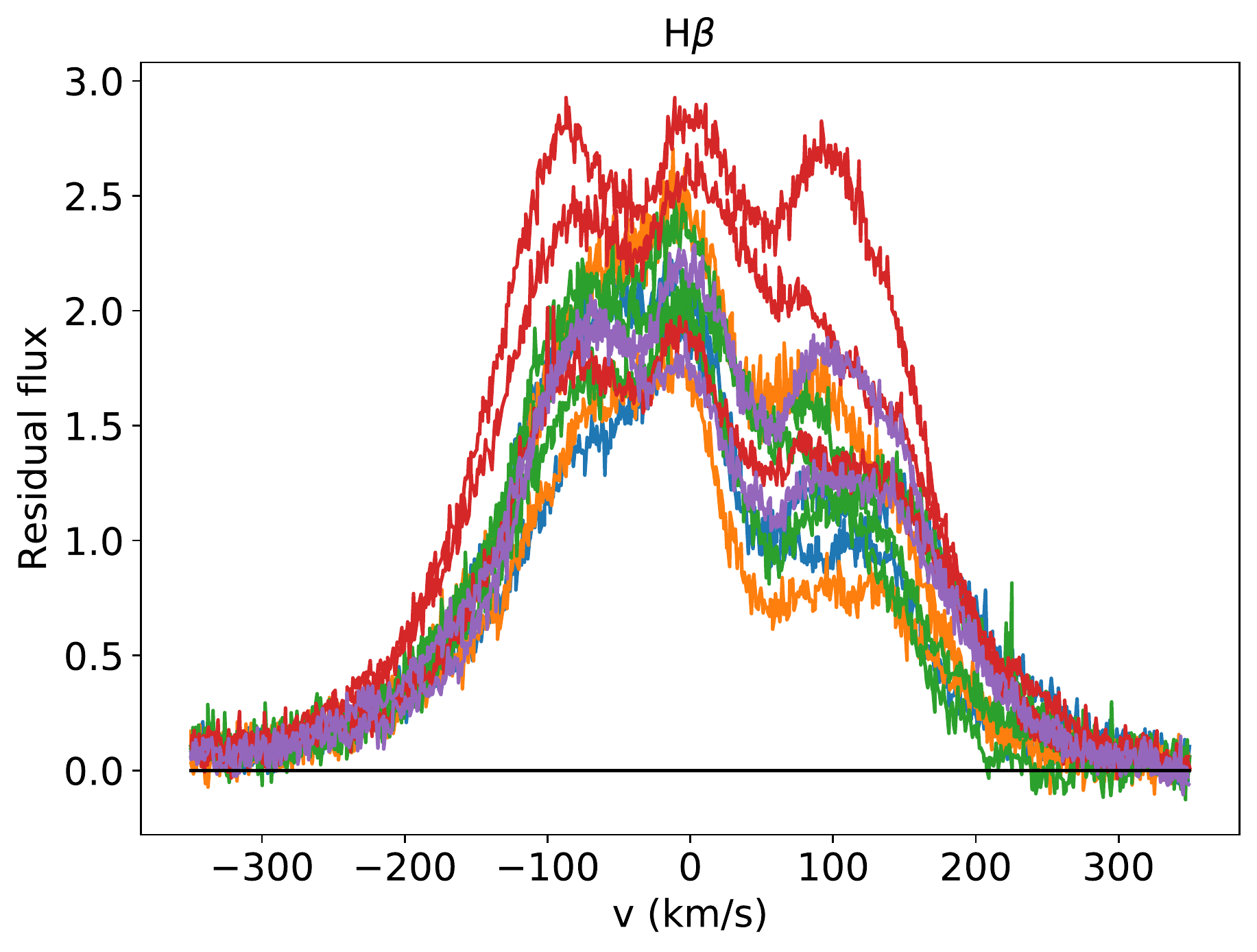}
   \caption{ Superimposed residual line profiles: H$\alpha$ ({\it left}) and H$\beta$ ({\it right}). The color code corresponds to different rotational cycles. Each line is covered by two spectral orders and both are shown.  }
              \label{hahbover}%
    \end{figure}
   \begin{figure}
   \centering
   \includegraphics[width=0.9\hsize]{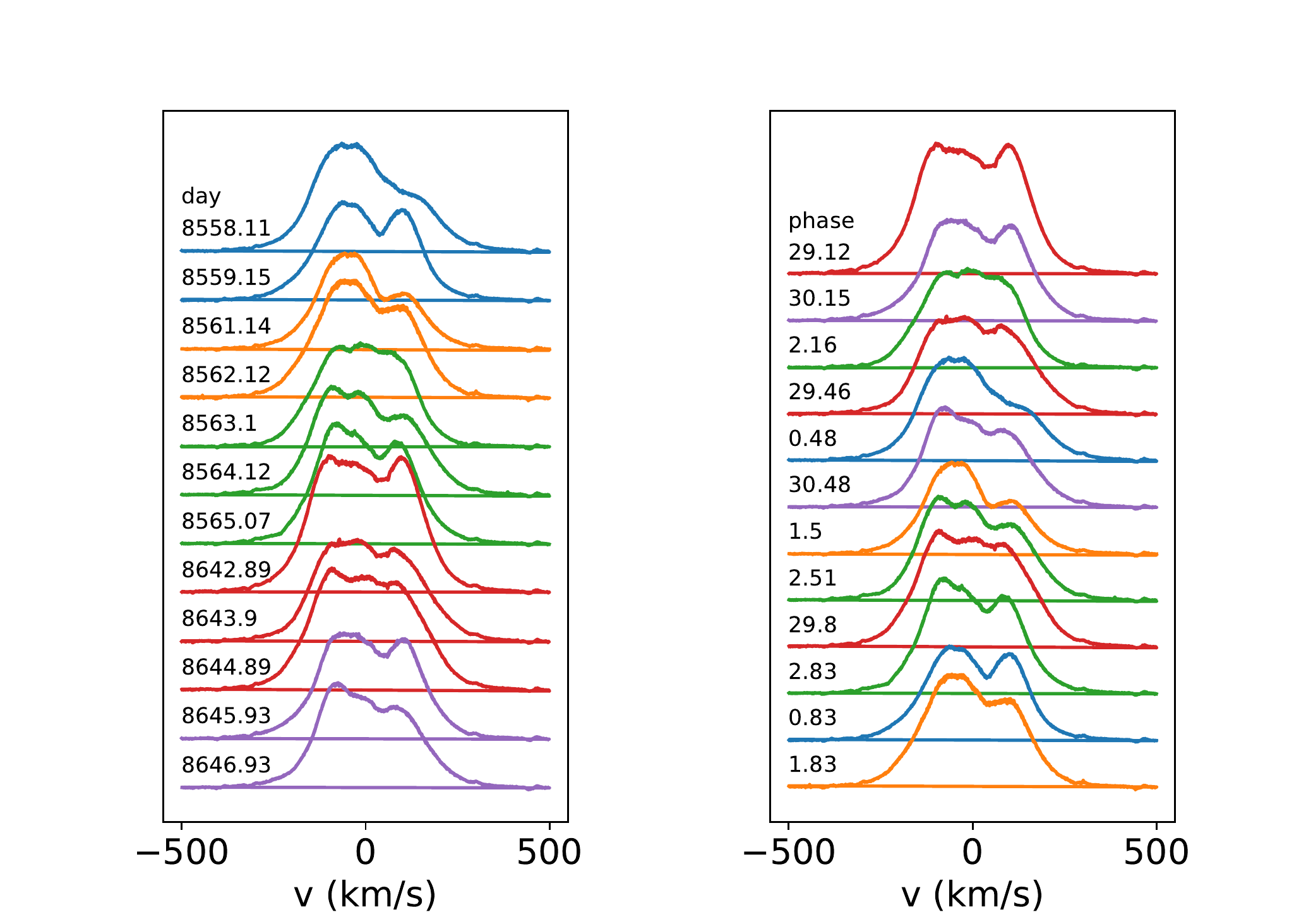}
            \includegraphics[width=0.9\hsize]{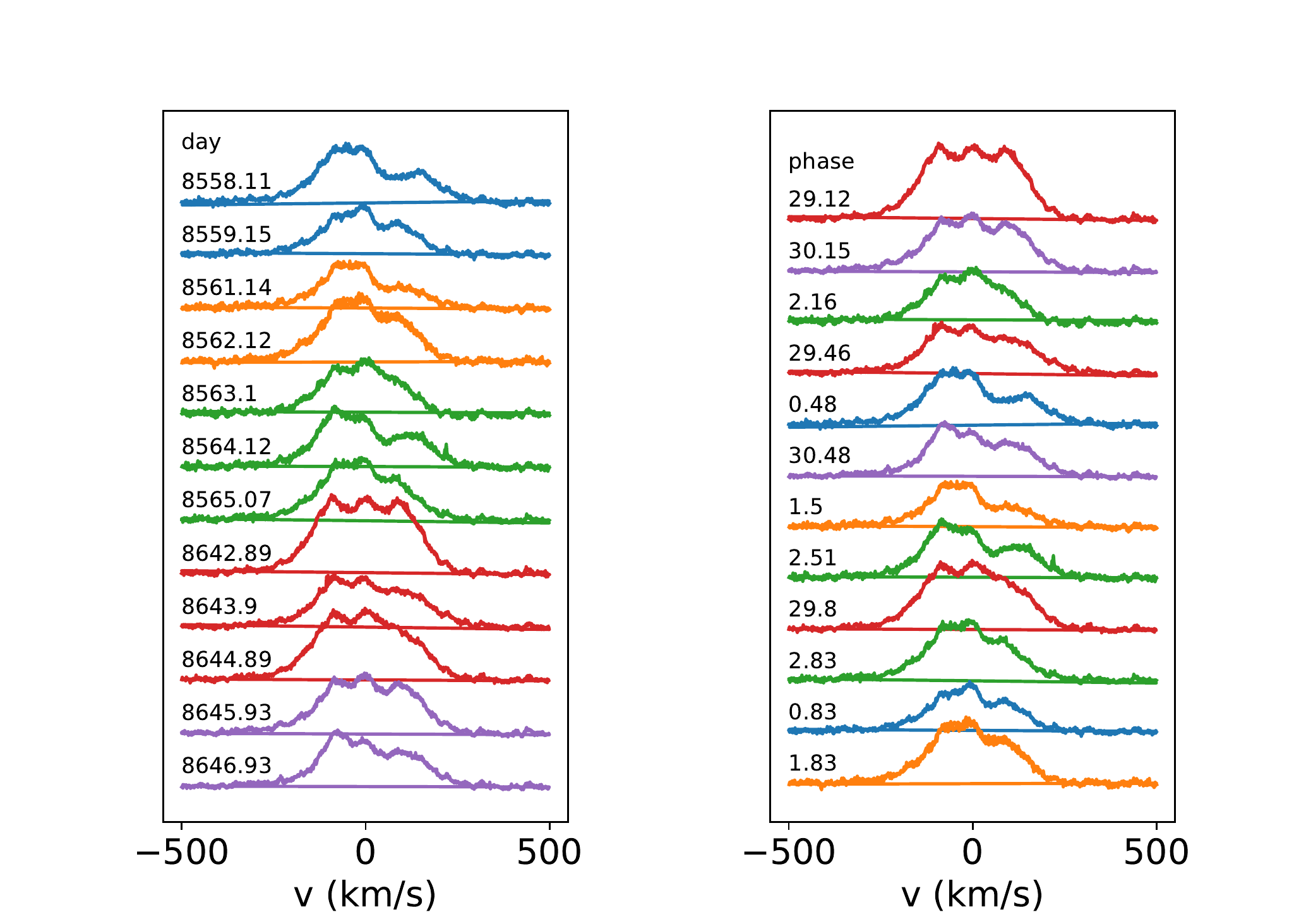}
   \caption{ H$\alpha$ ({\it top}) and H$\beta$ ({\it bottom}) residual line profile variability as a function of date and phase. The color code corresponds to different rotational cycles. We note that both profiles appear less asymmetric in June than in March 2019, with a more depressed red wing at the earlier epoch. }
              \label{hahbphase}%
    \end{figure}

   \begin{figure}
   \centering
      \includegraphics[width=0.9\hsize]{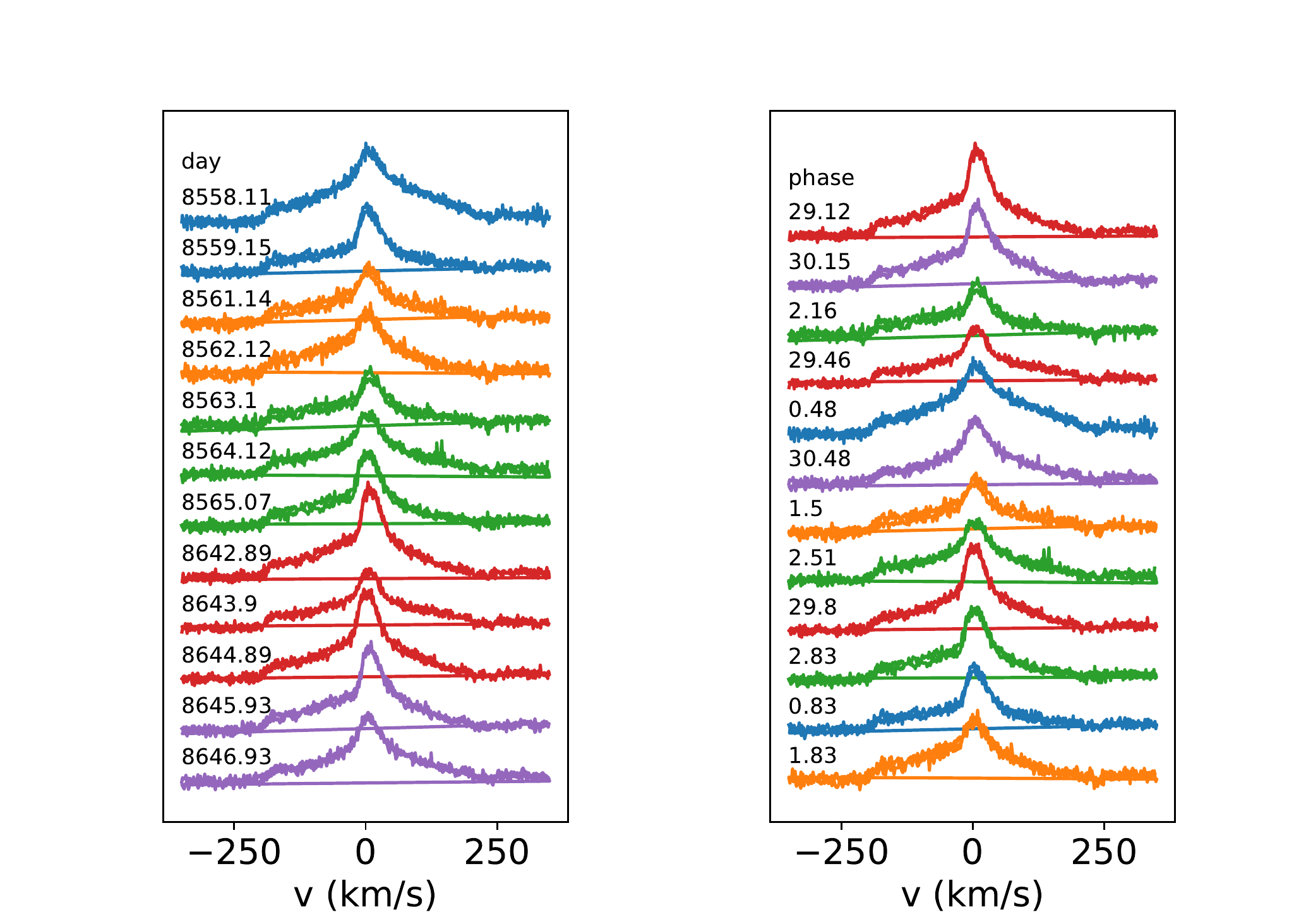}
   \caption{ Same as Fig.~\ref{hahbphase} for HeI 587.6~nm residual line profile. We note the presence of a narrow component superimposed on a broad pedestal. Similar line profiles are observed in March and June 2019. }
              \label{heiphase}%
    \end{figure}

\begin{table}
\caption{Photospheric radial velocity, \vrad, and equivalent width, W, of \ha, \hb, and \hei\ emission lines from ESPaDOnS spectra.}             
\label{ewtab}      
\begin{tabular}{c c c c c}        
\hline\hline                 
            HJD     & \vrad$^\dagger$  &  W(\ha)  & W(\hb) & W(\hei) \\
            (-2,450,000) & \kms &  (\AA) & (\AA) & (\AA)\\ 
\hline                        
  8558.10852 & -5.59 &  41.6 & 8.2 & 1.13\\
  8559.15082 & -5.78 & 38.2 & 5.8 & 0.62\\
  8561.13665 & -5.60 &  31.8 & 4.3 & 0.60\\
  8562.12015 & -5.80 &  48.1 & 7.7 & 0.85\\
  8563.09667 & -6.00 &  41.0 & 5.8 & 0.46\\
  8564.12225 & -5.63 &  44.1 & 6.8 & 0.76\\
  8565.06818 & -5.76 &  45.5 & 7.2 & 0.80\\
  8642.89230 & -6.10 &  60.5 & 12.5 & 1.06\\
  8643.89558 & -5.55 &  44.8 & 6.6 & 0.66\\
  8644.89415 & -5.66 &  53.3 & 10.0 & 1.04\\
  8645.93308 & -6.02 &  43.9 & 7.7 & 0.99\\
  8646.92779 & -5.58 &  39.6 & 6.1 & 0.77\\
\hline
\end{tabular}

$^\dagger$ The rms uncertainty on \vrad\ amounts to 0.05 km/s.
\end{table}

\subsection{Optical line profiles}

Deciphering the shape and variability of the emission line profiles seen in the spectra of cTTSs offers a window to the physical processes occurring close to the star, particularly the structure and dynamics of accretion and outflows \citep[e.g.,][]{Alencar00, Edwards06, Lima10, Esau14}.  
In the optical range, DoAr~44's spectrum shows line emission primarily the HI Balmer and Paschen series, in the CaII H\&K doublet and near-IR triplet, as well as in the \hei\ 587.6 nm, 667.8 nm, and 706.6 nm, [OI] 630.0 nm, and OI 777.3 nm lines. Residual emission line profiles were computed by subtracting the broadened photospheric spectrum of the Hyades cluster member Melotte 25-151, a K2-type star with \vrad = 37.98 \kms, \vsini = 4.83 \kms, and \teff = 4920~K \citep{Folsom18}, also observed with ESPaDOnS. 

\subsubsection{Line profile variability}

The spectral series we obtained on DoAr 44 with CFHT/ESPaDOnS allows us to investigate the spectral variability of the source on timescales ranging from days to months. 
Figure~\ref{meanvar} shows the mean and variance profile for the main emission lines. The Balmer line profiles are double peaked with a slightly depressed red peak, corresponding to the Type II(R) profiles in the classification of \cite{Reipurth96}. This profile shape is qualitatively accounted for by dipolar magnetospheric accretion models considering moderate to low inclinations, $i\simeq 30-45\degr$ \citep{Muzerolle98a, Muzerolle01}. Both the \ha\ and \hb\ profiles exhibit significant variance in the blue and red wings up to velocities of a few hundred \kms, and comparatively less variability in the line center. As is often observed in T Tauri stars \citep{Beristain01}, the \hei\ line profile is composite, consisting of a centrally peaked broad component, extending up to velocities of $\pm$200 \kms, and a narrow component located close to the line center with a FWHM of about 40~\kms. The profile's variability seems to be predominantly associated with the narrow component. According to \cite{Beristain01}, the properties of this type of \hei\ profile suggest the narrow component arises in the post-shock region close to the stellar surface, while the broad component is formed in the main accretion column above the accretion shock. 

Figure~\ref{hahbover} shows the superimposed H$\alpha$ and H$\beta$ line profiles recorded for DoAr~44 during the CFHT/ESPaDOnS runs of March and June 2019. Figure~\ref{hahbphase} is another representation of the same dataset, ordered by Julian date and by rotational phase. Similar plots are shown for the HeI 587.6 nm line in Fig.~\ref{heiphase}. We note that the Balmer line profiles appear more symmetric in June than in March, while the \hei\ line profile has remained the same, and it exhibits a broad component and a narrow component at both epochs. Remarkably, the \ha\ and \hb\ profiles do no exhibit high-velocity redshifted absorptions below the continuum at any rotational phase, contrary to what is often seen in other accreting sources as the funnel flow intercepts the LoS \citep[e.g.,][]{Bouvier07a, Alencar12, Alencar18}. The lack of inverse P Cygni profile signatures provides further support for a low inclination of the system \citep{Muzerolle01}, as independently derived for the stellar rotational axis and for the outer disk that both suggest $i\simeq$20-35\degr (see above and Paper I).

\begin{figure}
   \centering
   \includegraphics[width=0.48\hsize]{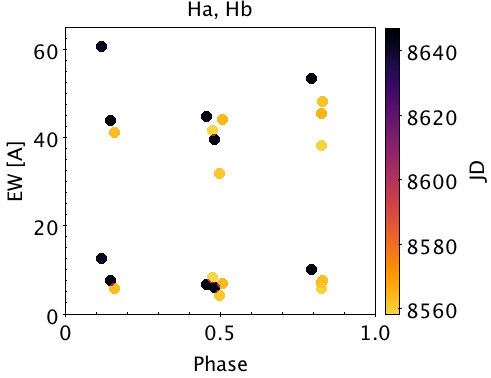}
   \includegraphics[width=0.48\hsize]{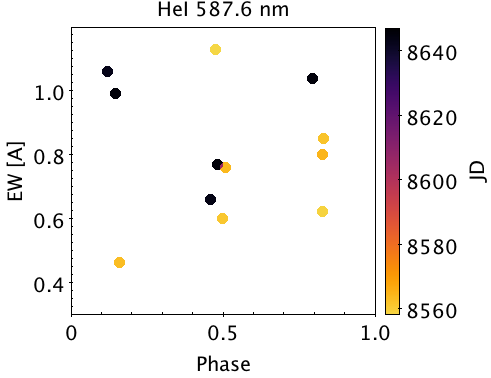}
   \caption{  Equivalent width (\AA) as a function of rotational phase. {\it Left panel:} \ha\ ({\it top}) and \hb\ ({\it bottom}); {\it right panel:} \hei. The color code corresponds to Julian date, mostly March (bright) and June (dark) 2019. }
              \label{ewphase}%
    \end{figure}

Equivalent width measurements for \ha, \hb, and \hei\ emission lines are reported in Table~\ref{ewtab}. As the continuum is well defined around these lines, the equivalent width measurement is accurate to within a few percent. The median equivalent widths over the two epochs amount to 44, 6.8, and 0.8~\AA, for \ha, \hb, and \hei\ lines, respectively. The lines have slightly larger median equivalent widths in June than in March. Their variation as a function of rotational phase is shown in Fig.~\ref{ewphase}. There is a hint of rotational modulation in the equivalent width variations of the three line profiles during the June run, with minimum values close to a rotational phase of 0.5, while no such modulation is seen in the March run. We estimate the mass accretion rate onto the star from the flux measured in the \ha\ and \hb\ lines, using relationships between accretion and line luminosity from \cite{Alcala14}. From the \ha\ line we thus derive \lacc/\lsun=0.06-0.13, taking into account \ewha\ variations, which yields \macc=4.0-8.5 10$^{-9}$ \msunyr\ assuming a magnetospheric truncation radius of 5~\rstar. Similarly, from the \hb\ line, we obtain \lacc/\lsun=0.05-0.17 and \macc=3-10 10$^{-9}$ \msunyr. These estimates are consistent with those reported by \cite{Manara14}, who obtained \macc=6.3 10$^{-9}$~\msunyr\  from slab models applied to XSHOOTER spectra.  

A periodogram analysis of the line profiles is detailed in Appendix A. The analysis does not reveal any significant periodicity in the intensity variations of the velocity channels across the \ha, \hb, and \hei\ line profiles when the full dataset is analyzed. This presumably results from the large temporal gap between the March and June ESPaDOnS runs. As described in Appendix~A, when each run is analyzed independently, we do find some evidence for the rotational modulation of the \ha\ and \hb\ line profiles at the stellar rotation period. More specifically, the intensity of the red wing (from about +50 to +150 \kms) of the \ha\ and \hb\ profiles appears to be modulated at the stellar rotation period during the March run. Since the veiling is nearly constant at these dates, we ascribe these variations to changes in the intrinsic line flux, as opposed to continuum variations. This result is of limited significance, due to the poor rotational phase sampling and the low number of spectra obtained during each run. Nevertheless, the examination of the shape of the line profiles plotted as a function of rotational phase in Fig.~\ref{profmarchjune} tends to support rotational modulation at both epochs, with a systematically depressed line flux in the red part of the profile around phase 0.5 for Balmer lines, as expected from accretion funnel flows as they cross the LoS \citep[e.g.,][]{Symington05}. 

\subsubsection{Line profile decomposition}

We further analyzed the line profiles attempting to reproduce their shape as a combination of Gaussian components. Details on the line profile decomposition are given in Appendix~B, and a few examples are illustrated in Fig.~\ref{decomp}. The Balmer line profiles are well fit with three Gaussian components, consisting of a broad central emission peak and two absorption components, one blueshifted, and the other redshifted, relative to the line center. The \hei\ line profile can be reproduced using only two emission components, a broad and a narrow one. The results are given in Table~\ref{decomptab}, which lists the equivalent width, peak intensity, radial velocity, and FWHM of each component.  

   \begin{figure}[t]
   \centering
   \includegraphics[width=0.49\hsize]{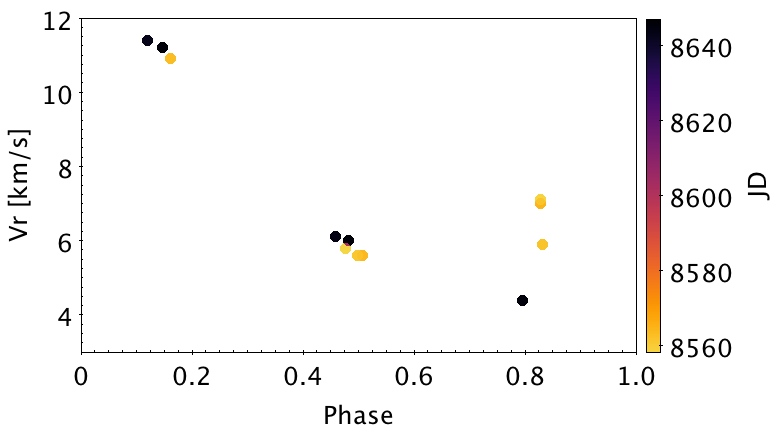}
      \includegraphics[width=0.49\hsize]{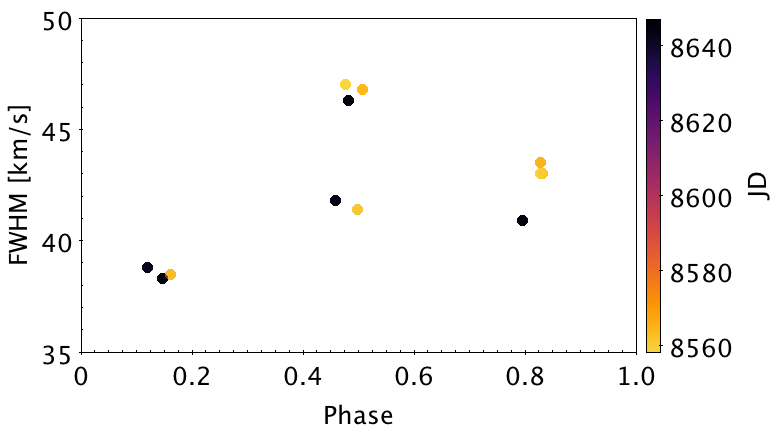}
         \includegraphics[width=0.49\hsize]{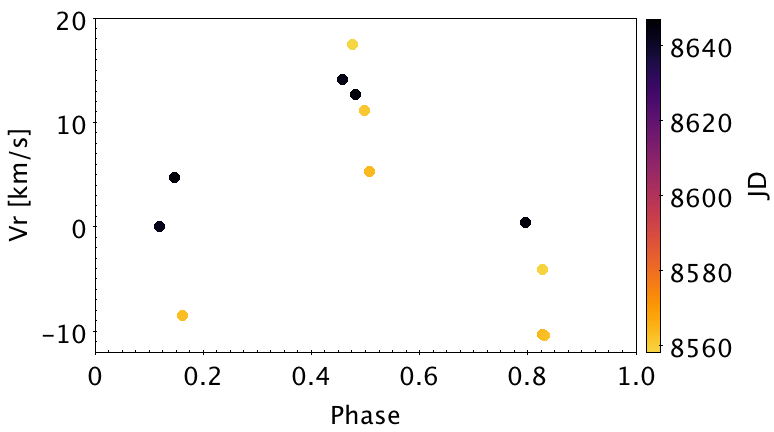}
   \includegraphics[width=0.49\hsize]{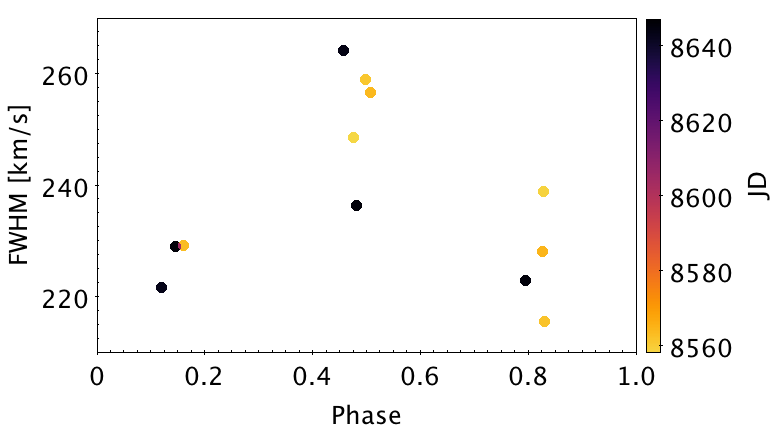}
   \caption{Radial velocity ({\it left}) and FWHM ({\it right}) variations of the narrow ({\it top}) and broad ({\it bottom}) components of the \hei\ line profile in the stellar rest frame as a function of rotational phase. The color code reflects the Julian date for the March (bright) and June (dark) ESPaDOnS runs.}
              \label{heivrad}%
    \end{figure}

The narrow component of the \hei\ line profile shows significant radial velocity variations between +4.4 and +11.4 \kms\ in the stellar rest frame, while its FWHM varies over the range 38-47~\kms. We find these variations to be modulated by stellar rotation with a strong phase coherence between the March and June ESPaDOnS runs (see Figure~\ref{heivrad}). Assuming that the \hei\ line narrow component arises from the accretion post-shock (Beristain et al. 2001), as suggested by its slightly redshifted velocity, and covers a small area on the stellar surface, one can relate the radial velocity amplitude $\delta$\vrad\ to the hot spot latitude $\theta$, with $\delta V_r = 2 \cdot $\vsini$ \cdot \cos \theta$ \citep{Bouvier07a}. We deduce a latitude of 78$\pm$2\deg\ for the accretion shock at the stellar surface, assuming an uncertainty of 0.5~\kms\ on $\delta$\vrad. 

Interestingly, the radial velocity of the wide component of the \hei\ line profile seems to be rotationally modulated as well, albeit with increased scatter. If this component arises from a hot wind close to the star, the modulation may result from the geometric projection of the wind velocity pattern onto the LoS. However, its slightly redshifted velocity relative to the star instead suggests it arises from the accretion funnel flow, located above the post-shock region probed by the narrow component \citep{Beristain01}. 
The FWHM of the narrow and wide components appears to be rotationally modulated as well. They reach a maximum at phase 0.5 of 259 \kms\ for the broad component and 47 \kms\ for the narrow one, and both vary by about 15\% along the rotational cycle. This variation could conceivably be related to the projection on the LoS of the velocity gradients of the emitting regions responsible for the narrow and broad \hei\ components. The vastly different FWHM of the narrow and broad components, and the fact that the radial velocities of both components appear to be modulated but reach a maximum value at different rotational phases (see Fig.~\ref{heivrad}), suggest the two components probe different regions along the funnel flow. 

Considering the two absorption components in the \ha\ profile, one being blueshifted, the other redshifted, we find that their FWHMs are rotationally modulated and strongly anti-correlated (see Table~\ref{decomptab}). The equivalent width and peak absorption intensity of the blueshifted component are the largest around phase 0, while the equivalent width of the redshifted absorption component is the largest around phase 0.5. As the former is probably related to an outflow, while the latter probes accretion, this suggests that both flows are physically related and occur at opposite rotational phases. We also notice a hint of a correlation between the radial velocity variations of these two absorption components in the \ha\ line profile, as reported earlier for AA Tau \citep{Bouvier03} and LkCa 15 \citep{Alencar18}.

As illustrated in Fig.~\ref{decomp}, the decomposition of the CaII 854.2 nm emission line reveals the presence of a highly redshifted absorption component during the March run, around \vrad$\simeq$+120 \kms. It is observed at phases 0.83 (JD 8559.15) and 2.83 (JD 8565.07), but, surprisingly enough, not at phase 1.83 (JD 8562.12). Similar high-velocity redshifted absorptions are also seen in the NaI D line profiles at the same phases. This suggests transient accretion episodes onto the stellar surface or, at least, mass accretion rate variations occur within the funnel flow. No redshifted absorption components are seen in this line profile during the June run. 
 
 \subsubsection{Correlation matrices}

 The observed emission lines of DoAr~44 are highly variable, with emission
and absorption components superposed, as can be seen in Fig~\ref{hahbover}.
The different components may come from various circumstellar regions,
like the chromosphere, the accretion funnel, and the wind, which can vary
over different timescales. To investigate how the variations across a line 
or between different lines are related, we calculated correlation matrices
consisting of 2D plots of the linear correlation coefficients
of the line profile intensity variations at different velocity bins.
A good correlation between two components indicates a common origin
of the components, for example.

\begin{figure}[t]
   \centering
   \includegraphics[width=0.48\hsize]{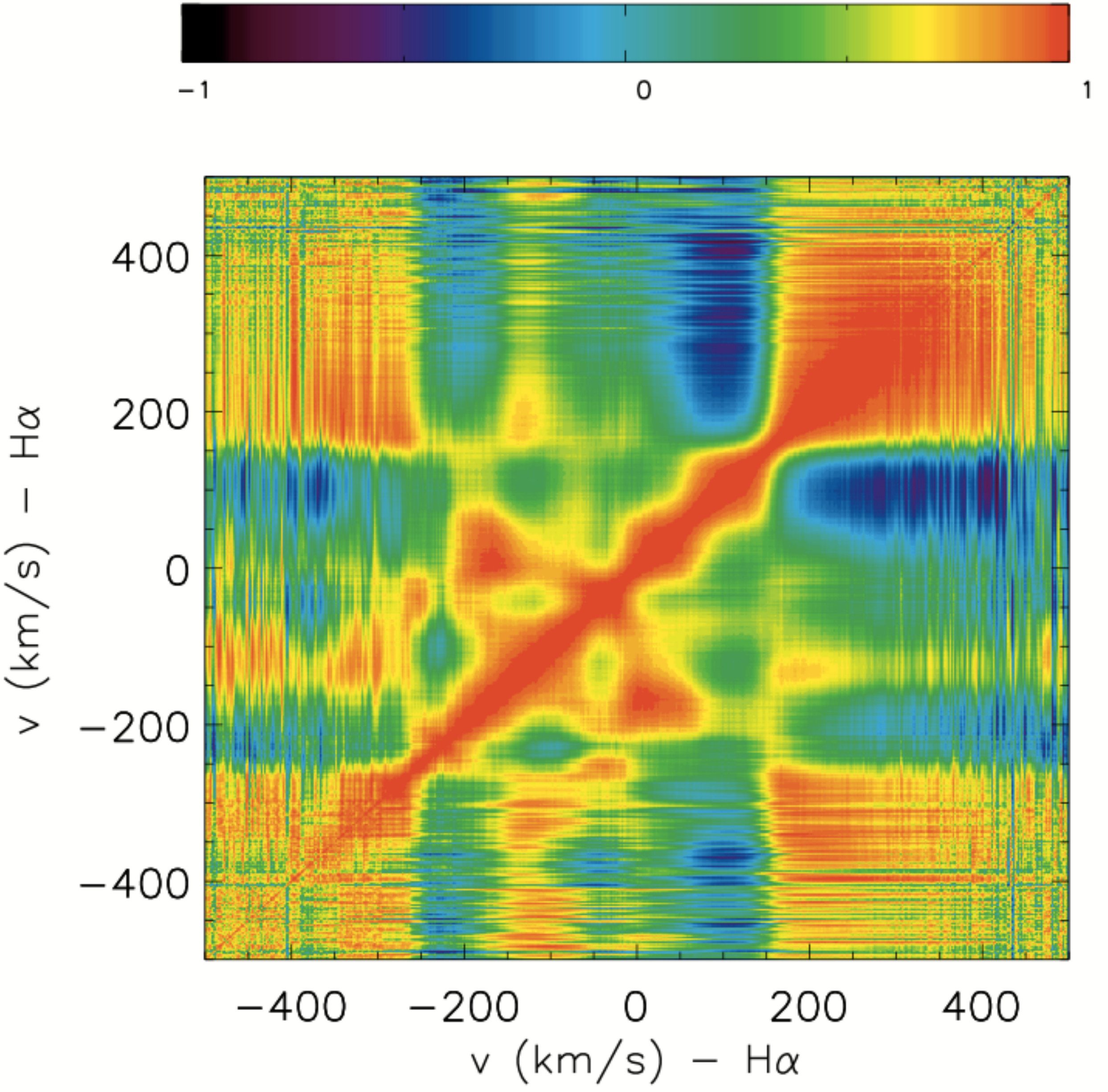}
      \includegraphics[width=0.48\hsize]{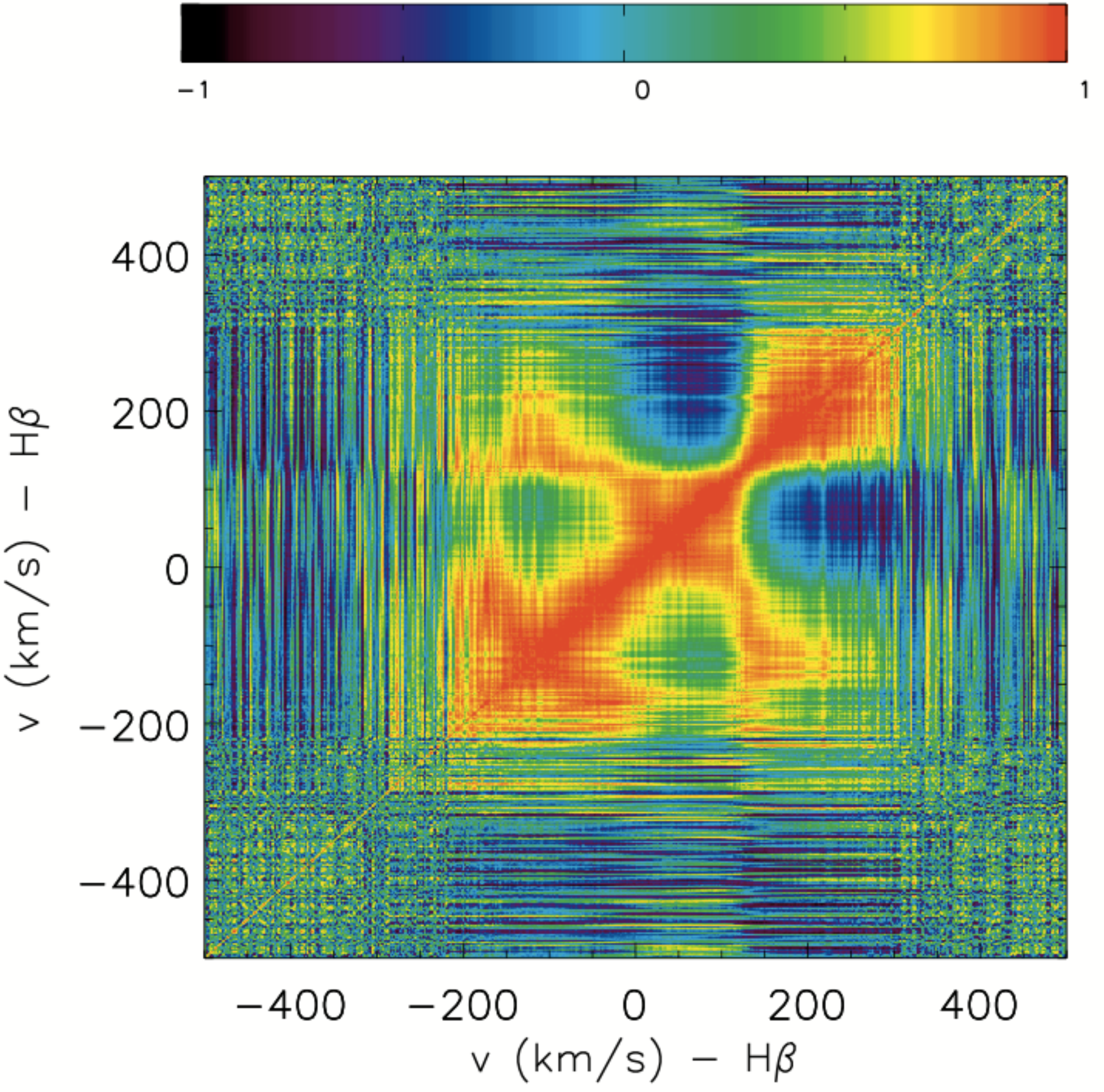}
         \includegraphics[width=0.48\hsize]{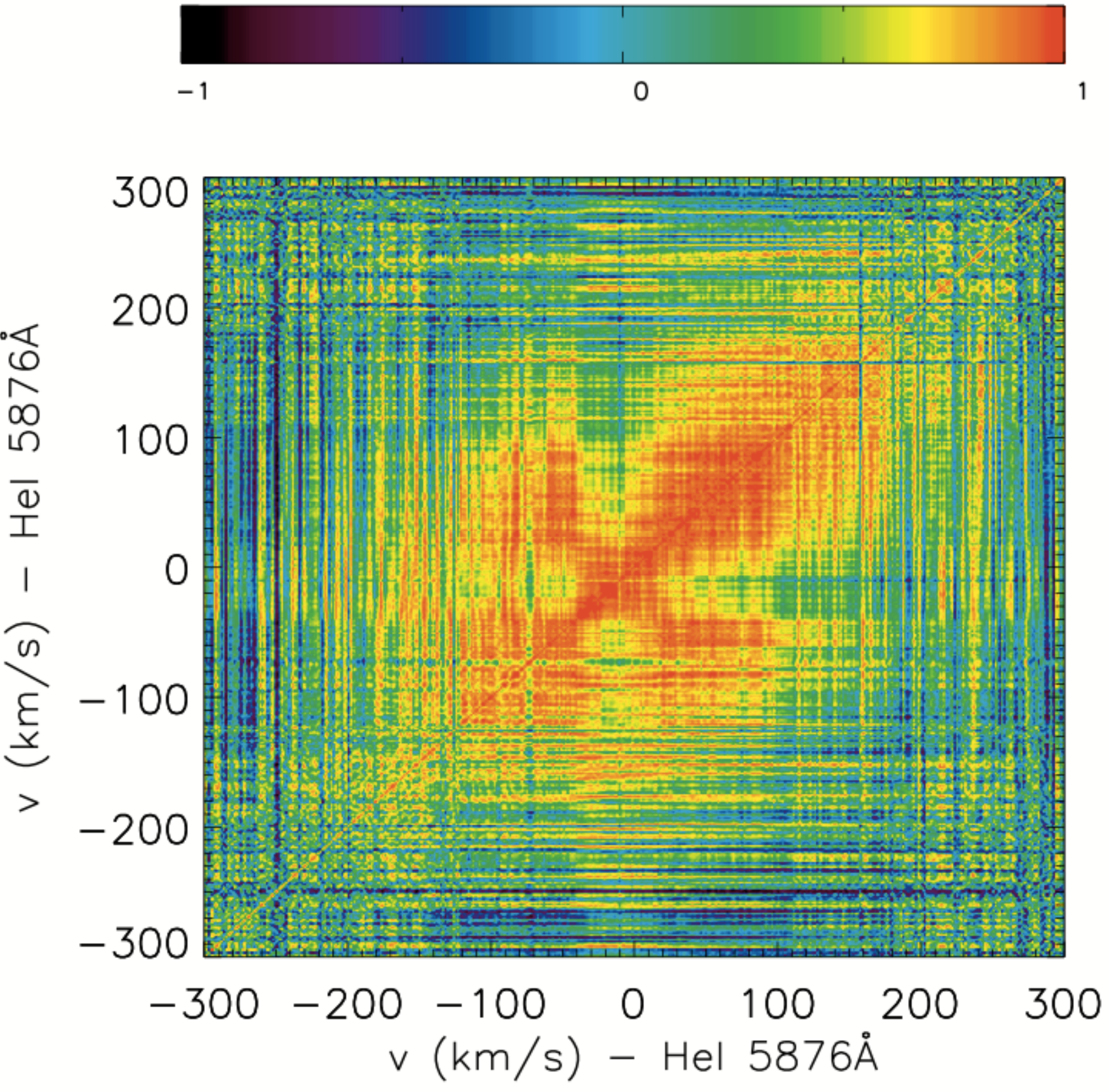}
   \includegraphics[width=0.48\hsize]{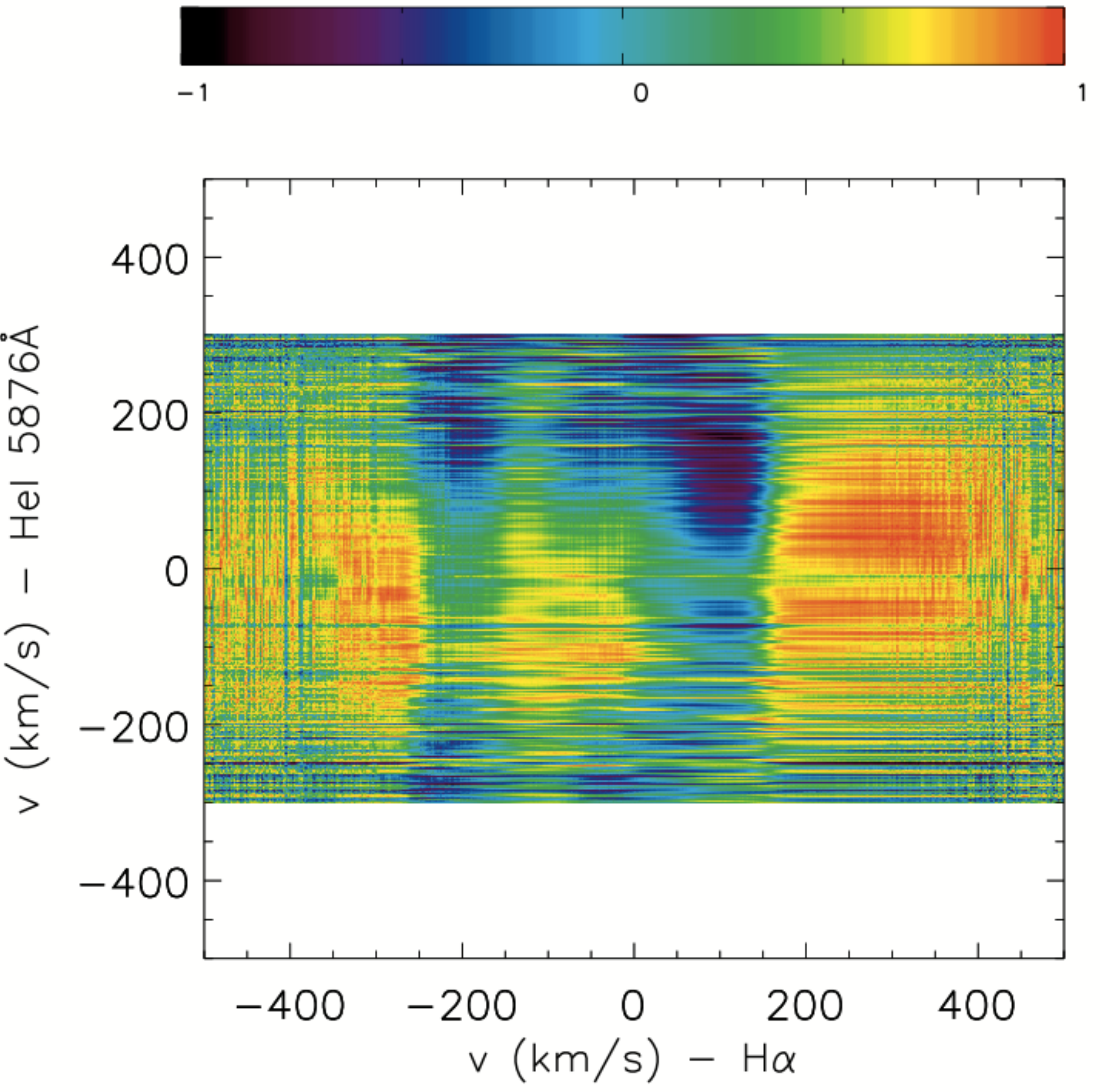}

   \caption{{\it Top:} auto-correlation matrices of \ha\ ({\it left}) and \hb\ ({\it right}) line profile variation for ESPaDOnS March run. {\it Bottom:} auto-correlation matrices of \hei\ ({\it left}) and cross-correlation matrices of \hei\ vs. \ha\ ({\it right}).}
              \label{cormat}%
    \end{figure}

The H$\alpha$ and H$\beta$ lines are similar, and their correlation matrices 
also show the same characteristics (Fig.~\ref{cormat}). There is little correlation between the blue and red part of the profile, which suggests different physical processes at the origin of their variability. The red wing of the profile varies coherently beyond +200~\kms\ and is strongly anti-correlated with the red emission peak around 100 \kms. This red emission peak is strongly variable, as can be clearly seen in Figs.~\ref{hahbover} and \ref{hahbphase}. As the velocity channels corresponding to the peak appear to be modulated at the stellar rotation period, we interpret its anti-correlation with the profile's far red wing as the signature of periodic redshifted absorptions due to the funnel flow crossing the LoS. 

The \hei\ line 
is composed of a broad and
a narrow component.  The correlation matrix is shown in Fig.~\ref{cormat}. It indicates that the BC varies coherently and independently of the NC component. 
The \ha\ vs. \hei\ cross correlation matrix reveals that the \hei\ BC component is correlated with the high velocity wings of the Balmer profile beyond 200~\kms. The red wing of the \hei\ BC is, however, strongly anti-correlated with the red emission peak seen in the \ha\ profile around 100~\kms. This suggests that redshifted absorption in the \ha\ profile occurs as the BC of the \hei\ line profile becomes stronger, which is consistent with both the redshifted \ha\ absorption and the \hei\ BC emission originating from an accretion funnel flow crossing the LoS. 

\begin{figure*}
   \centering
   \includegraphics[width=0.33\hsize]{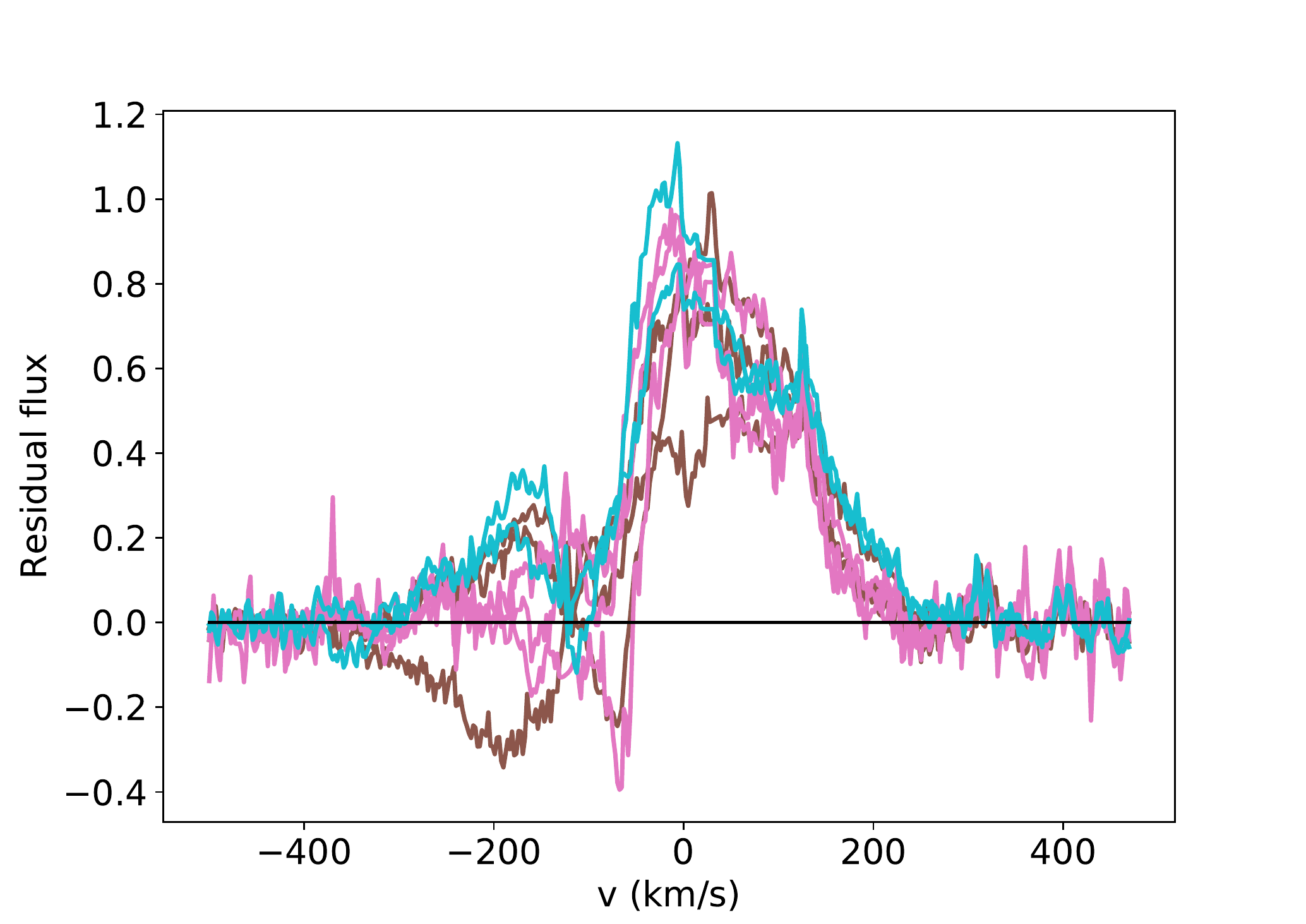}
      \includegraphics[width=0.33\hsize]{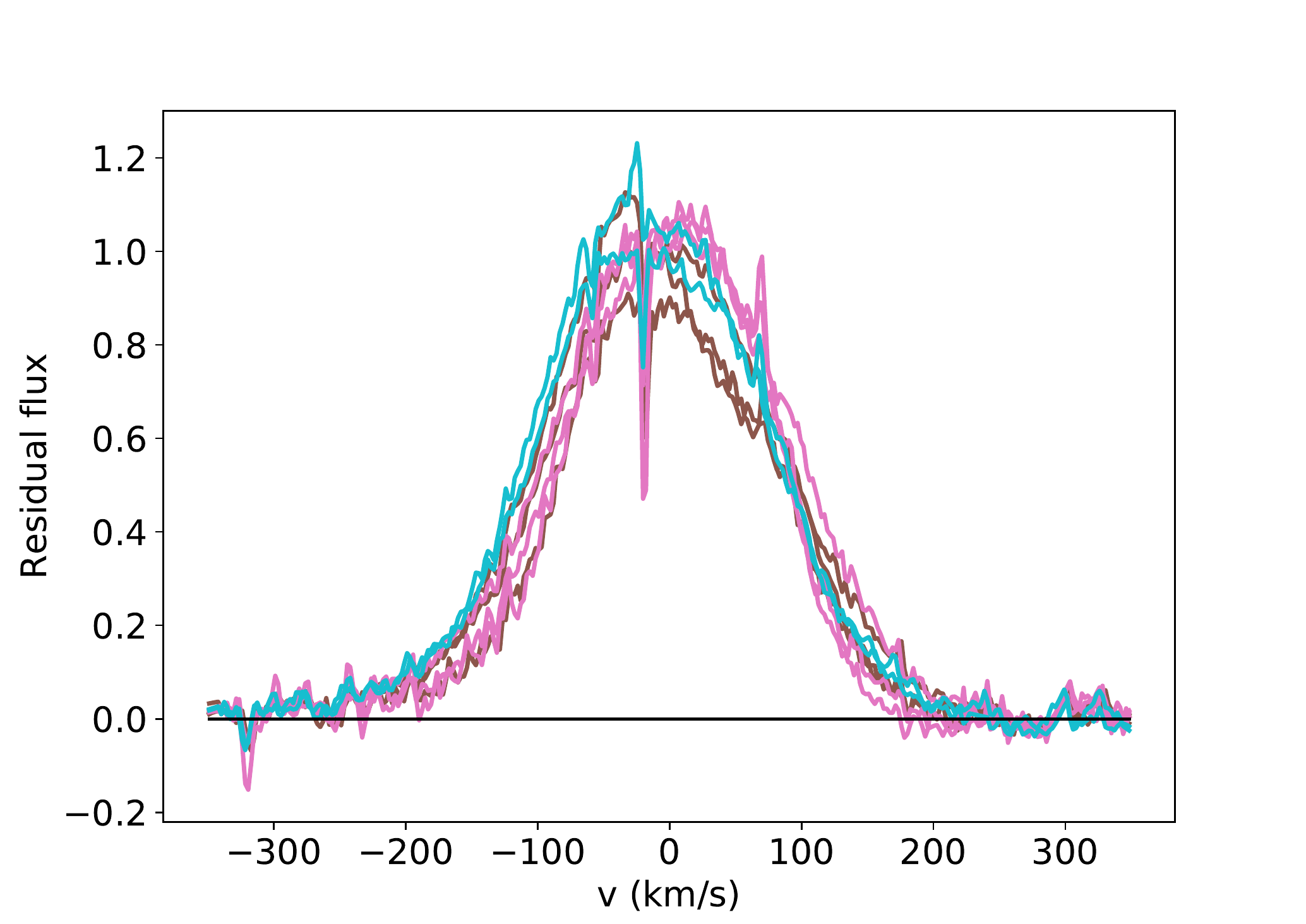}
            \includegraphics[width=0.33\hsize]{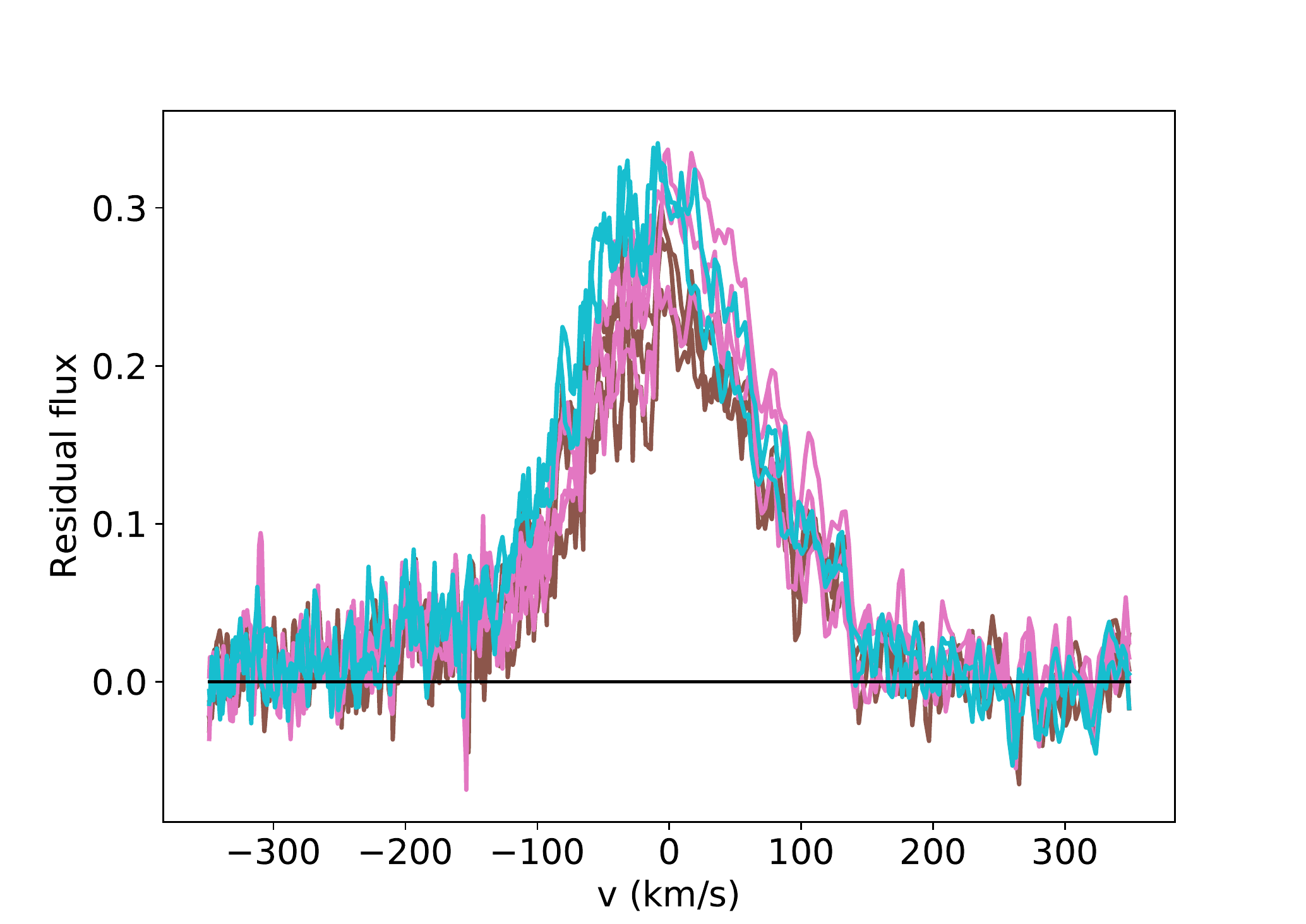}
   \caption{\heiir\ 1083.3~nm ({\it left}), \pab\ ({\it middle}), and \brg\  ({\it right}) residual line profiles. The narrow, slightly blueshifted absorption component close the center of the \pab\ line profile is due to imperfect telluric line correction. The color code corresponds to successive rotational cycles. }
              \label{irprof}%
    \end{figure*}
   \begin{figure*}
   \centering
   \includegraphics[width=0.33\hsize]{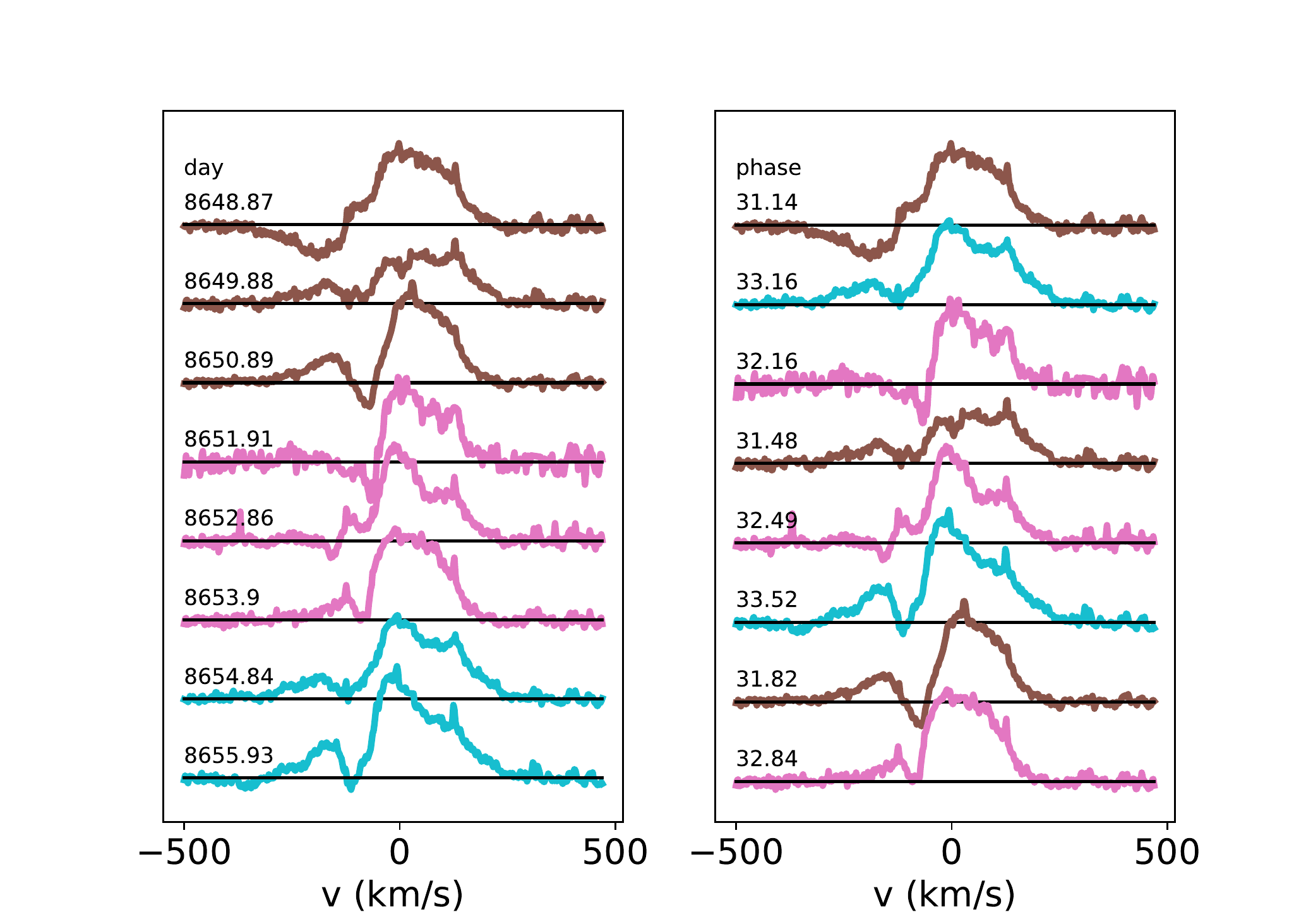}
         \includegraphics[width=0.33\hsize]{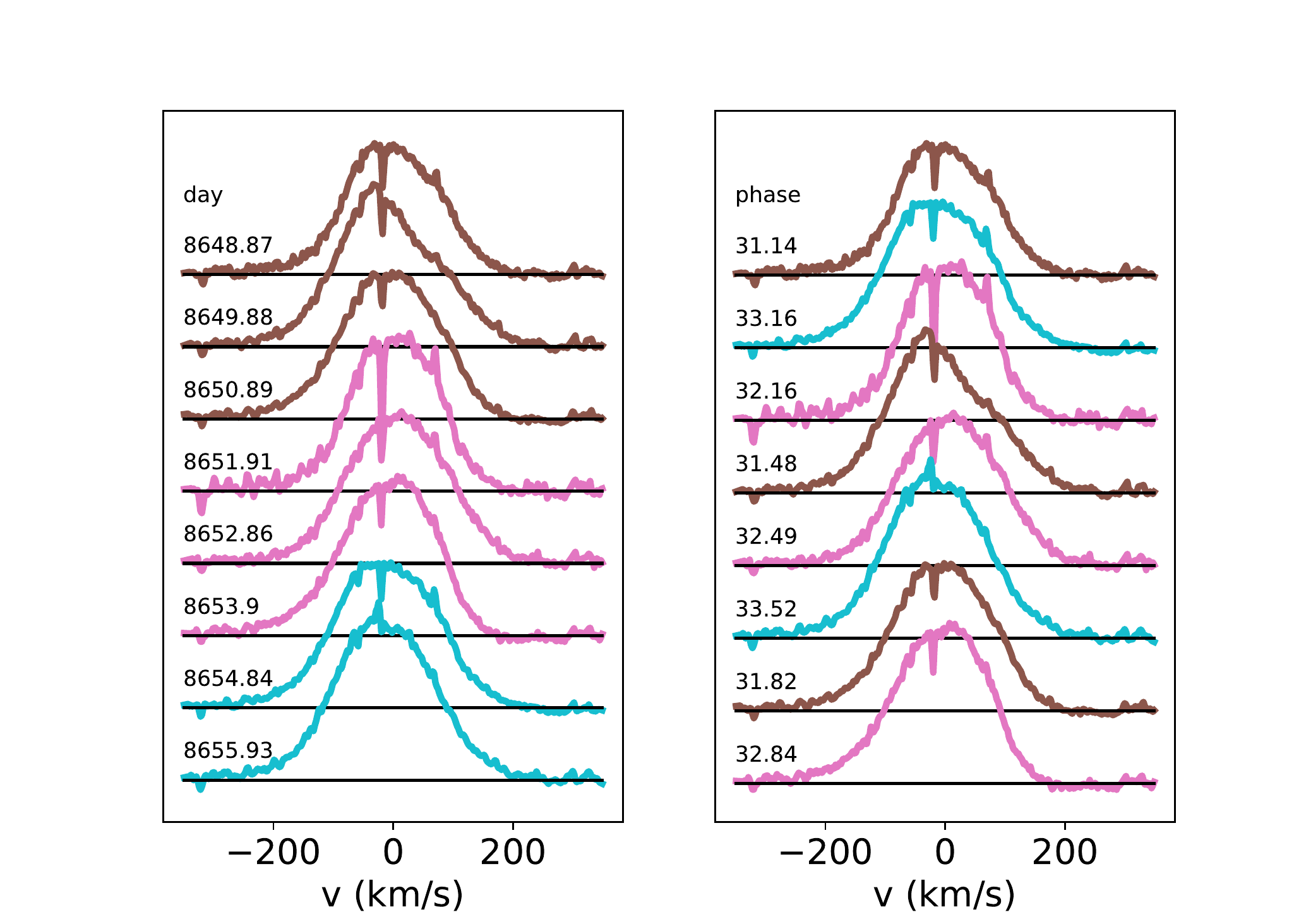}
                        \includegraphics[width=0.33\hsize]{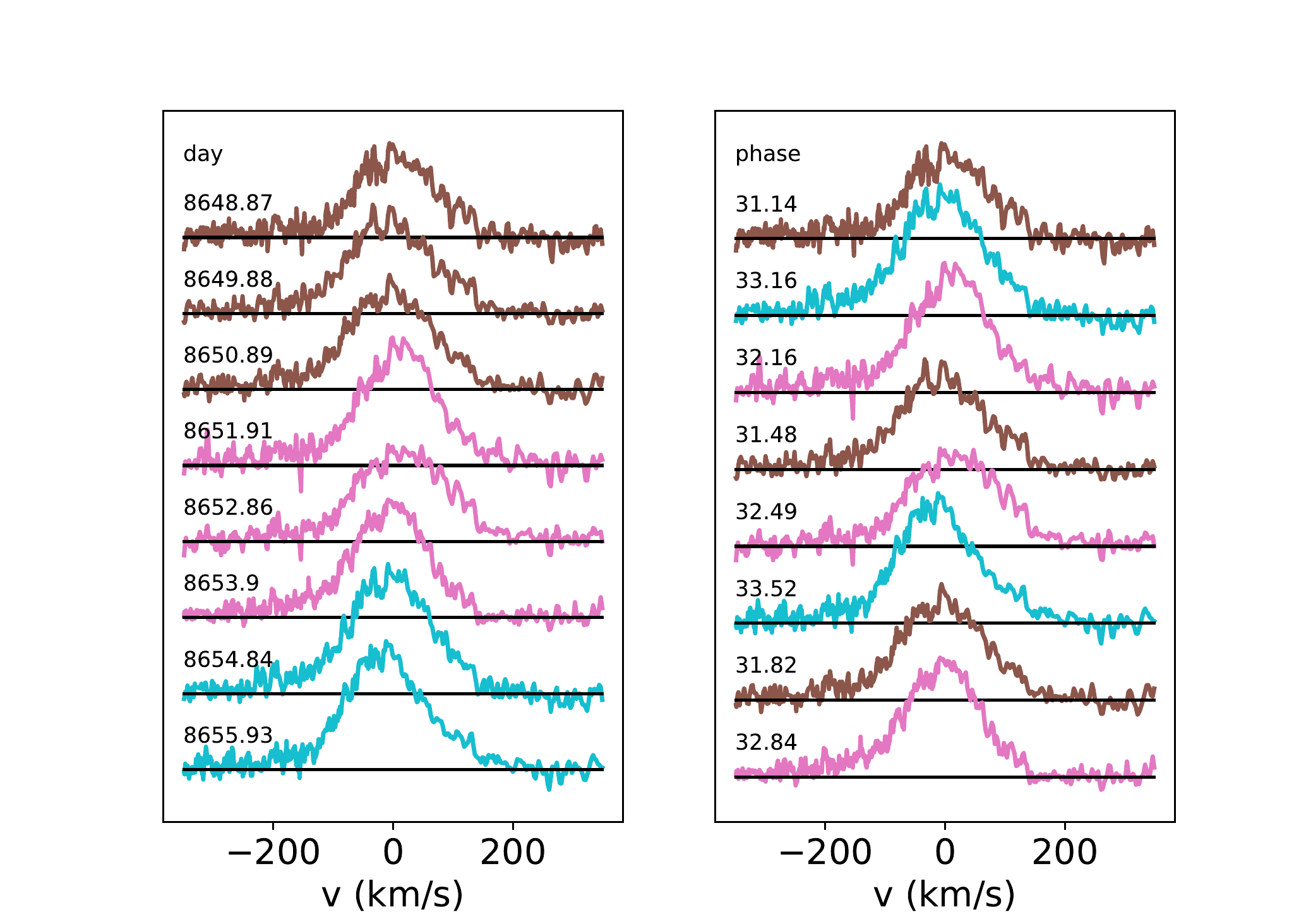}
   \caption{\heiir\ 1083.3~nm ({\it left}), \pab\ ({\it middle}), and \brg\  ({\it right}) residual line profile variability as a function of date and phase. The narrow, slightly blueshifted absorption component close to the center of the \pab\ line profile is due to imperfect telluric line correction. The color code corresponds to different rotational cycles.}
              \label{irphase}%
     \end{figure*}

\subsection{Infrared line profiles}

\begin{table}
\caption{Equivalent width (EW) of \hei\ 1083.3~nm, \brg, and \pab\ emission lines from SPIRou spectra. The uncertainty on EW is of the order of 0.1-0.2~\AA\ depending on the SNR. }             
\label{ewir}      
\begin{tabular}{c  c c c}        
\hline\hline                 
            HJD     &  EW(\hei)  & EW(\brg) & EW(\pab) \\
            (-2,450,000)  & (\AA) & (\AA) & (\AA)\\            
\hline                        
8648.86809 & 4.0 & 2.7 & 6.9 \\
8649.87590 & 4.7 & 3.0 & 8.6 \\
8650.89169 & 5.5 & 3.2 & 8.4 \\
8651.90703 & 4.0 & 3.4 & 8.1 \\
8652.85790 & 5.1 & 2.8 & 8.6 \\
8653.89859 & 6.8 & 3.3 & 8.5 \\
8654.84281 & 6.7 & 3.8 & 8.9 \\
8655.93314 & 8.3 & 4.0 & 9.3 \\
\hline\end{tabular}
 \end{table}

    \begin{figure}
   \centering
      \includegraphics[width=0.49\hsize]{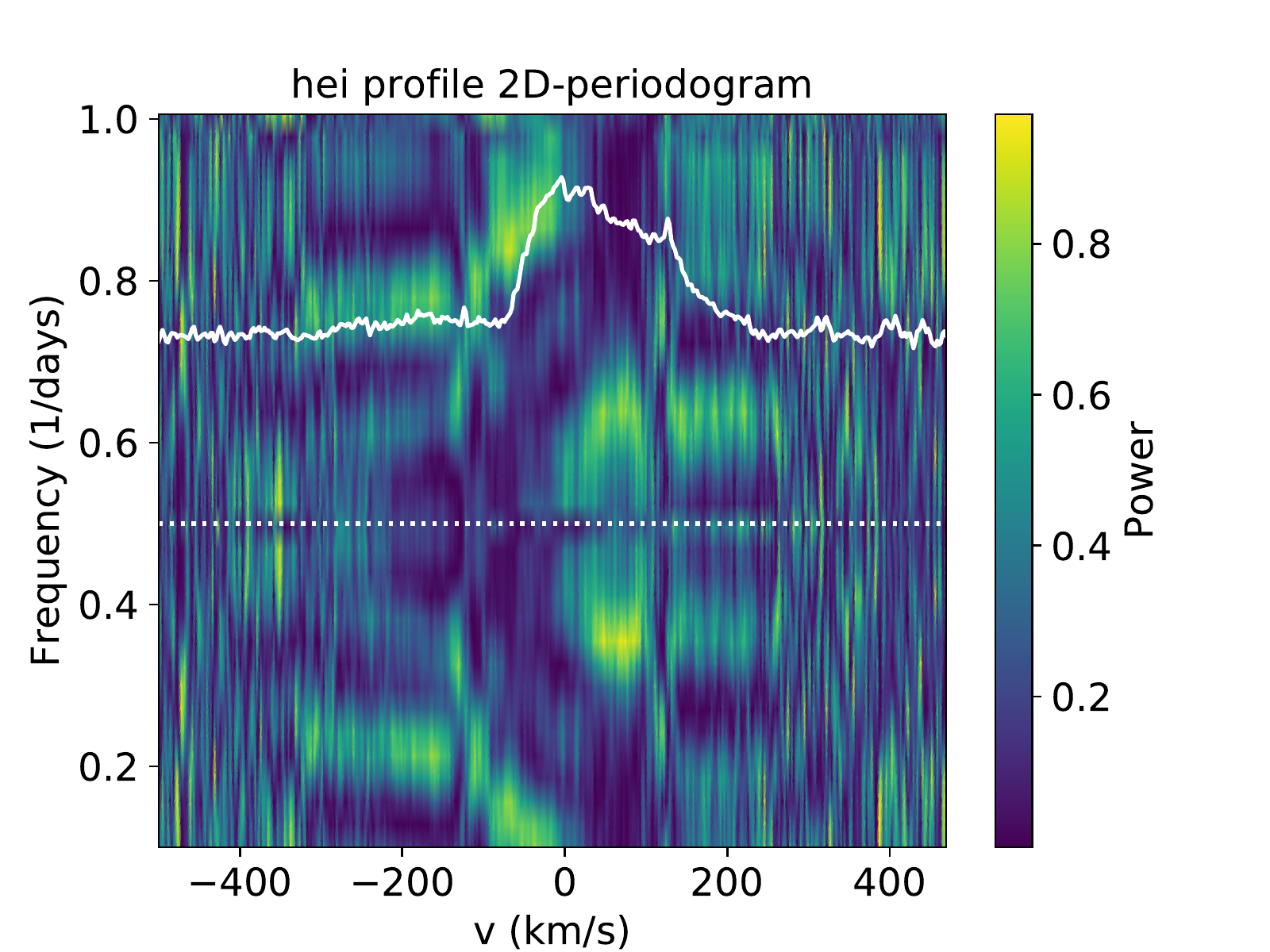}
         \includegraphics[width=0.49\hsize]{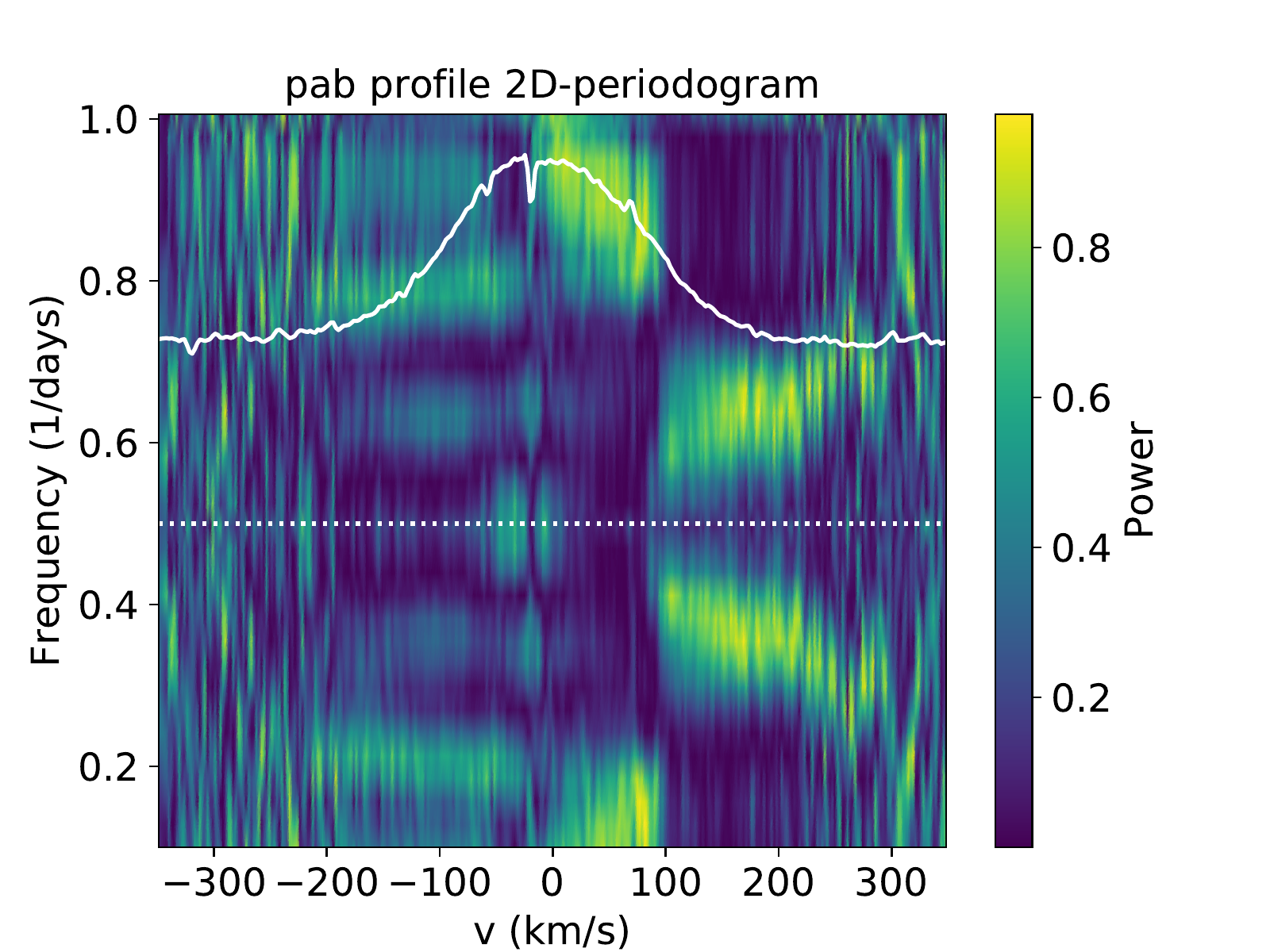}
   \caption{2D periodogram of the \hei\ 1083.3~nm ({\it left}) and \pab\ ({\it right}) line profiles computed from the eight SPIRou spectra obtained on consecutive nights. The color code represents the periodogram power, and the superimposed white curve is the average line profile. We note the symmetry of the periodogram around the frequency 0.5 d$^{-1}$ due to the one-day alias resulting from the night-to- night sampling. } 
              \label{perheipab}%
     \end{figure}
     
         \begin{figure}
   \centering
      \includegraphics[width=0.45\hsize]{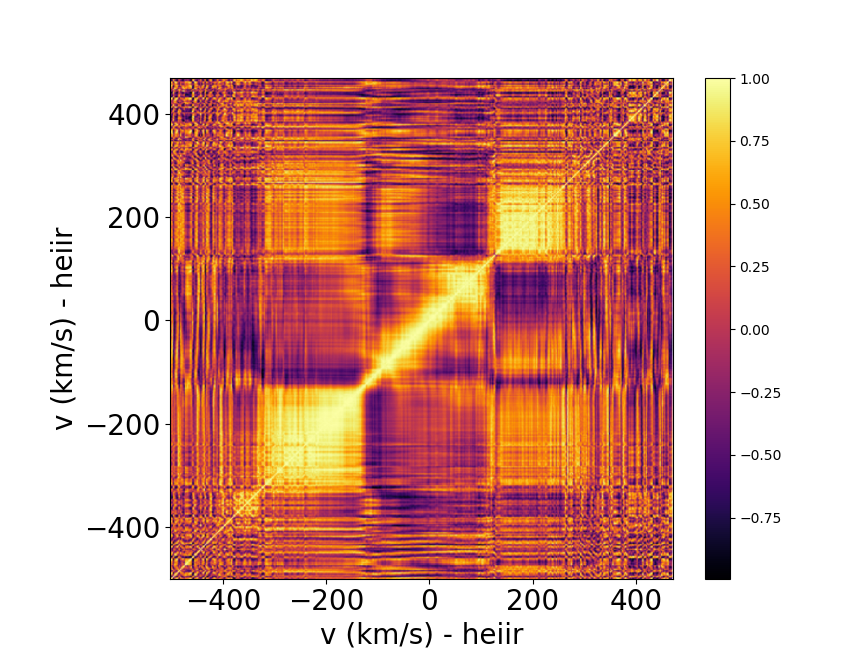}
                \includegraphics[width=0.45\hsize]{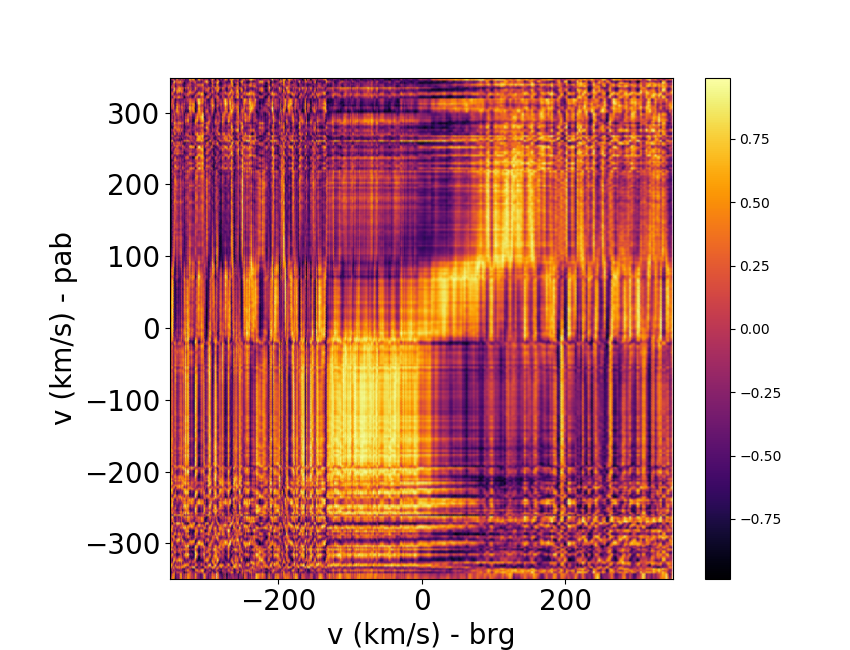}
         \includegraphics[width=0.45\hsize]{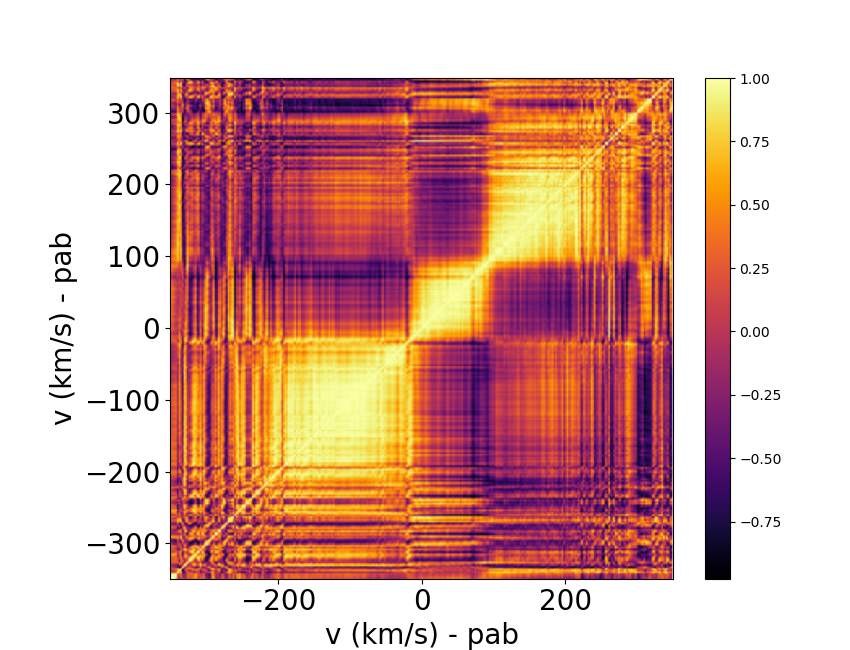}
                             \includegraphics[width=0.45\hsize]{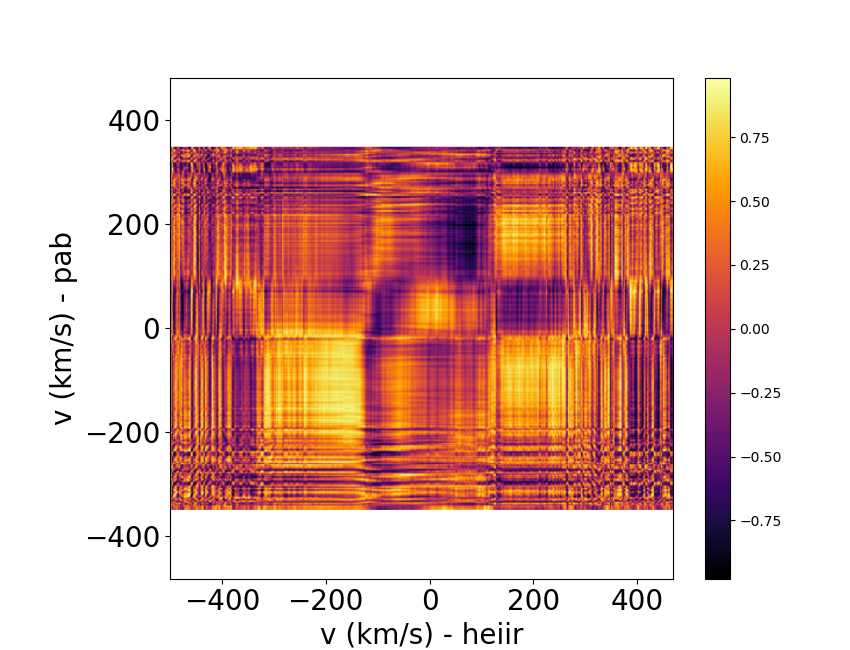}
         \includegraphics[width=0.45\hsize]{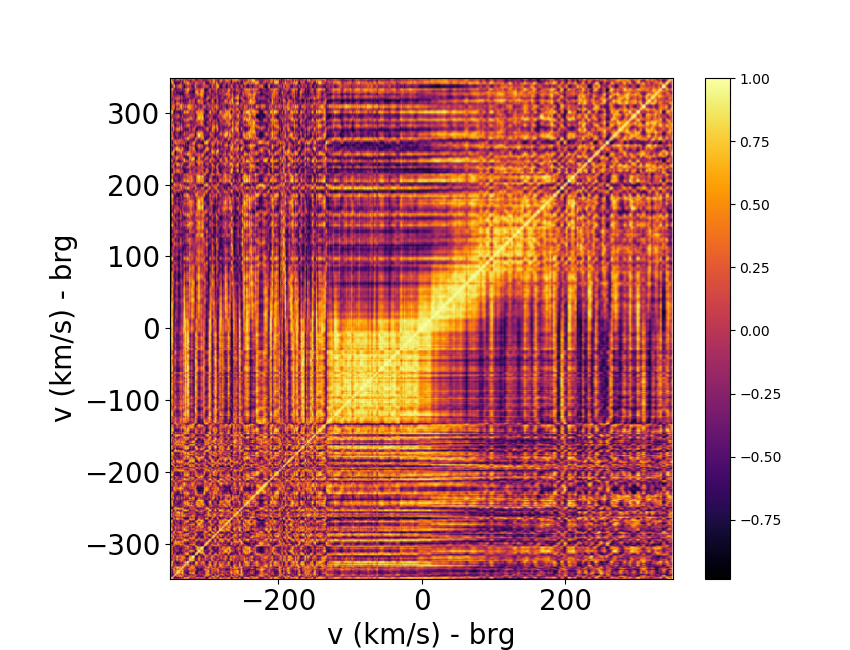}
          \includegraphics[width=0.45\hsize]{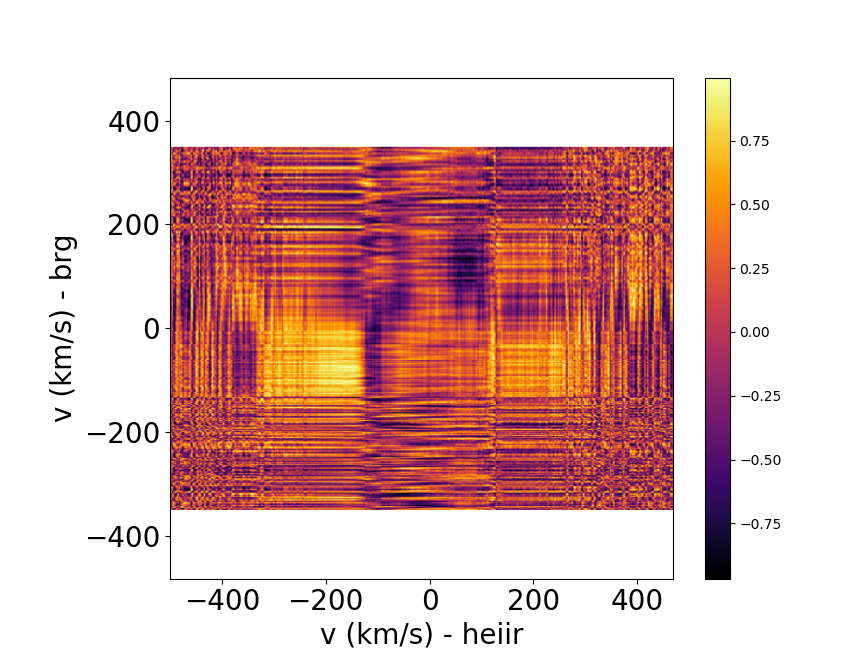}
   \caption{Auto-correlation matrices for the \hei, \pab, and \brg\  line profiles ({\it left}), and cross-correlation matrices for the three line profiles ({\it right}). } 
              \label{corrir}%
     \end{figure}

The wide spectral range covered by SPIRou includes a number of circumstellar diagnostics, such as \hei\ 1083.3~nm, \pab\ 1282.2~nm, and \brg\ 2166.1~nm. These lines are seen in emission in all the spectra acquired during the SPIRou observing run, and their equivalent widths are listed in Table~\ref{ewir}. From the equivalent widths of the \pab\ and \brg\ line profiles measured on the mean spectrum, EW(\pab) = 8.75 $\pm$ 0.15~\AA\ and EW(\brg) = 3.5 $\pm$ 0.2~\AA,  and using \cite{Alcala14} relationship between line and accretion luminosities, we derived a mass accretion rate \macc = 6.5 $\pm$ 0.5 10$^{-9}$ \msunyr, which is consistent with the estimate obtained above from the luminosity of the optical emission lines. 

Residual line profiles were computed following the same procedure as described above for optical spectra, using the weak-line T Tauri star V819 Tau (SpT K4) as a template, also observed with SPIRou. Line profiles are shown in Figure~\ref{irprof}. The \hei\ profile exhibits a strongly asymmetric shape, with significant blueshifted absorption components reaching down to velocities of -350~\kms\  with a depth that varies on a timescale of days, while the red wing appears in emission and also exhibits significant variability. In contrast, the \pab\ and \brg\ profiles are much more symmetric, with only low-level variability in intensity over the course of the observations. The profiles are plotted as a function of date and rotational phase in Fig.~\ref{irphase}.  

We performed a 2D periodogram analysis across the line profiles. The 2D periodogram of the \pab\ line profile is shown in Fig.~\ref{perheipab}. A strong signal appears at the star's rotational period in the redshifted wing of the profile over a velocity range extending from +100 to +200~\kms. Radiative transfer models of magnetospheric accretion onto an inclined dipole \citep[e.g.,][]{Kurosawa08} predict such a rotationally modulated intensity of the high velocity redshifted wing of the \pab\ profile, as funnel flows periodically cross the LoS. 
A strong signal is also seen in the 2D periodogram from the line center up to +100~\kms\ at frequencies around 0.9~d$^{-1}$, with its alias at 0.1~d$^{-1}$. The former frequency is close to the sampling rate, and the latter is beyond the observation time span. The signal at these frequencies is therefore probably spurious, being induced by the observing window. Nevertheless, the variability seen in this part of the \pab\ profile is real, and is also seen in the \brg\ line profile over the same velocity channels, where the line intensity seems to first increase then decrease smoothly over the time span of the observations (see Fig.~\ref{irprof}). 

The 2D periodogram of the \hei\ line profile displays the strongest signal at a frequency corresponding to the stellar rotation period, extending over redshifted velocity channels from +30 to +100 \kms\  (see Fig.~\ref{perheipab}). This periodic variability of the red part of the line profile is most likely induced by the rotational modulation of accretion funnel flow as the star rotates \citep[e.g.,][]{Fischer08}. The other parts of the \hei\ profile are extremely variable, especially around the line center and throughout the blueshifted wing down to velocities of -350~\kms. The blue wing exhibits a rapidly changing blueshifted absorption component sometimes reaching below the continuum. Yet, this outflow component shows no sign of modulation. Instead, we observe significant change in \hei\ line profiles taken at the same rotational phase over successive cycles, for example, at $\phi$=31.14, 32.16, 33.16,  and at $\phi$=31.48, 32.49 (see Fig.~\ref{irphase}), which suggests intrinsic variations of the outflow component over a timescale of days. The velocity of the peak of the blueshifted absorption component appears to be the most extreme at the beginning of the observations (v$_{blue}\simeq$ -200~\kms), shifting toward the line center up to v$_{blue}\simeq$  -60~\kms\  within three days, then returning to much bluer channels ($v_{blue}\simeq$ -130~\kms) over the same timescale. Its width changes as well, the most blueshifted absorption having the largest FWHM. 

Radiative transfer models assign the presence of moderately blueshifted, narrow absorption components in the profile of the \hei\ 1083.3~nm line to a disk wind \citep[e.g.,][]{Kurosawa11}, while broader absorptions extending to large blueshifted velocities are usually considered as the signature of accretion powered, hot inner winds originating close to the star \citep[e.g.,][]{Edwards03, Edwards06}.  We are apparently witnessing here a mix of disk and ``stellar'' winds, with a predominance of the former except for the first spectrum of the series, whose relative strength varies on a timescale of days. 

Indeed, the gallery of emission line profiles gathered from the quasi-simultaneous optical and near-IR high-resolution spectra obtained for DoAr 44 during this campaign, including \hei\ 587.6~nm, \hei\ 1083.3~nm, \ha, \hb, \pab, and \brg, bears strong resemblance with synthetic line profiles derived from the combined magnetospheric accretion and disk wind, low-inclination (i=20$\degr$) model presented in Fig.7 of \cite{Kurosawa11}. It is therefore likely that most of the emission seen in these profiles comes from magnetospheric funnel flows controlled by a misaligned dipole, with additional wind components best seen in the \hei\ near-IR line profile. The qualitative agreement between observations and models supports this interpretation. The low inclination of the system is further supported by the maximum edge velocity at which the blueshifted \hei\ absorption component meets the continuum, which amounts to -350~\kms. According to the edge velocity versus the system's inclination relationship reported for this component by \cite{Appenzeller13}, here the system's inclination should lie in the range 20-35$\degr$. 

The auto-correlation matrices of the three profiles are shown in Fig.~\ref{corrir} and share the same behavior: different parts of the line profile vary independently of each other. In the \pab\ line for instance, the blue wing varies independently of both the low-velocity and high-velocity redshifted  channels, and the latter also varies independently. The cross-correlation matrix between \pab\ and \brg\ reveals the same three components: the blue wing, the low-velocity red channels, and the high-velocity red channels. While each of these components varies independently of the other, they are well correlated between the \pab\ and \brg\ line profiles. 

A more complex behavior is seen in the cross-correlation matrices of \hei\ with \pab\ and \brg.   
The high-velocity blue wing of the \hei\ line is clearly correlated with that of the hydrogen lines. 
More surprisingly perhaps, there is a hint of a correlation, albeit relatively weak, between the high-velocity redshifted wing of \hei\ and the blueshifted wing of the \pab\ profile. Another feature is the apparent anti-correlation between the low-velocity redshifted \hei\ channels and the high-velocity redshifted hydrogen wings. This occurs over the interval of velocities where both profiles are rotationally modulated, namely around 30-100~\kms\ for \hei\ and beyond 100~\kms\ for \pab.  
If real, this anti-correlation is somewhat surprising as the redshifted absorptions in both profiles, though appearing at different velocities, are expected to occur simultaneously at the time the funnel flow crosses the LoS. This feature is also present in the autocorrelation matrices
of \hei\ and \pab. In fact, the \hei\ vs. \pab\ matrix closely mimics the structures seen in the \hei\ auto-correlation matrix. While these features are intriguing, the value of the correlation coefficient remains of the order of 0.5 or less, and would therefore have to be confirmed from a longer time series. 

\subsection{Zeeman-Doppler analysis}

\begin{table*}
\begin{center}
\caption{Longitudinal magnetic field strength (B$_l$) expressed in Gauss and computed from photospheric Stokes LSD profiles, from each line of the CaII IR triplet (CaII$_{1,2,3}$), and from the \hei\ 587.6~nm line appearing on two successive spectral orders (HeI$_{1,2}$).}
\begin{tabular}{lllllllllllll}
\hline
\hline
\multicolumn{13}{c}{Longitudinal magnetic field strength B$_l$ (G)}\\
\hline
HJD & LSD & $\sigma$ & CaII$_1$ &   $\sigma$ & CaII$_2$ & $\sigma$ & CaII$_3$ & $\sigma$ & HeI$_1$ &  $\sigma$ & HeI$_2$ &  $\sigma$\\
\hline
  8558.10852 & -119 & 14 & 270 & 80 & 290 & 70 & 310 & 90 & 890 & 640 & 280 & 620\\
  8559.15082 & -- & -- & 540 & 60 & 650 & 70 & 750 & 80 & 3020 & 470 & 4070 & 450\\
  8561.13665 & -122 & 14 & 310 & 60 & 300 & 70 & 370 & 90 & 2340 & 720 & 1210 & 630\\
  8562.12015 & -14 & 15 & 460 & 70 & 690 & 90 & 390 & 110 & 1780 & 880 & 420 & 910\\
  8563.09667 & -85 & 14 & 310 & 70 & 490 & 90 & 410 & 100 & 920 & 860 & 2700 & 770\\
  8564.12225 & -136 & 15 & 170 & 80 & 300 & 90 & 330 & 110 & 1380 & 640 & 2720 & 620\\
  8565.06818 & -50 & 14 & 480 & 60 & 550 & 60 & 520 & 70 & 2450 & 390 & 2480 & 370\\
  8642.8923 & -58 & 15 & 350 & 70 & 550 & 80 & 690 & 90 & 2110 & 380 & 1830 & 340\\
  8643.89558 & -123 & 14 & 60 & 70 & 140 & 90 & 140 & 110 & 400 & 520 & 2190 & 580\\
  8644.89415 & -43 & 14 & 490 & 70 & 790 & 80 & 730 & 90 & 2150 & 360 & 1860 & 350\\
  8645.93308 & -62 & 14 & 510 & 70 & 440 & 70 & 630 & 80 & 2280 & 470 & 1430 & 420\\
  8646.92779 & -116 & 14 & 340 & 80 & 310 & 80 & 320 & 120 & 1510 & 570 & 2500 & 610\\
\hline\end{tabular}
\end{center}
\end{table*}

  \begin{figure*}
   \centering
      \includegraphics[width=0.3\hsize]{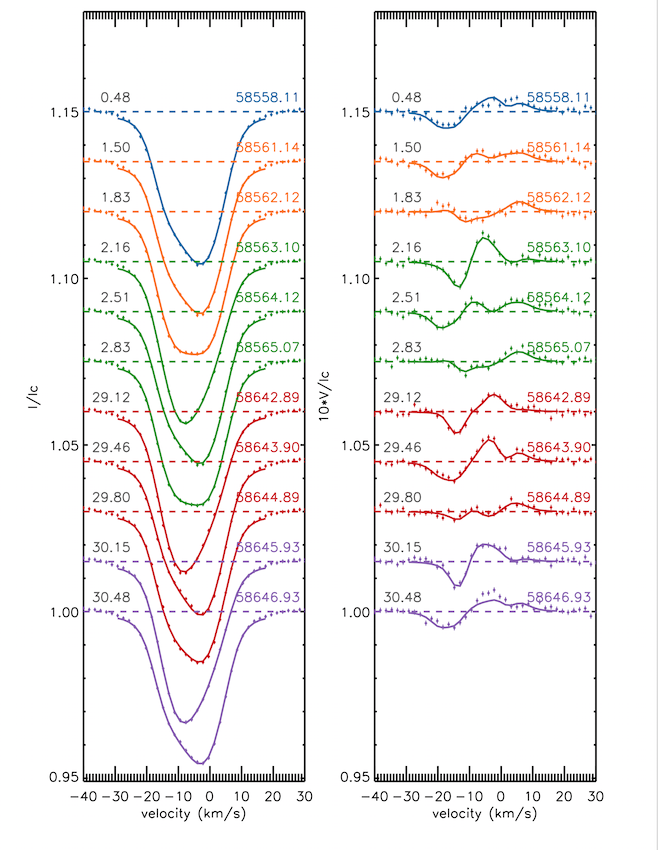}
   \includegraphics[width=0.3\hsize]{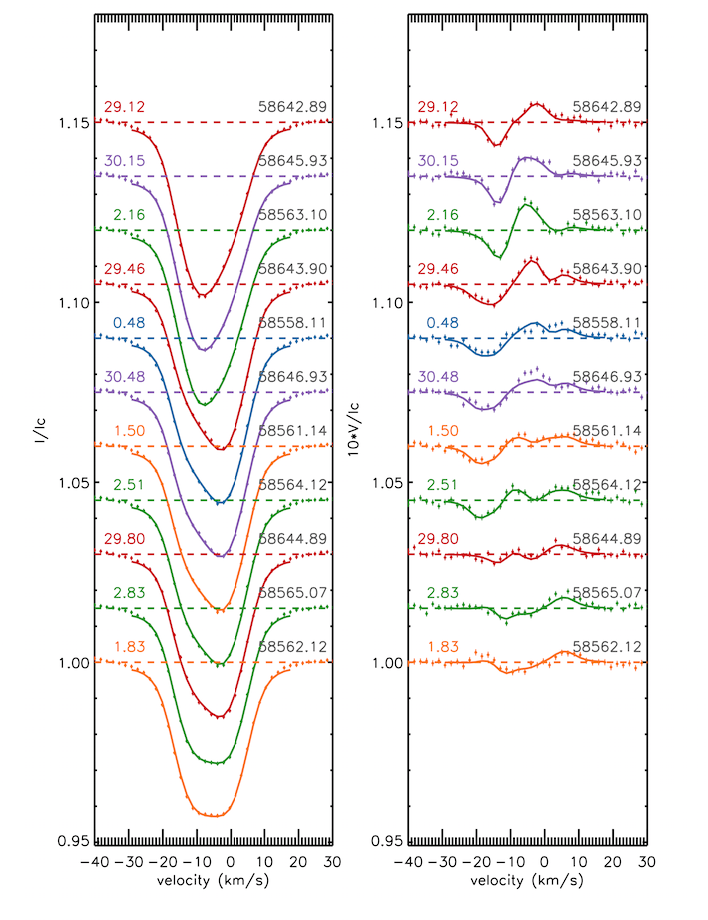}
      \includegraphics[width=0.3\hsize]{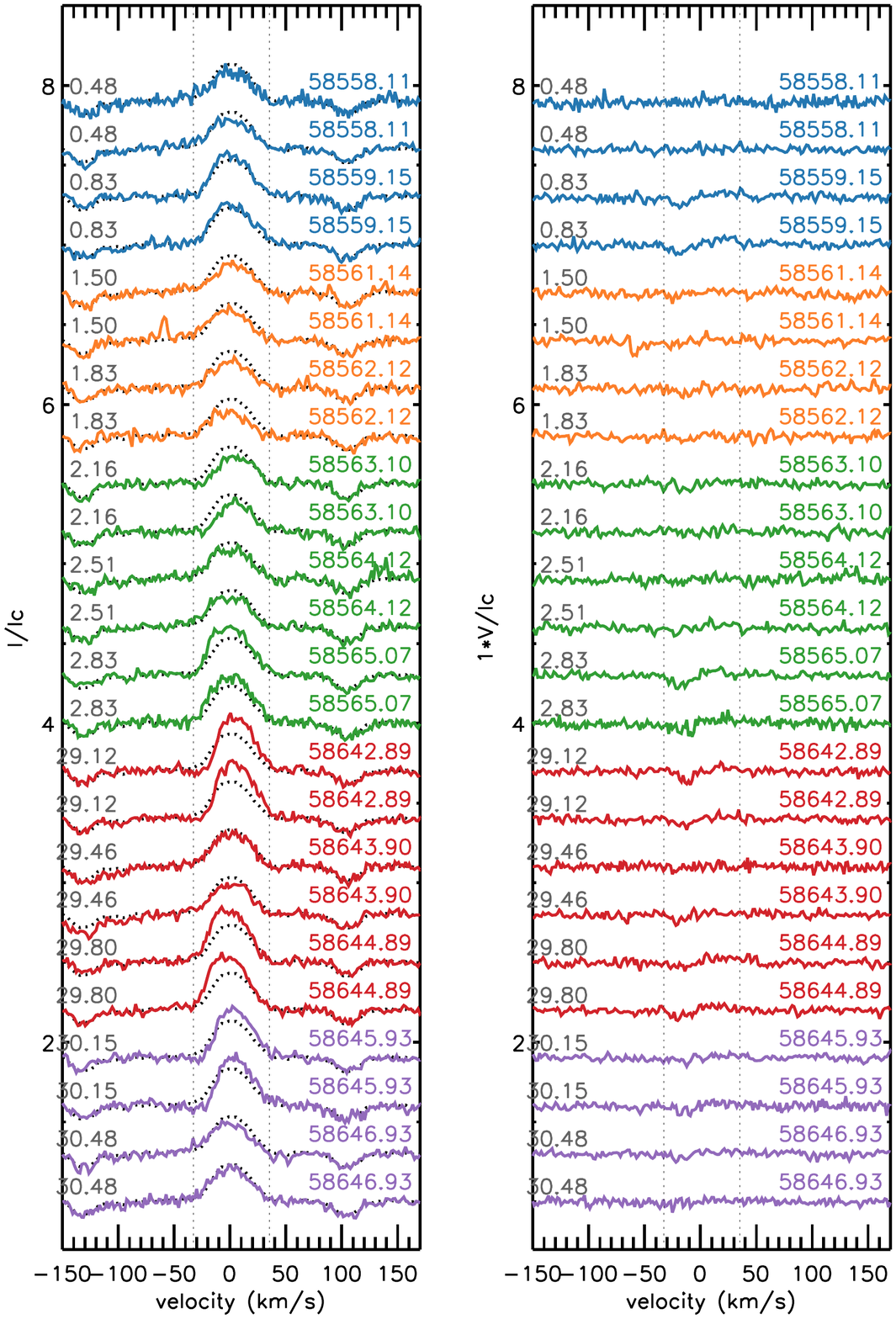}
   \caption{{\it Left:} Stokes I and Stokes V LSD profiles ordered as a function of Julian date. {\it Middle}: Stokes I and Stokes V LSD profiles ordered as a function of rotational phase. {\it Right}: Stokes I and Stokes V profiles of the \hei\ 587.6~nm line's narrow component ordered as a function of rotational phase. The broad component of the \hei\ line profile has been fitted by a Gaussian and subtracted. The color code corresponds to successive rotational cycles.   }
              \label{stokes}%
    \end{figure*}
   \begin{figure*}
   \centering
      \includegraphics[width=0.3\hsize]{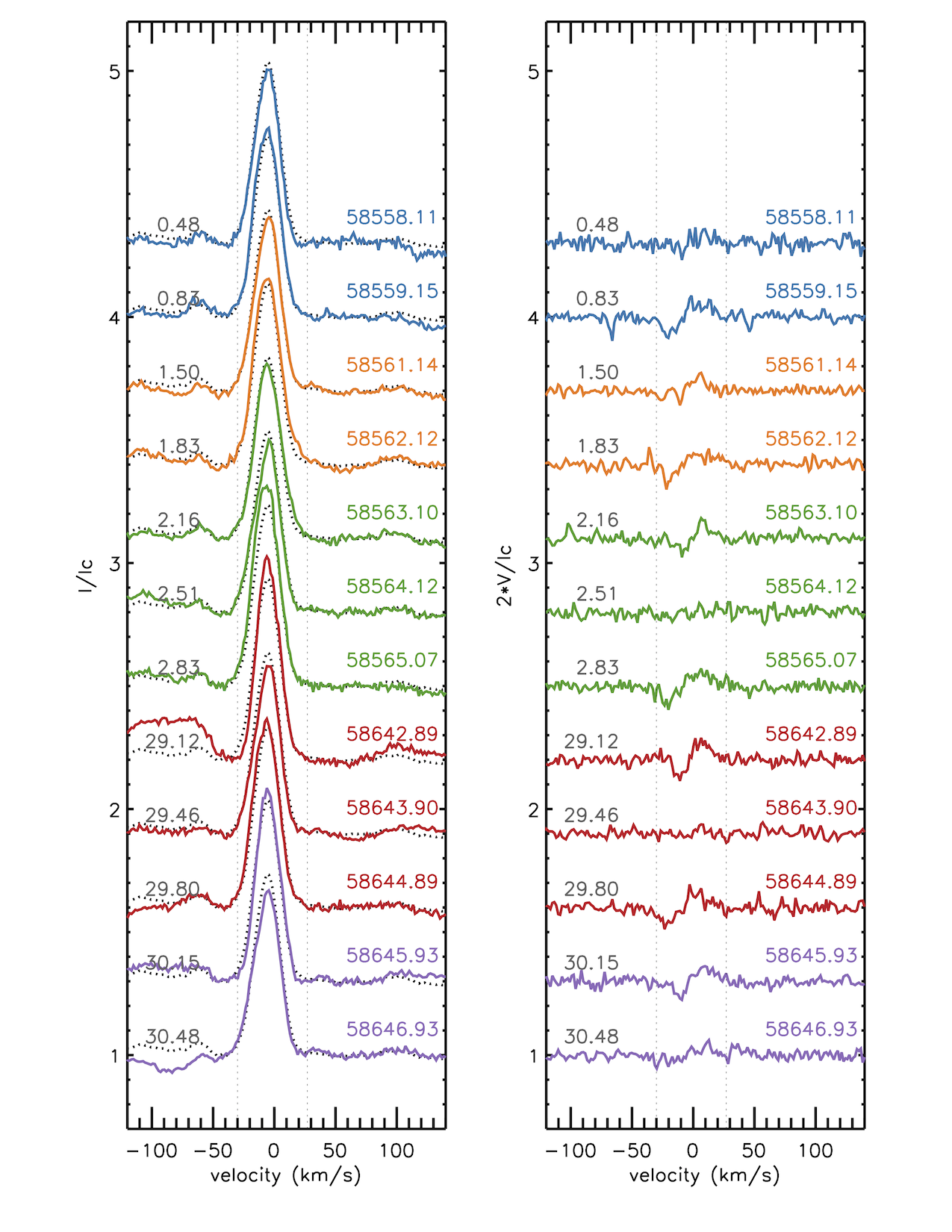}      
      \includegraphics[width=0.3\hsize]{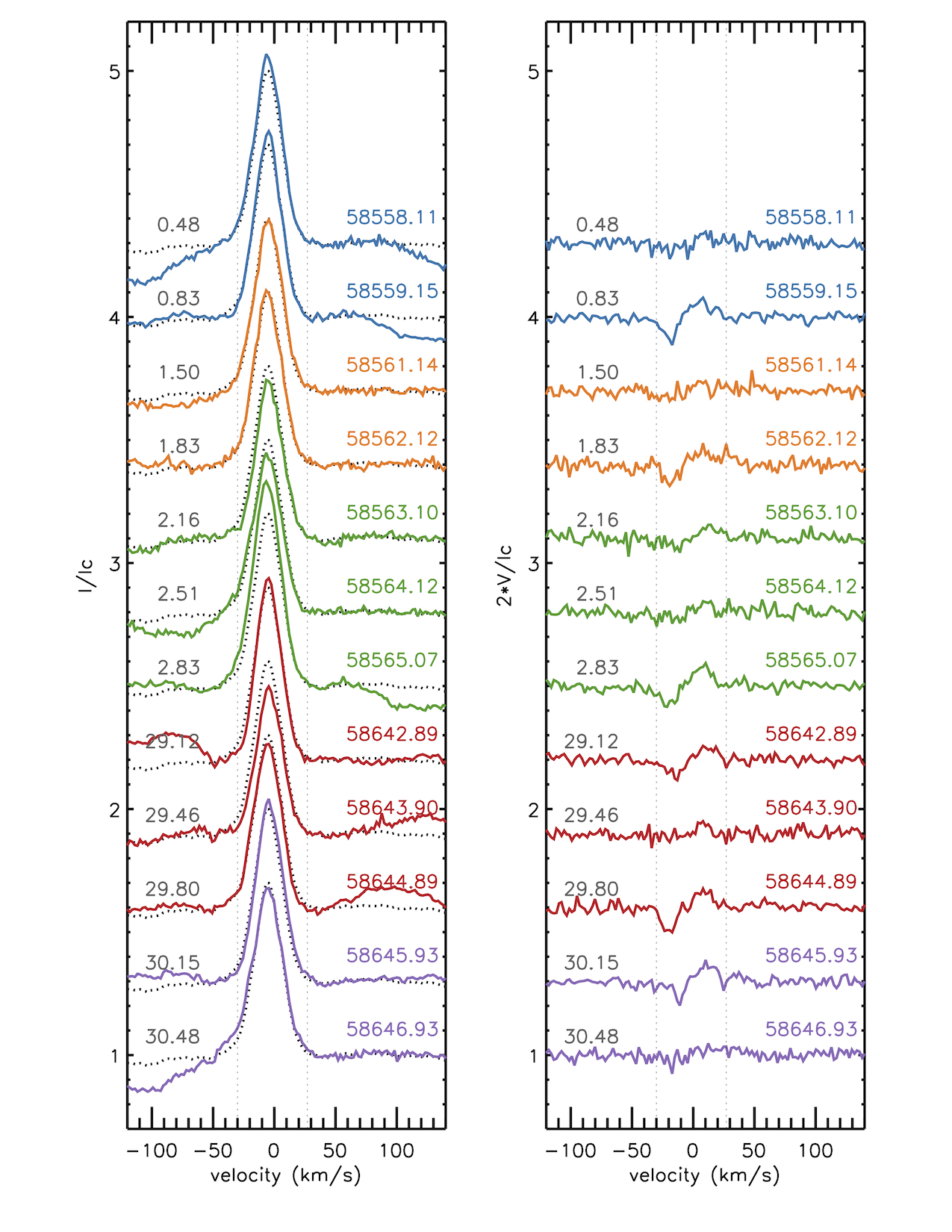}      
      \includegraphics[width=0.3\hsize]{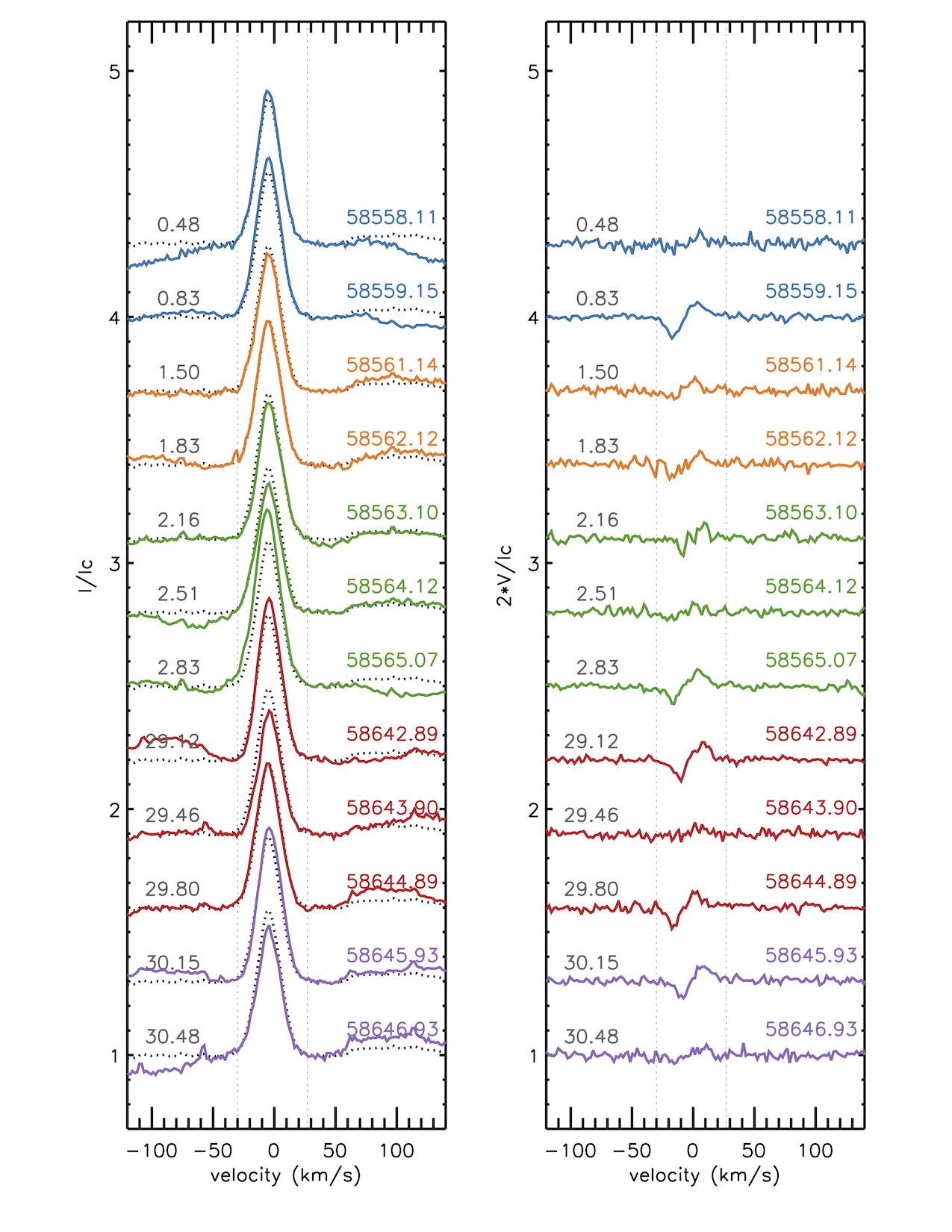}      
   \caption{Stokes I ({\it left panels}) and V ({\it right panels}) profiles ordered as a function of Julian date for the three emission components of the \caii\ IR triplet: 849 nm ({\it left}), 854 nm ({\it middle}), 866 nm ({\it right}). The phase of the corresponding rotational cycles is indicated on the left of the profiles. The color code corresponds to successive rotational cycles.  }
              \label{caii}%
    \end{figure*}

   \begin{figure}
   \centering
  \includegraphics[width=0.9\hsize]{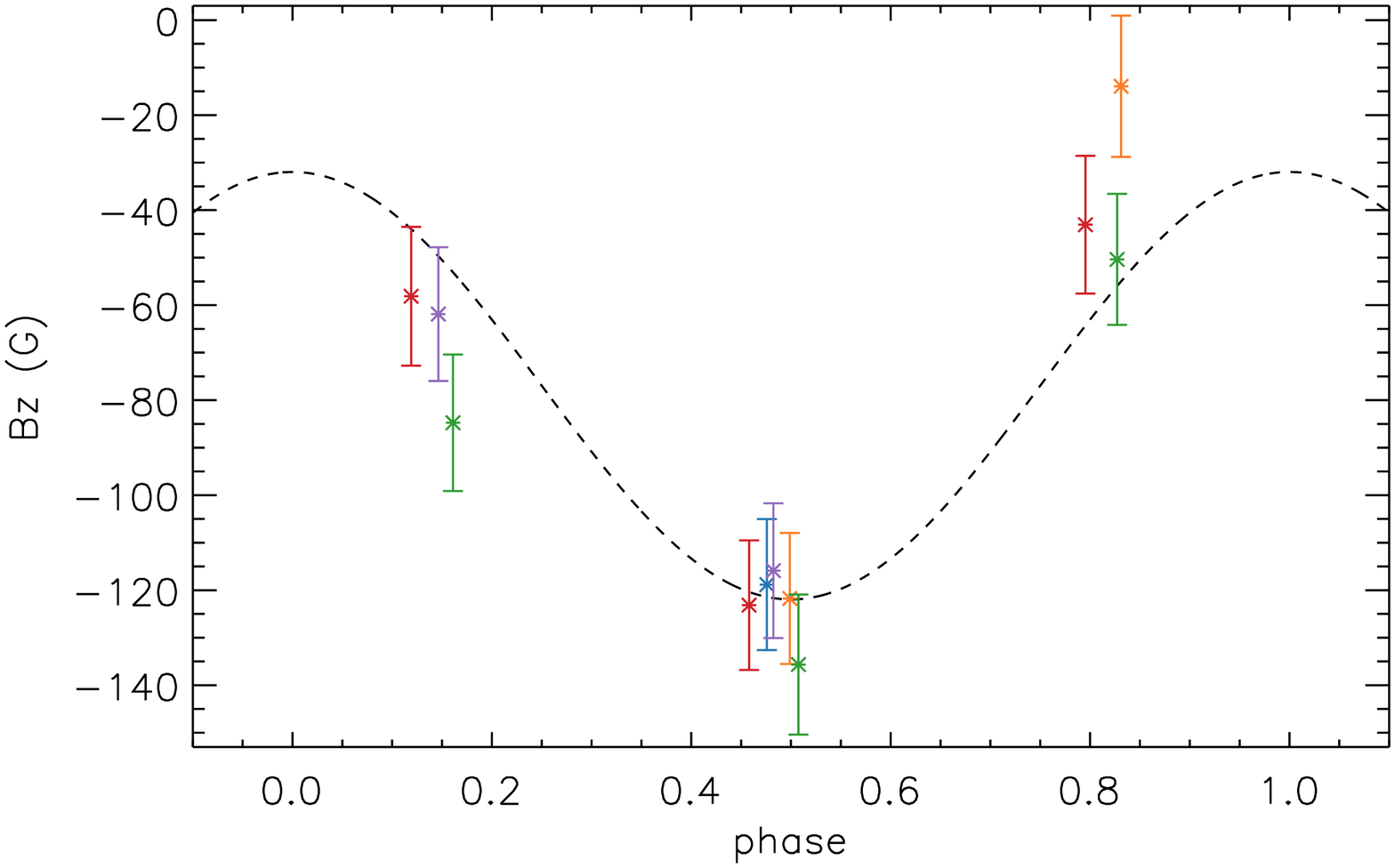}
                \includegraphics[width=0.9\hsize]{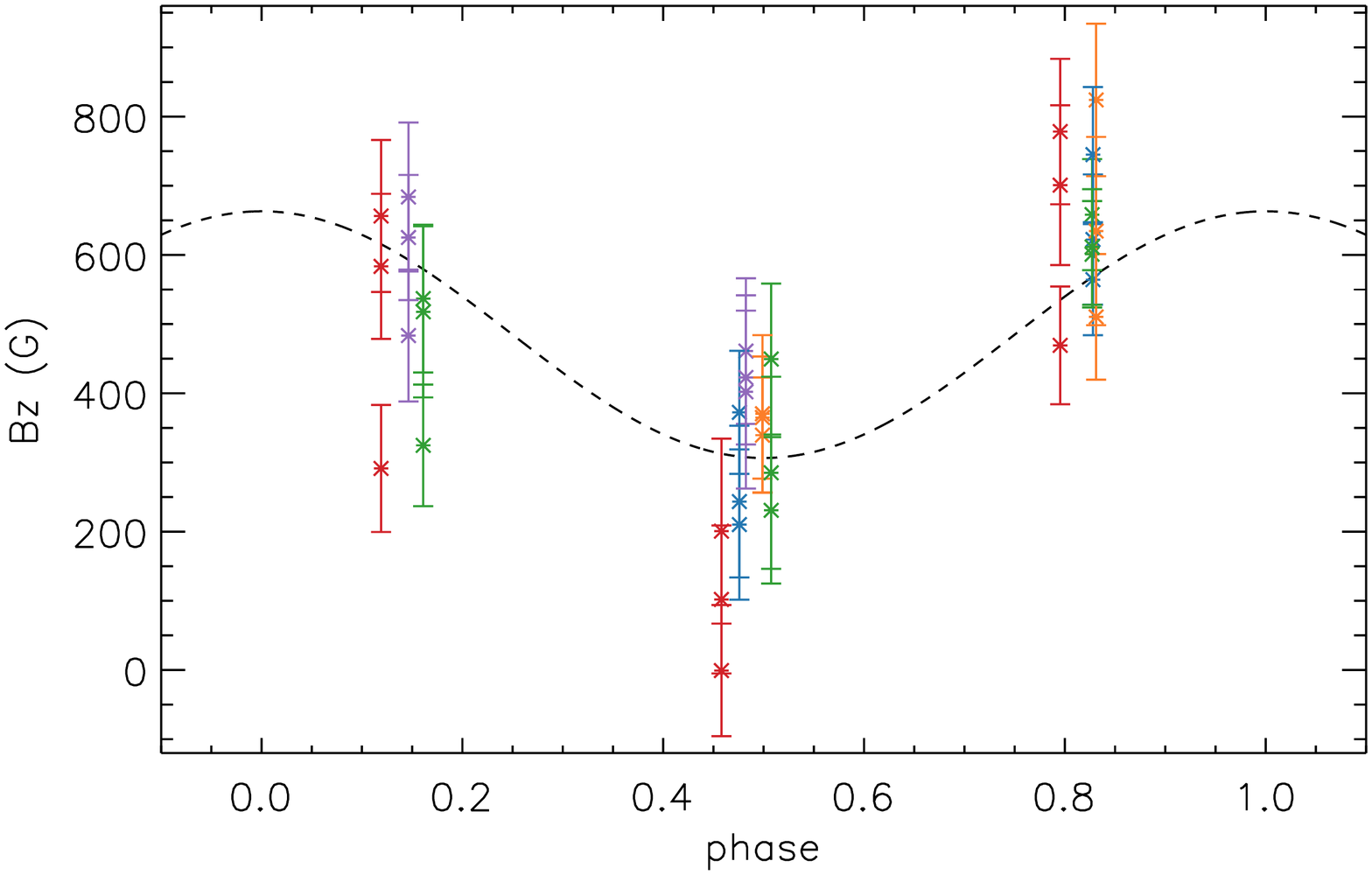}
                \includegraphics[width=0.9\hsize]{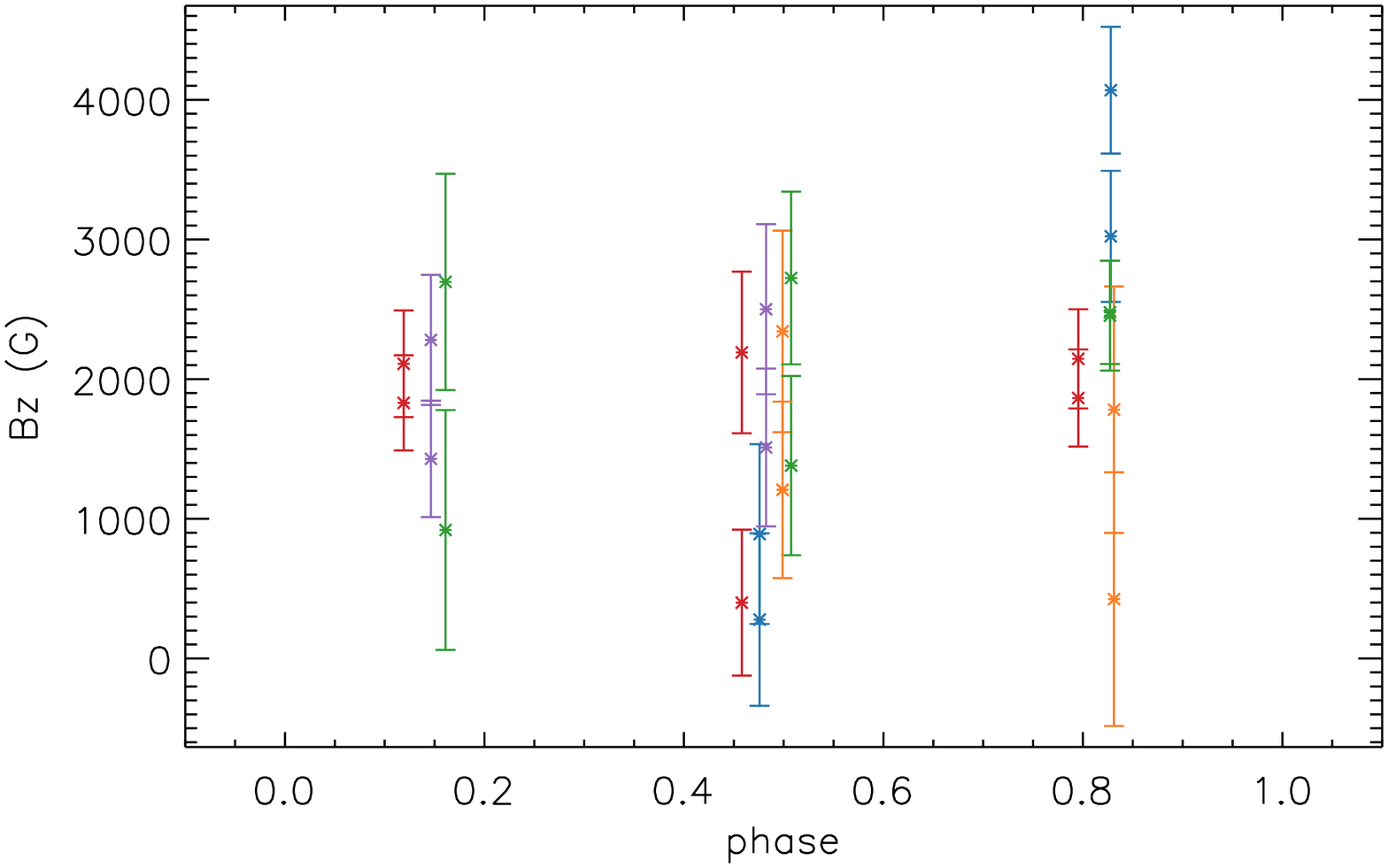}

   \caption{Rotational modulation of the longitudinal component of the large-scale magnetic field, derived from the LSD profiles ({\it top}), from the \caii\ IR triplet ({\it middle}), and from the narrow component of the \hei\  587.6~nm lines profile ({\it bottom}). The color code corresponds to successive rotational cycles. We emphasize the phase consistency between the March and June datasets, as well as the apparently constant amplitude of the field between the two epochs. In the top panel, the sine curve has been fit with the period and origin of phase of the photometric ephemeris. It shows that the maximum intensity of the photospheric field occurs at phase 0.5, that is, at the time of minimum brightness (see Fig.~\ref{lc}).  In the middle panel, the three measurements at each phase are obtained for each line of the CaII IR triplet. In the lower panel, there are two measurements per spectrum, as the \hei\ line appears on successive spectral orders. The three plots are not on the same scale.}
              \label{blong}%
    \end{figure}

   \begin{figure}
   \centering
           \includegraphics[width=\hsize]{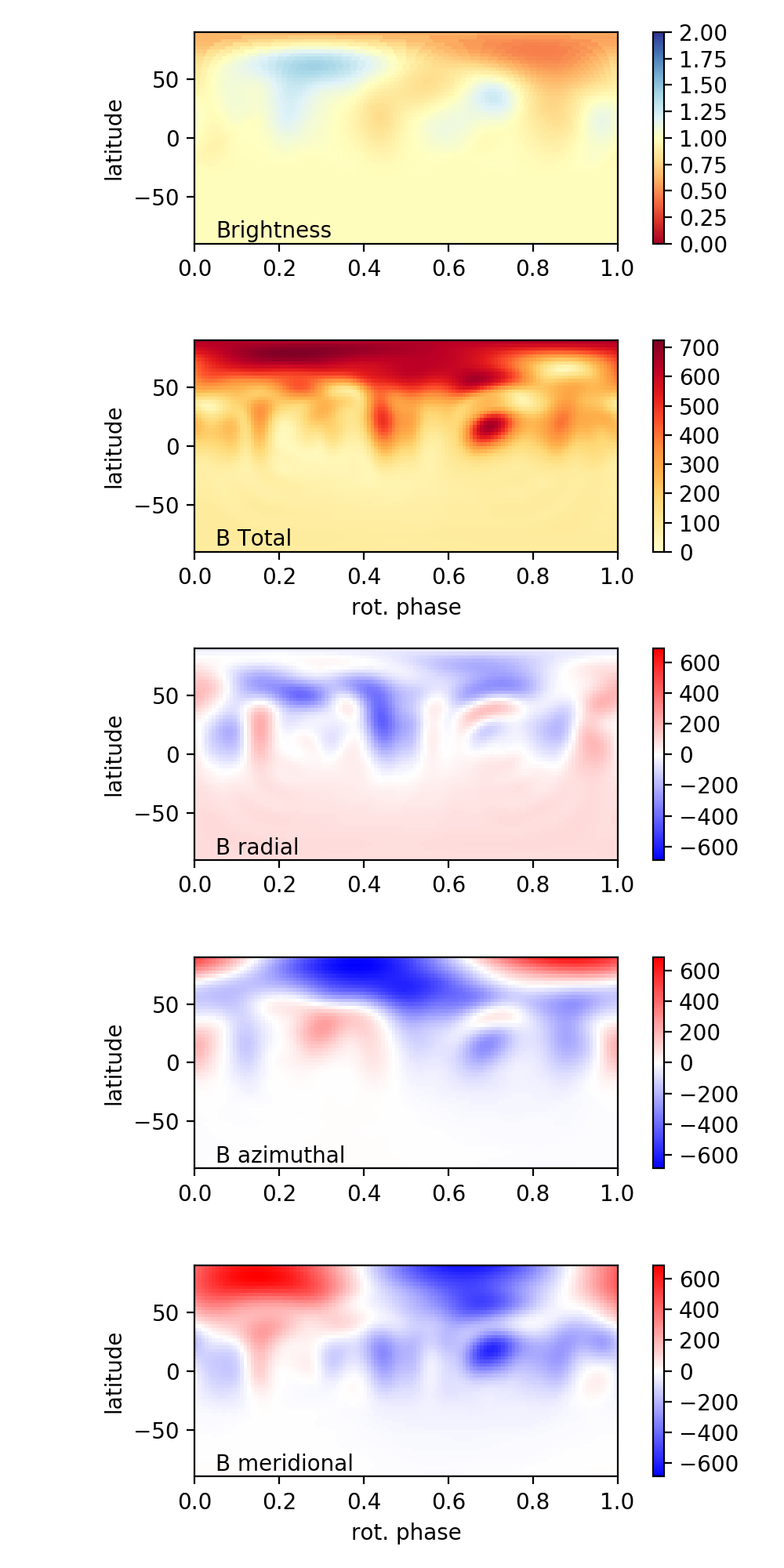}
   \caption{{\it Top:} brightness map derived from the variation of the LSD Stokes I profiles. The color code is 1 for the photospheric temperature, increases for hot spots, and decreases for cold ones.  {\it Bottom:} the large-scale total, radial, longitudinal, and latitudinal surface magnetic field components, as derived from the variations of the LSD Stokes V profiles. The color code indicates the strength of the surface magnetic field in Gauss.  }
              \label{zdi}%
    \end{figure}

The polarimetric capabilities of CFHT/ESPaDOnS allow us to investigate the large-scale structure of the magnetic field at the surface of the star. 
Figure~\ref{stokes} shows the least-square deconvolution \citep[(LSD)][]{Donati97} Stokes I and Stokes V profiles over the two ESPaDOnS runs, computed using a mean Land\'e factor of 1.2 and a central wavelength of 500~nm. We did not use the spectrum obtained on JD 8559.15 ($\phi$=0.83) for the ZDI analysis as the LSD profiles show contamination by the morning twilight. A clear and relatively stable magnetic signature is present in all Stokes V profiles. The variation of the longitudinal component of the large scale magnetic field, B$_l$, derived from the Stokes I and V signatures \citep{Donati97, Wade00}, are shown in Fig.~\ref{blong}. It varies between  -14 and -136 Gauss (see Table~\ref{ewtab}) and appears to be modulated by stellar rotation. Remarkably, the magnetic field intensity does not exhibit significant change between the March and June runs, nor does the phase of its modulation. This suggests a fairly stable magnetic configuration between the two epochs. 

A similar analysis has been performed in the \caii\ near-IR triplet. 
The three lines of the \caii\ triplet exhibit consistent magnetic signatures in their Stokes V profiles as shown in Fig.~\ref{caii}.   
The longitudinal component of the magnetic field projected onto the LoS is still modulated at the stellar rotation period (see Fig.~\ref{blong}) and now reaches a maximum value of 800~G, which is about four times larger than the intensity deduced from the LSD profiles, as is often the case for young accreting stars \citep[e.g.,][]{Donati19}. It is worth mentioning here that the intensity of B$_l$, as well as its sign, are widely different when using LSD (photospheric) profiles and the \caii\ IRT (chromospheric) lines.  Indeed,  these spectral features probe different regions at the stellar surface.  The Stokes V signal in the optical \hei\ line was also analyzed, as it provides the most direct estimate of the magnetic field strength at the base of the funnel flow, close to the accretion shock. Unfortunately, the Stokes V profile in the narrow component of the line is weak (see Fig.~\ref{stokes}), which translates into large uncertainties in the derivation of the longitudinal field strength. Figure~\ref{blong} shows the absence of modulation within error bars, with an average field strength of 2$\pm$0.8~kG. The lack of significant modulation may result from the low inclination of the system, where the high latitude accretion shock remains in view during the whole rotational cycle. 
 
We used the Doppler imaging (DI) and Zeeman-Doppler imaging (ZDI) implementation of \cite{Folsom18} to respectively map the surface brightness and magnetic field of DoAr~44. A full description of the code can be found in \cite{Folsom18}. Briefly, the code uses a time series of Stokes I and V LSD profiles to construct a surface brightness map and the large-scale vector magnetic field. The local line models are calculated using a Voigt profile, with a Gaussian width of 1.2~\kms and a Lorentzian width of 1.9~\kms. The local line models for Stokes V are calculated from the local I line profile, using the weak-field approximation \citep{Landi04}, and the LSD mean Land\'e factor and central wavelength. The magnetic field is expressed as a combination of spherical harmonics as in \cite{Donati06}. The integration over the surface of the star is done using a linear limb-darkening coefficient of 0.8 \citep{Claret11}, adapted to the temperature and gravity of DoAr~44. The fit to the data is performed by both minimizing $\chi^2$ and maximizing the entropy, using the maximum entropy fitting routine of \cite{Skilling84}. We have adopted the photometric ephemeris defined above, and fixed the inclination of the rotation axis at 35$^{\circ}$, which is consistent with the inclination of the inner disk seen in interferometry on a scale of 0.14 au (see Paper I).

The resulting brightness and magnetic maps are shown in Fig.~\ref{zdi}. One large cool spot is observed on the brightness map, located close to the rotational axis and facing the observer around phase 0.83. Two bright spots are present at lower latitudes, facing the observer around phases 0.30 and 0.70. Due to the low inclination of the system, little is known of the southern hemisphere of the star. We used this brightness map to perform ZDI and reconstruct the magnetic map. We fixed the maximum value of the Legendre polynomial degree $l$ to 15 in the  spherical harmonics decomposition. The resulting distribution of the magnetic vectors over the stellar surface is shown in Fig.~\ref{zdi}. It indicates a dominant dipolar component, encompassing 75\% of the total flux, while only 8\% and 4\% of the flux is contained in the quadrupolar and octupolar components, respectively. The negative pole of the dipole is facing us at the phase 0.5, which is consistent with the maximum absolute value of the longitudinal field. It is located at the latitude of 70\deg, and thus inclined by 20\deg\ onto the rotational axis.

\section {Discussion}

The variability analysis of the DoAr 44 system is complicated by its nearly integer rotational period of 2.96~d. While we observed the system over several rotational periods, the resulting phase coverage remains limited, with essentially the same three phases being sampled over the observations, at every third of the rotational cycle. This prevents us from having a complete picture of the circumstellar environment of this young stellar object, and also severely  limits the longitudinal resolution we have on the stellar surface for image reconstruction. With the data at hand we are nevertheless able to derive the main physical and geometrical properties of the system in the region where the stellar magnetic field interacts with the inner disk.  

\begin{figure}
   \centering
      \includegraphics[width=0.98\hsize]{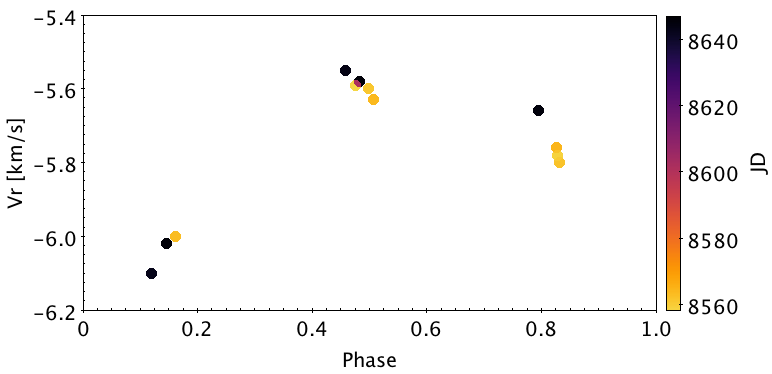}      
   \caption{Photospheric radial velocity variations are plotted as a function of rotational phase.  The color code reflects the Julian date for the March (bright) and June (dark) ESPaDOnS runs. }
              \label{vrad}%
    \end{figure}

The combination of the rotational period derived above and the \vsini\ we measure suggests that the star is seen at a low inclination of  30$\pm$5$\degr$. In Paper I, we derived a similar inclination of 34$\pm$2$\degr$ for the inner edge of the circumstellar disk from long baseline near-IR interferometry. That the inner system seen at low inclination is further supported by the lack of strong photometric modulation. While we do recover a photometric period of 2.96~d, assigned to the stellar rotational period, the photometric amplitude of $\leq$0.1 mag in the V-band is significantly lower than that of most accreting T Tauri stars \citep[e.g.,][]{Grankin07}. The photospheric radial velocity is modulated over the same 2.96~d period (see Fig.~\ref{vrad}), which confirms the presence of a long-lived, large, dark spot at the stellar surface. According to the Doppler image, the high-latitude spot faces the observer at phase 0.83, which yields symmetric LSD profiles (see Fig.~\ref{stokes}). Indeed, the photospheric radial velocity lies midway from the extrema of the \vrad\ curve at this phase, as expected when the spot is on the LoS (see Fig.~\ref{vrad}). These concordant results confirm that the 2.96~d period is the rotational period of the star whose surface, seen at a low inclination, is covered by a high-latitude, large, cold spot. In contrast, the photometric minimum appears to occur at phase 0.5, which suggests a more complex distribution of bright and dark spots at the stellar surface drives the photometric modulation. 

In spite of the low inclination, clear modulation signals are also detected in several accretion diagnostics. This is the case of the redshifted wings of emission lines formed in the funnel flow (e.g., \hb, \ha, \pab, \hei\ 1083.3~nm) that smoothly vary along the rotational cycle, even though they do not exhibit inverse P Cygni components, that is, absorptions reaching below the continuum, which are the hallmark of accretion funnel flow passing onto the LoS in high-inclination systems. The radial velocity and the intensity of the narrow component of the \hei\ 587.6~nm line profile are also  modulated at the stellar rotational period and indicate a high-latitude hot spot at the stellar surface, presumably the location of the accretion shock close to the magnetic pole. Indeed, the magnetic surface map reconstructed from the ZDI analysis features a dominant dipolar field whose axis is moderately inclined onto the rotational axis, thus producing an accretion shock at high latitudes. These results suggest that the inner disk is truncated at some distance above the stellar surface and that the accretion onto the star proceeds along magnetospheric funnel flows. 

If we assume that the disk's truncation radius lies close to the disk's corotation radius, we get
$$ r_{tr} \simeq r_{co} = (GM_\star)^{1/3} (P_{rot}/2\pi)^{2/3} = 0.043~\rm{au} = 9.2~R_\odot = 4.6~R_\star,  $$
and according to \cite{Bessolaz08}, the truncation radius occurs at
$$r_{tr}/R_\star =  2 \cdot B_\star^{4/7} \dot M_{acc}^{-2/7} M_\star^{-1/7} R_\star^{5/7}, $$
where the stellar field at the equator B$_\star$ is normalized to 140 G, the mass 
accretion rate \macc\ to 10$^{-8}$ \msunyr, the stellar mass \mstar\ to 0.8 \msun, and
the stellar radius \rstar\ to 2 \rsun.

Thus, the condition $r_{tr} \simeq r_{co}$ requires a dipolar magnetic field of the order of 537~G at the equator, which translates into 1.1~kG at the pole. The average value of the longitudinal component of the magnetic field we measure in the narrow component of the \hei\ 587.6~nm line amounts to 2$\pm$0.8~kG. Assuming this is the large scale component that interacts with the disk, we would derive a truncation radius that is about twice the corotation radius. However, the large error bars associated with the field strength and derived from the \hei\ NC component are not inconsistent with the disk being truncated close to the corotation radius. 
In Paper I, we measured the size of the \brg\ line emitting region, thought to reflect the extent of the magnetospheric cavity, from long baseline near-IR interferometry, and we found it to be less than 0.047 au, that is, 5~\rstar. 

All the results summarized above suggest that DoAr 44 was observed during a phase of stable magnetospheric accretion with a magnetic topology that shares some similarities with other young accreting systems that we previously studied, such as \object{V2129~Oph} \citep{Alencar12}. Some of these systems exhibit dipper light curves at times, with deep periodic eclipses of the central object being produced by circumstellar dust: for example, \object{AA~Tau} \citep{Bouvier03, Bouvier07a} and \object{LkCa~15} \citep{Alencar18}, as well as periodic redshifted absorptions  reaching below the continuum appearing in the wings of emission lines formed in the funnel flow. These specific properties are not shared by DoAr~44 in spite of a qualitatively similar magnetospheric structure.  We estimate that the dust sublimation radius lies at a distance of R$_{sub}$ = 4.7 \rstar = 0.044~au, as derived from Eq.1 of \cite{Monnier02} for T$_{sub}$ = 1500~K and Qr=1.  Dust might thus to be present at the disk truncation radius, even though the half-flux radius of the inner dusty disk we report in Paper~I is of the order of 0.14 au. While a photometric modulation is observed, the light curve is spot dominated, with no evidence for significant dips. Similarly, while the red wing of Balmer line profiles appear modulated at the stellar rotation period, there is no sign of high-velocity redshifted absorptions reaching below the continuum, even at phases where the magnetic pole is facing the observer.  

Indeed, dipper characteristics and conspicuous inverse P Cygni profiles are expected to occur primarily in systems seen at a moderate to high inclination \citep[e.g.,][]{Mcginnis15, Sousa16, Fonseca14}. This is clearly not the case in DoAr~44, and the low inclination of this system accounts for the low-amplitude, smooth light curve; the mildly varying red wings of the emission line profiles; and more generally, the global shape of the emission line profiles as suggested by radiative transfer models of magnetospheric accretion onto a misaligned dipole. This clearly highlights the importance of taking into account inclination when investigating the variability of young stellar systems.  Systems with otherwise similar properties, notably regarding the magnetospheric accretion process, might exhibit quite different characteristics depending on the viewing angle under which they are seen.    

DoAr 44 is a pre-transitional disk system. On the large scale, from mm observations, \cite{Andrews09, Andrews11} and \cite{Vandermarel16} reported the existence of a large cavity extending from the central star up to 25-30~au in the circumstellar disk. They showed that cold dust emission arises from an outer, relatively narrow  ring (25 to 35 au wide), inclined by 20 to 35\degr\ on the LoS. On much smaller scales, the existence of a compact, inner disk close to the star is supported by the near-infrared excess flux measured in the spectral energy distribution of the system \citep{Espaillat10} and significant CO emission arising from the inner disk edge \citep{Salyk09, Salyk11}. Furthermore, a misalignment between the inner and outer disk was assumed to be responsible for shadows seen in the outer dusty ring \citep{Casassus18}. In Paper I, we reported the direct detection of the inner disk from long baseline near-IR interferometry. The half-flux radius is 0.14 au, and the compact disk is indeed found to be inclined relative to the outer ring, as predicted by the \cite{Casassus18} model to account for the outer shadows. The pre-transitional status of DoAr 44's disk was a strong motivation to include this specific target in our long-term monitoring program of T Tauri stars. The goal here was to investigate whether the highly structured disk, including a large cavity and inner-outer disk misalignment, both possibly betraying ongoing planetary formation, may impact the accretion process on the central star.  All the results obtained here suggest that, at the star-disk interaction scale of less than 0.1 au, this pre-transitional disk system is not qualitatively different from other young stellar objects surrounded by smoother protoplanetary disks. 

\section{Conclusion} 

We monitored the variability of the young stellar object, DoAr 44, over several rotational periods using several complementary observational techniques to investigate the physics of accretion onto the star in a pre-transitional disk system. The specific geometry of pre-transitional disks, including large cavities and possible inner- or outer-disk misalignment, raises issues regarding the accretion process in these systems. How does accretion proceed from the outer disk to the inner scales through the gap? Is the star-disk interaction process fed in the same way or differently from what is observed in classical T Tauri stars with continuous circumstellar disks? If the pre-transitional nature of disks is the signature of ongoing planet formation in the disk, does this process affect accretion on any scale? 

In an attempt to link the outer scales, best studied with mid-IR and radio interferometry, to the inner scales, we focused here on the characterization of the DoAr 44 system at the star-disk interaction level, meaning on a scale extending from the stellar surface up to a few 0.1 au, combining photometric and spectropolarimetric monitoring with long baseline interferometry. The accretion diagnostics studied here and the direct interferometric measurements we reported in Paper I \citep{Bouvier20} suggest the existence of an inner magnetospheric cavity extending from a distance of about 4.6 stellar radii (0.043 au) down to the stellar surface. A large-scale, mainly dipolar stellar magnetic field of a few hundred Gauss over the photosphere, but reaching up to 2$\pm$0.8~kG close to the accretion shock, is sufficient to disrupt the inner-disk region at the corotation radius or beyond in this moderately accreting system. The reported line profile variability is consistent with the expected behavior of funnel flows rotating at the stellar period. Yet, due to the low inclination of the inner system, of the order of 30$\degr$, the modulation signatures are not as pronounced as in other, more inclined systems. Importantly, the combination of contemporaneous optical and near-IR high-resolution spectroscopic monitoring provides us with a unique means to investigate diagnostics that simultaneously probe accretion funnel flows and stellar or disk winds. 

Overall, over our three-week observation period, DoAr 44 appeared to be in a state of stable magnetospheric accretion, simultaneously driving outflows, in most aspects similar to what we had previously witnessed in other systems we monitored in the same way. This suggests that the physics of accretion from the inner disk to the stellar surface is not significantly altered by the complex large-scale structure of the circumstellar disk specific to pre-transitional systems.   
 
\begin{acknowledgements} 
We gratefully acknowledge the help from Tim Brown at LCOGT for obtaining test images and from LCOGT support service for scheduling and data reduction. We thank the very efficient QSO team at CFHT. This study is based in part on observations obtained at the Canada-France-Hawaii Telescope (CFHT) which is operated by the National Research Council of Canada, the Institut National des Sciences de l'Univers of the Centre National de la Recherche Scientifique of France, and the University of Hawaii. The observations at the Canada-France-Hawaii Telescope were performed with care and respect from the summit of Maunakea, which is a significant cultural and historic site. This publication is based in part on data obtained under CNTAC program CLN2019B-004. This publication is based in part on observations collected at the European Organisation for Astronomical Research in the Southern Hemisphere under ESO programme 0103.C-0097. This work has made use of data from the European Space Agency (ESA) mission
{\it Gaia} (\url{https://www.cosmos.esa.int/gaia}), processed by the {\it Gaia}
Data Processing and Analysis Consortium (DPAC,
\url{https://www.cosmos.esa.int/web/gaia/dpac/consortium}). Funding for the DPAC
has been provided by national institutions, in particular the institutions
participating in the {\it Gaia} Multilateral Agreement.
SHPA acknowledges financial support from CNPq, CAPES, Fapemig, and Cofecub.  A.B. acknowledges support from FONDECYT grant 1190748 and from ICM (Iniciativa Cient\'ifica Milenio) via the N\'ucleo Milenio de Formaci\'on Planetaria grant. This project has received funding from the European Research Council (ERC) under the European Union's Horizon 2020 research and innovation programme (grant agreement No 742095; {\it SPIDI}: Star-Planets-Inner Disk-Interactions, http://www.spidi-eu.org, and grant agreement No 740651 NewWorlds). 
\end{acknowledgements}

\bibliographystyle{aa} 
\bibliography{doar44_rev0} 

\begin{appendix} 
\section {Line profile variability}

\begin{figure}[h]
   \centering
   \includegraphics[width=0.48\hsize]{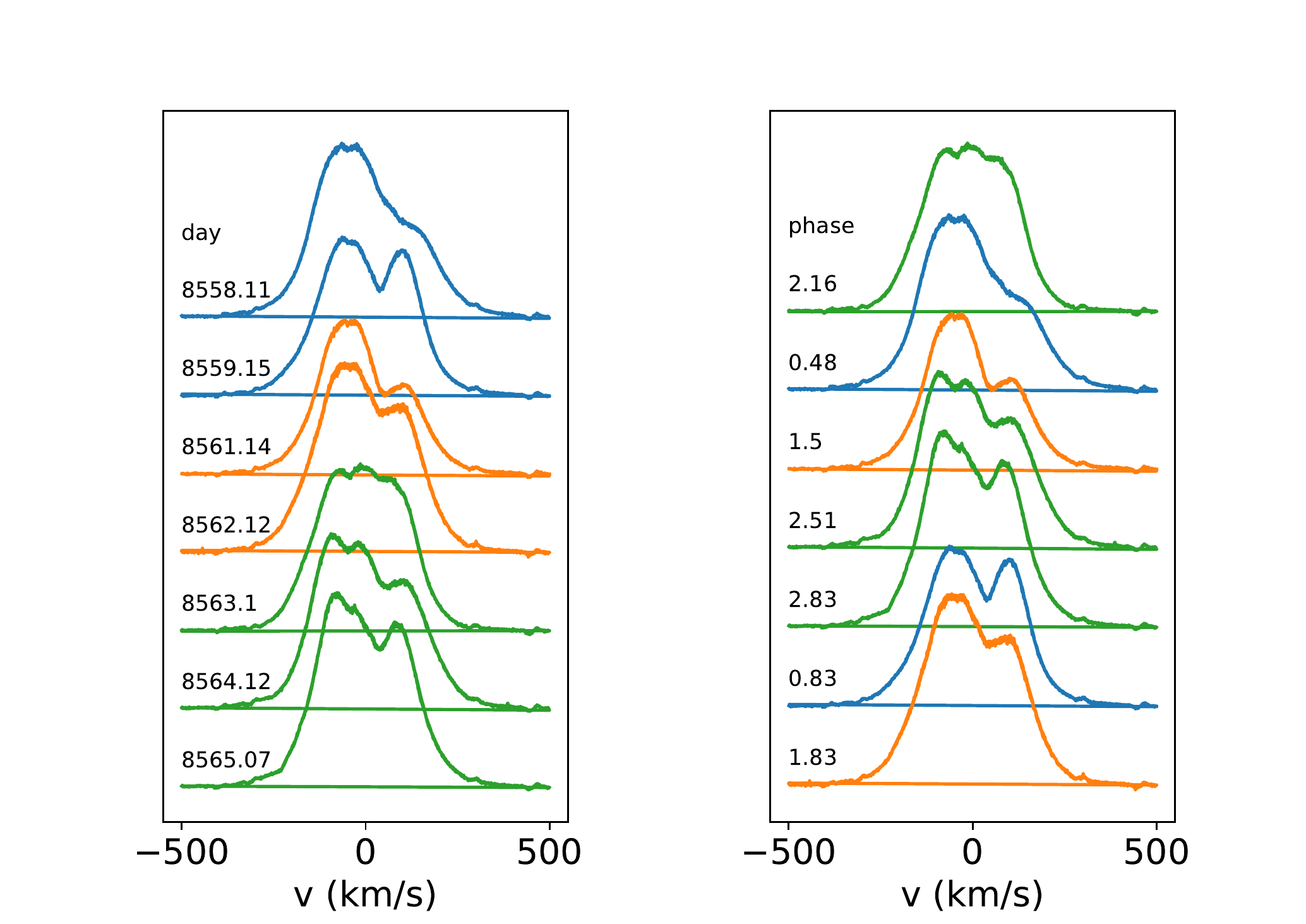}
   \includegraphics[width=0.48\hsize]{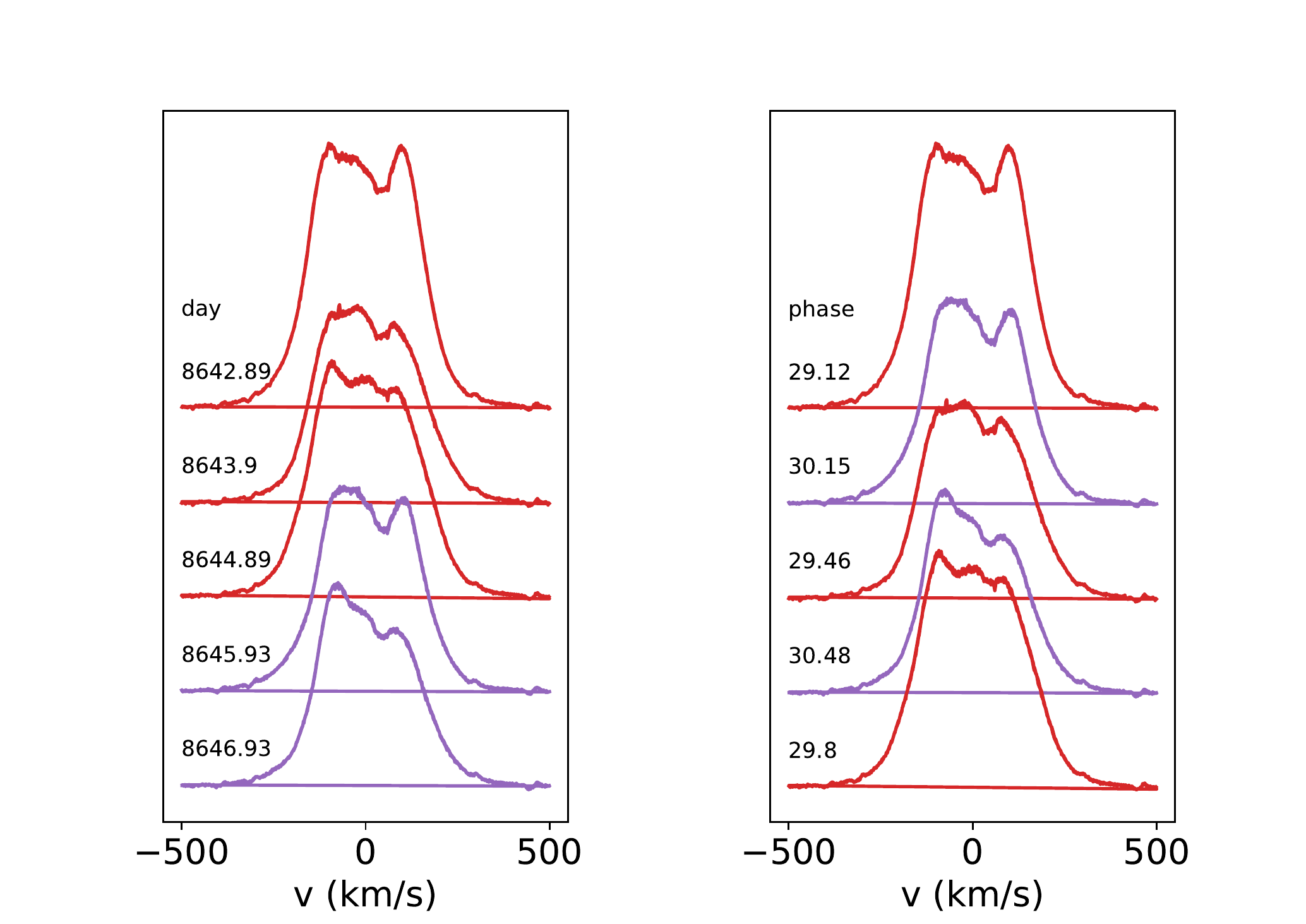}
   \includegraphics[width=0.48\hsize]{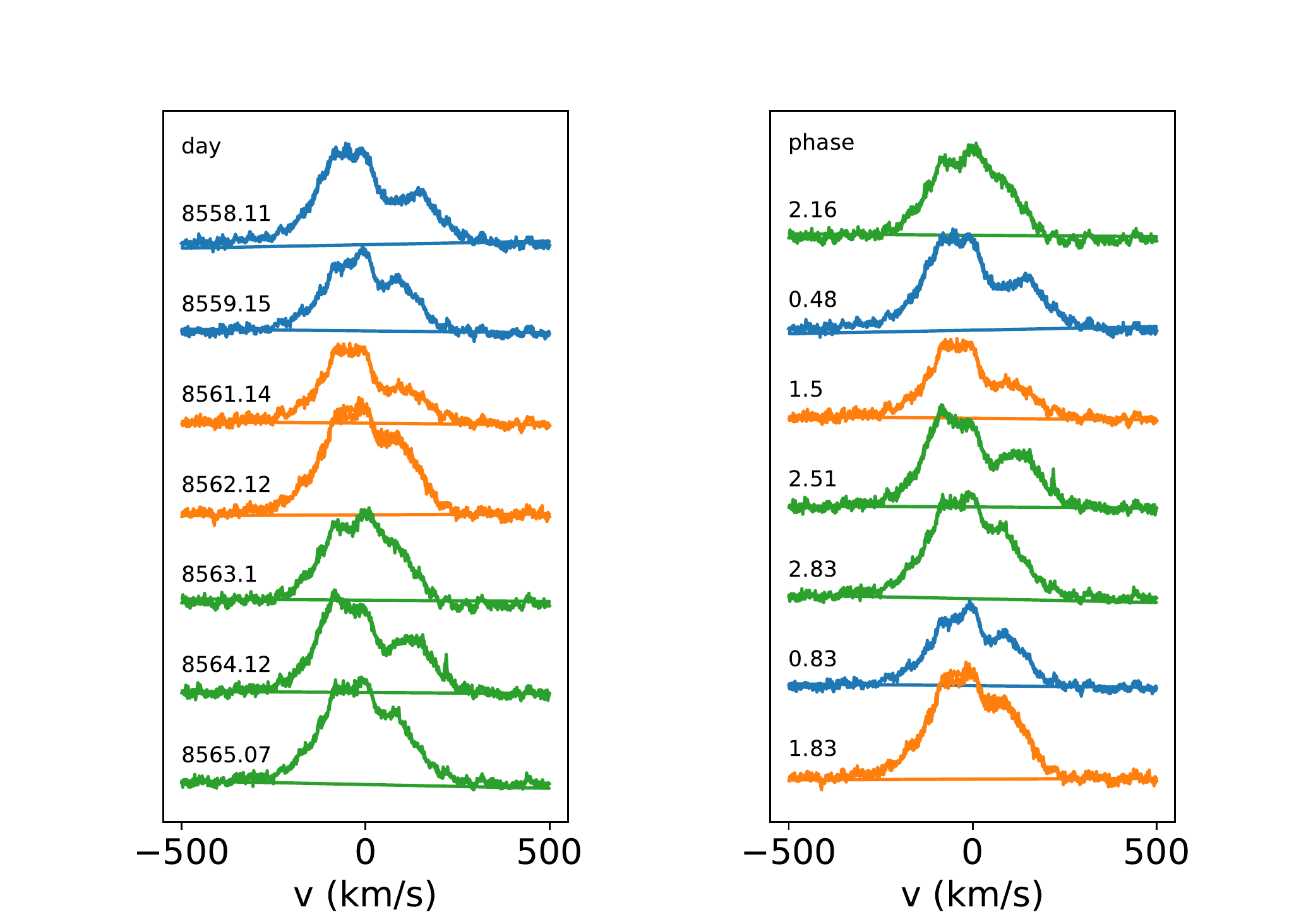}
   \includegraphics[width=0.48\hsize]{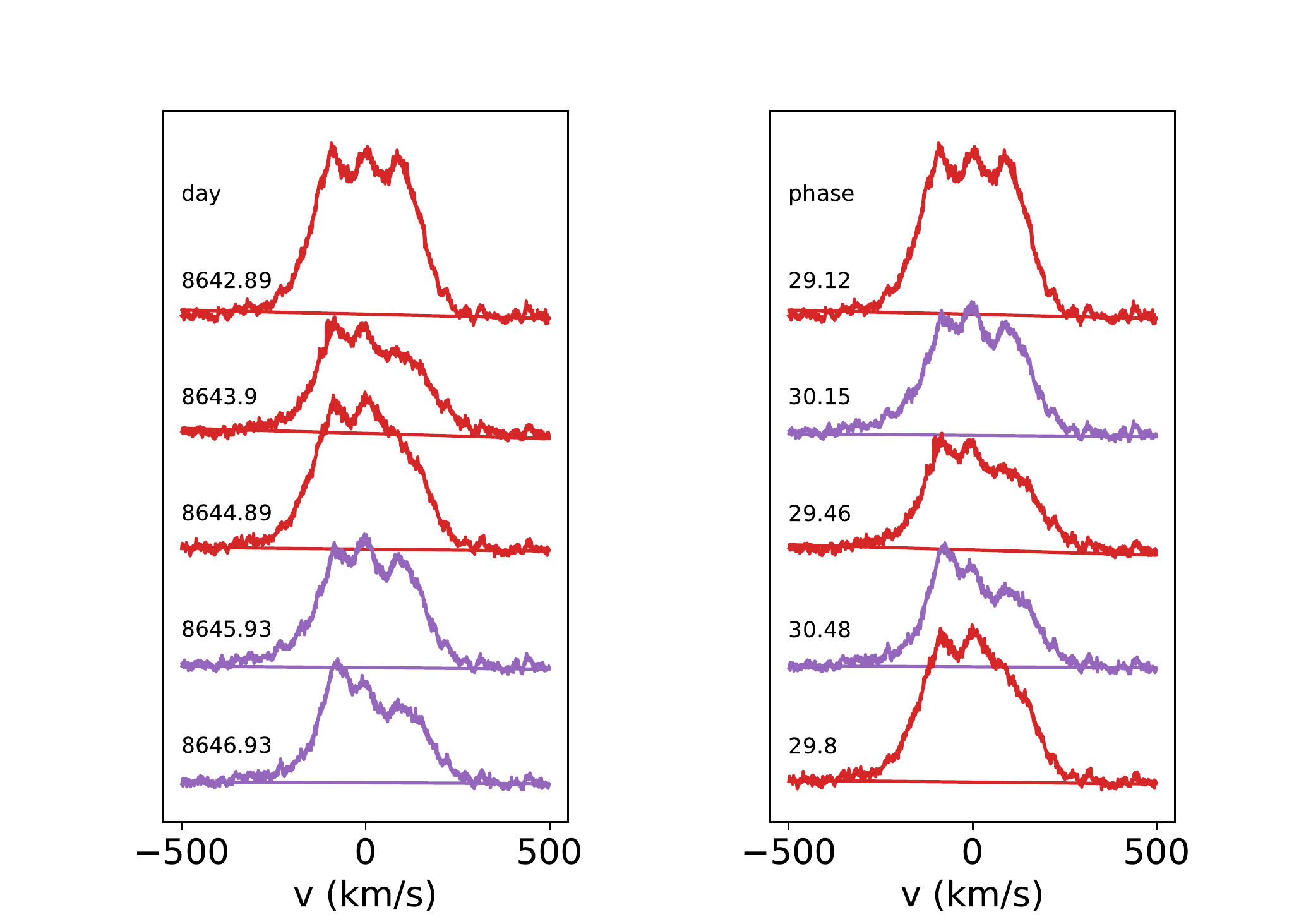}
   \includegraphics[width=0.48\hsize]{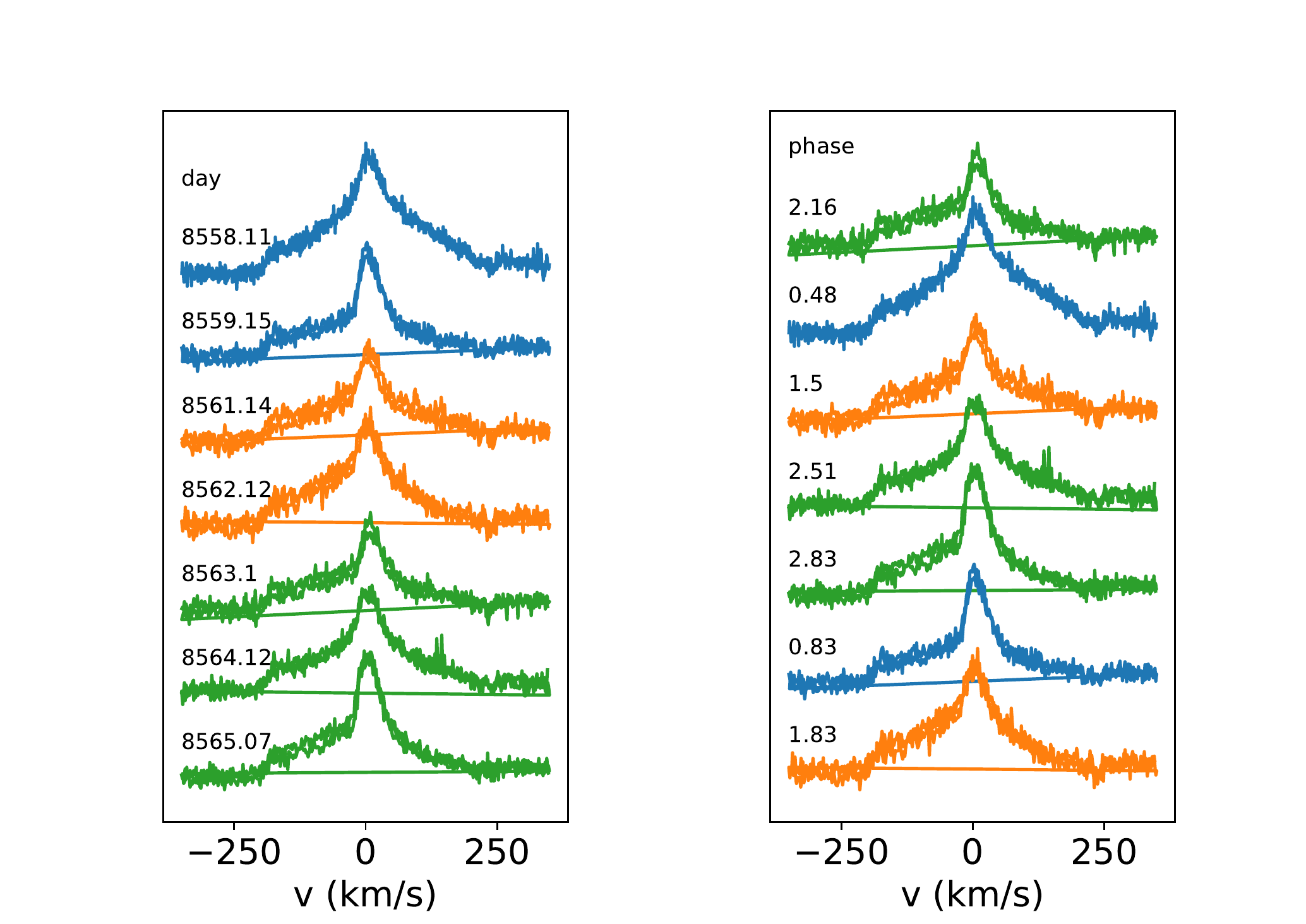}
   \includegraphics[width=0.48\hsize]{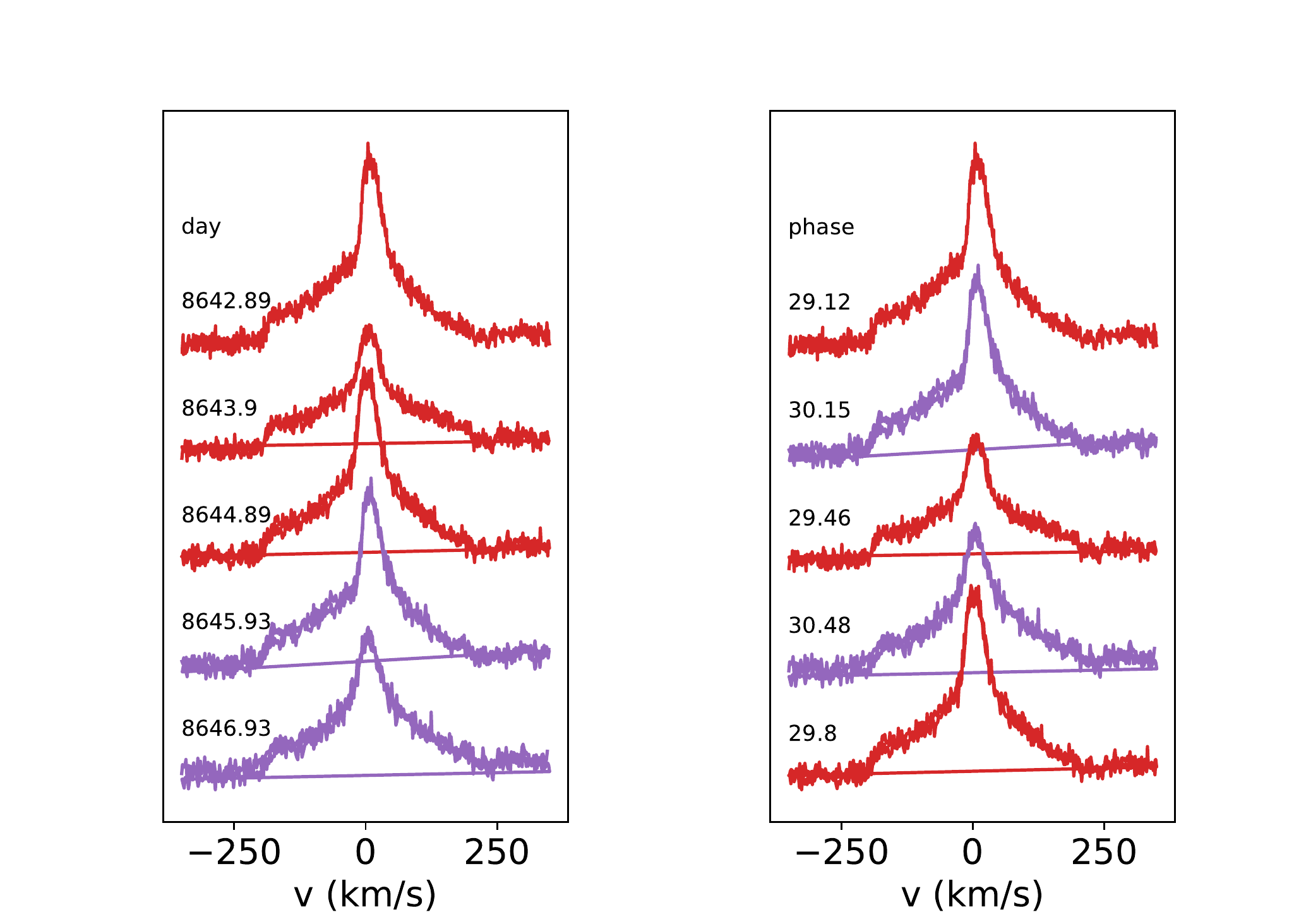}
   \caption{\ha, \hb, and \hei\ line profiles from the seven spectra obtained during the ESPaDOnS March run ({\it left panels}) and the five spectra obtained during the June run ({\it right panels}). In each panel, the profiles are plotted in order of increasing date ({\it left}) and phase ({\it right}). Each color corresponds to a specific rotational cycle.}
              \label{profmarchjune}%
    \end{figure}

Since the shape of the emisson line profiles has varied between the ESPaDOnS March and June 2019, runs, we present in Fig.~\ref{profmarchjune} their variability over timescales of days, independently for each run. 

\begin{figure}[h]
   \centering
   \includegraphics[width=0.45\hsize]{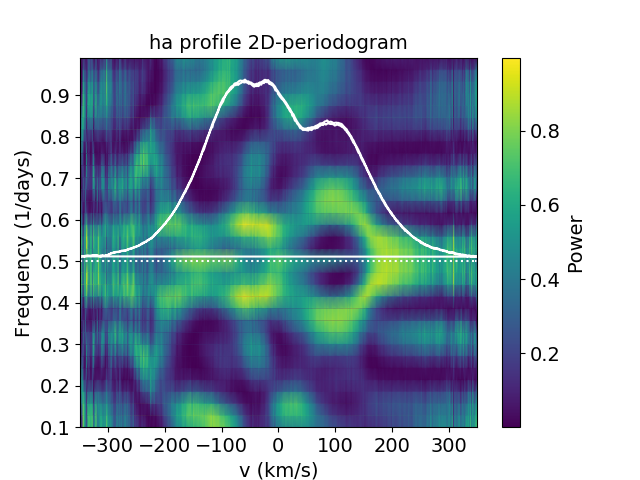}
   \includegraphics[width=0.45\hsize]{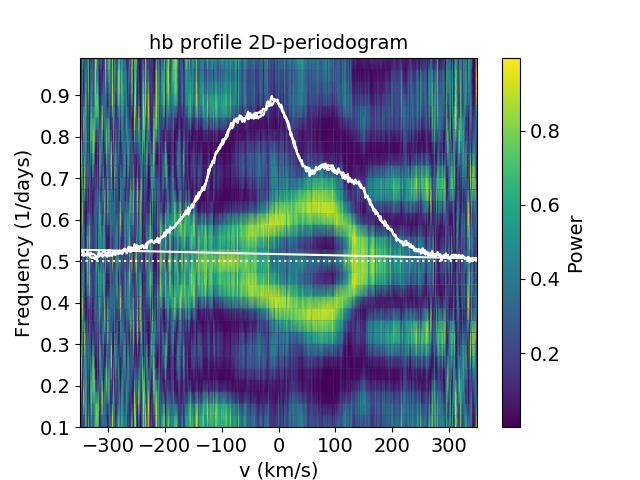}
   \includegraphics[width=0.45\hsize]{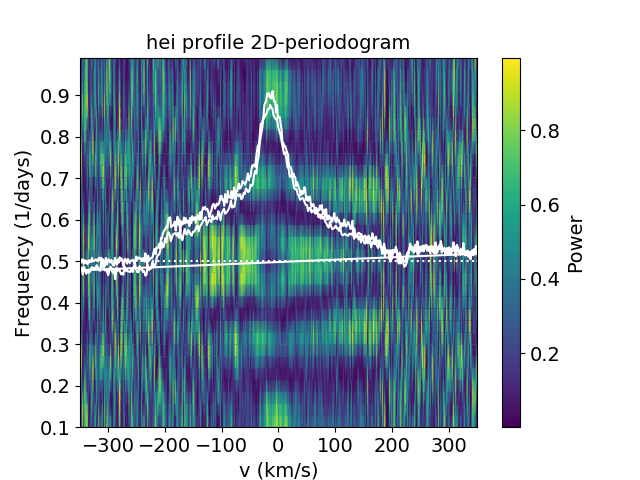}
   \caption{2D periodogram of \ha, \hb, and \hei\ line profiles from the seven spectra obtained during the ESPaDOnS March run. The color code represents the periodogram power, and the superimposed white curve is the average line profile. We note the symmetry of the periodogram around the frequency 0.5~d$^{-1}$ due to the 1~d alias resulting from the night-to-night sampling. }
              \label{period2dmarch}%
    \end{figure}

\begin{figure}[h]
   \centering
   \includegraphics[width=0.45\hsize]{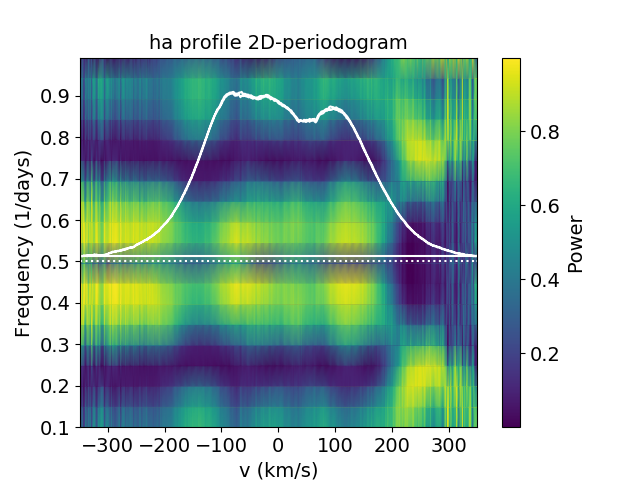}
   \includegraphics[width=0.45\hsize]{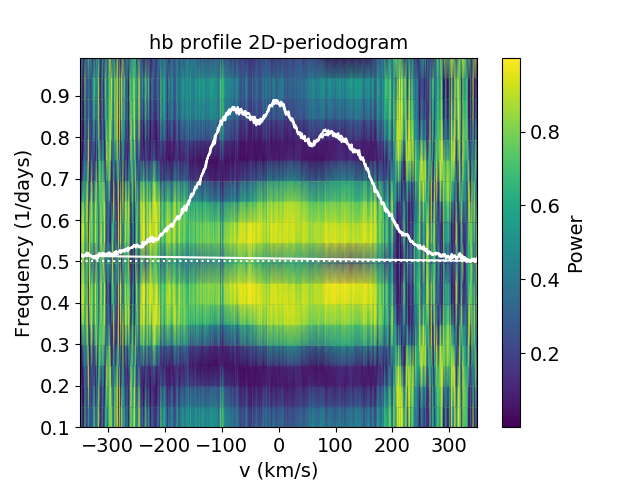}
   \includegraphics[width=0.45\hsize]{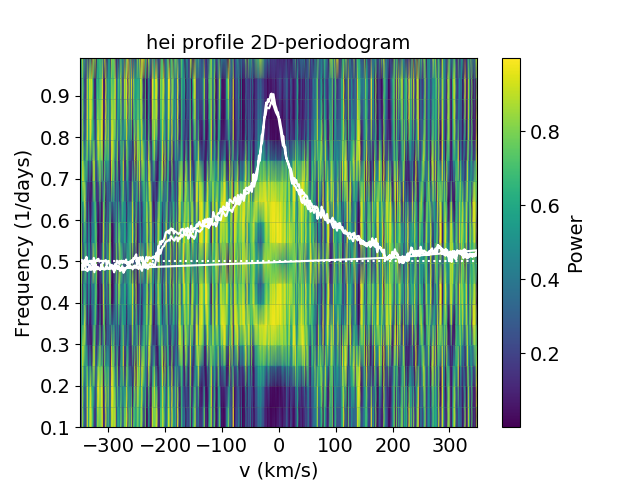}
   \caption{2D periodogram of \ha, \hb, and \hei\ line profiles from the five spectra obtained during the ESPaDOnS June run. The color code represents the periodogram power, and the superimposed white curve is the average line profile. We note the symmetry of the periodogram around the frequency 0.5~d$^{-1}$ due to the 1~d alias resulting from the night-to-night sampling.}
              \label{period2djune}%
    \end{figure}

The periodogram analysis of line profiles over the two ESPaDOnS runs is inconclusive. We therefore proceed to analyzing the March and June 2019 runs independently. Figures~\ref{period2dmarch} and~\ref{period2djune} show the periodogram analysis of the \ha, \hb, and \hei\ line profiles for the two epochs, respectively. During the March run, seven spectra were obtained over eight nights, while during the June run, five spectra were obtained over five nights. Moreover, since the stellar period of 2.96~d is close to an integer, the same rotational phase repeats every third night. The short duration of each run combined with the poor phase sampling makes the periodogram analysis fragile. Nevertheless, we find marginal evidence for the rotational frequency of 0.34~d$^{-1}$ in the red wing of the \ha\ and \hb\ line profiles around +100~\kms\ during the March run, and in the narrow central component of the \hei\ line during the June run. While of limited significance, the similar shapes exhibited by the various line profiles at specific rotational phases (see Fig.~\ref{profmarchjune}) lends credit to this result. For instance, the Balmer line profiles appear much more symmetric around phases 0.12-0.16 and 0.83-0.93 than around phases 0.46-0.51, where the red wing is consistently depressed. This suggests additional redshifted absorption occurs at phases around 0.5.  

\section {Line profile decomposition}

Here, we illustrate examples of line profile decomposition by multiple Gaussian fits for the \ha, \hb, \hei, and CaII emission line profiles observed on JD 8565.07 and JD8644.89. The \ha\ and \hb\ line profiles are well reproduced with a central Gaussian emission peak onto which a blueshifted and a redshifted absorption components are superimposed. The \hei\ line is well fit by the combination of two Gaussian components in emission: a broad one and a narrow one, respectively. This is also the case for  the CaII line profile except for two occurrences, on JD 8559.15 and JD 8565.07, where an additional redshifted absorption component is required to reproduce the observed line profile. 

      \begin{figure}[h]
   \centering
   \includegraphics[width=0.45\hsize]{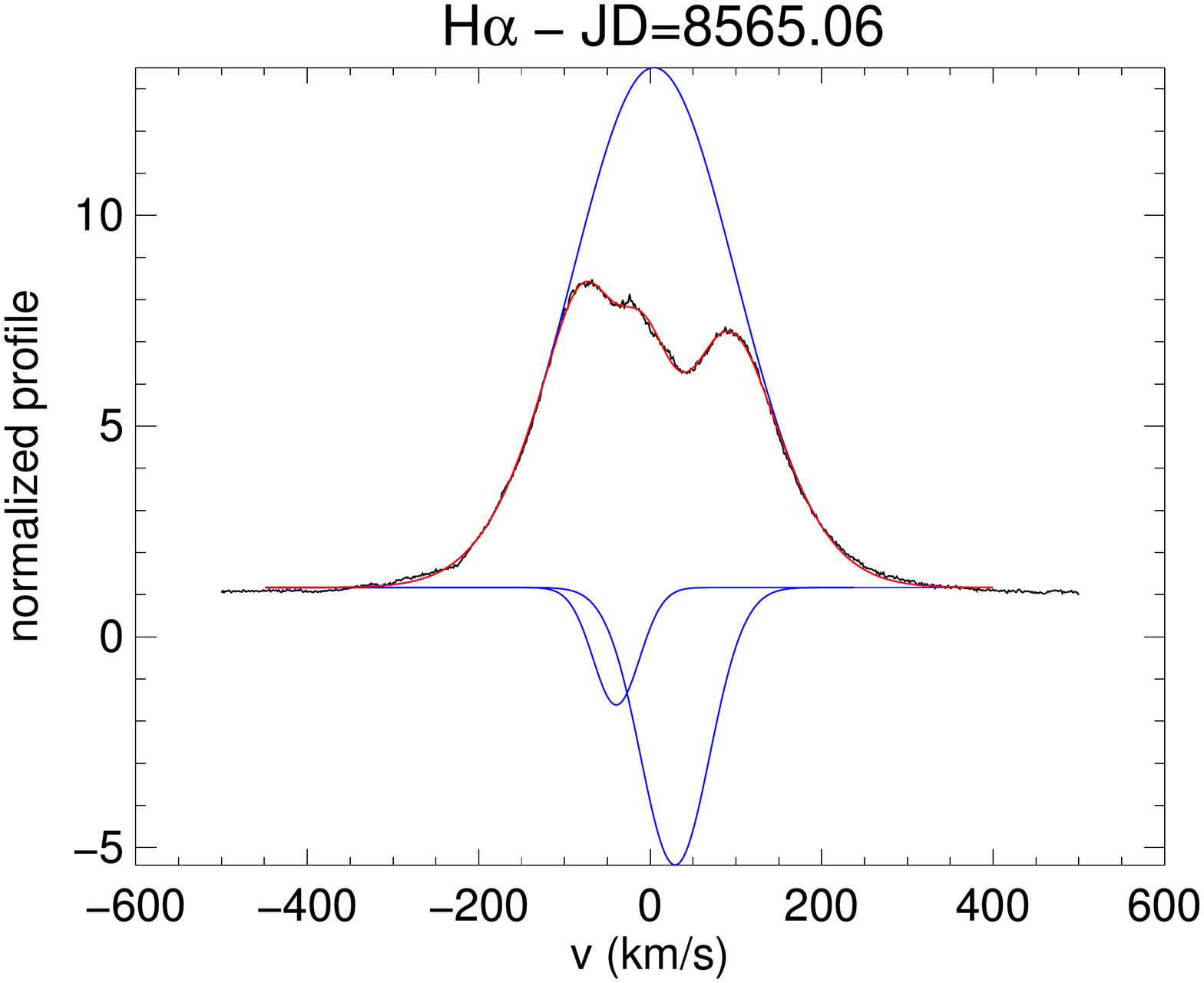}
   \includegraphics[width=0.45\hsize]{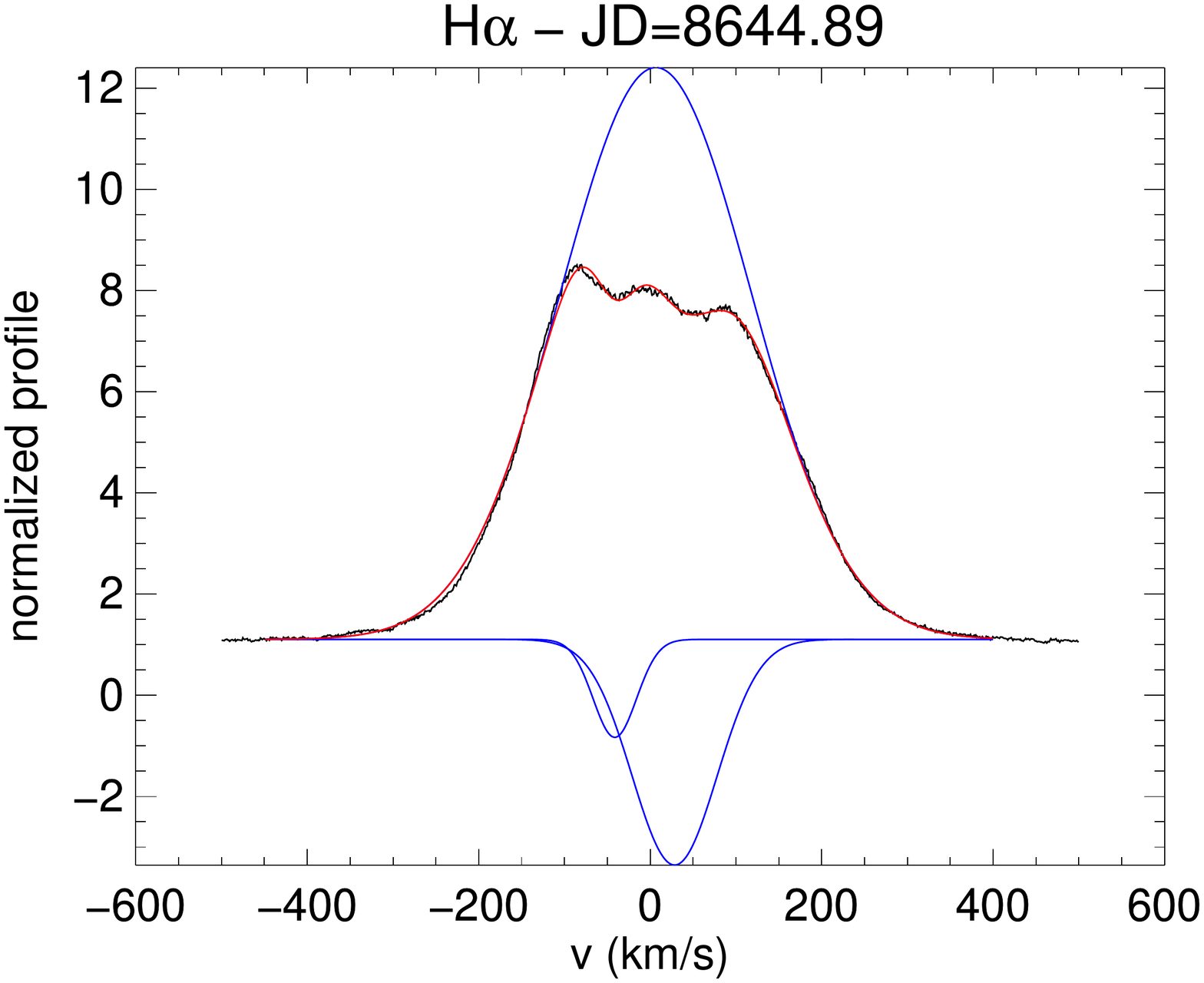}
   \includegraphics[width=0.45\hsize]{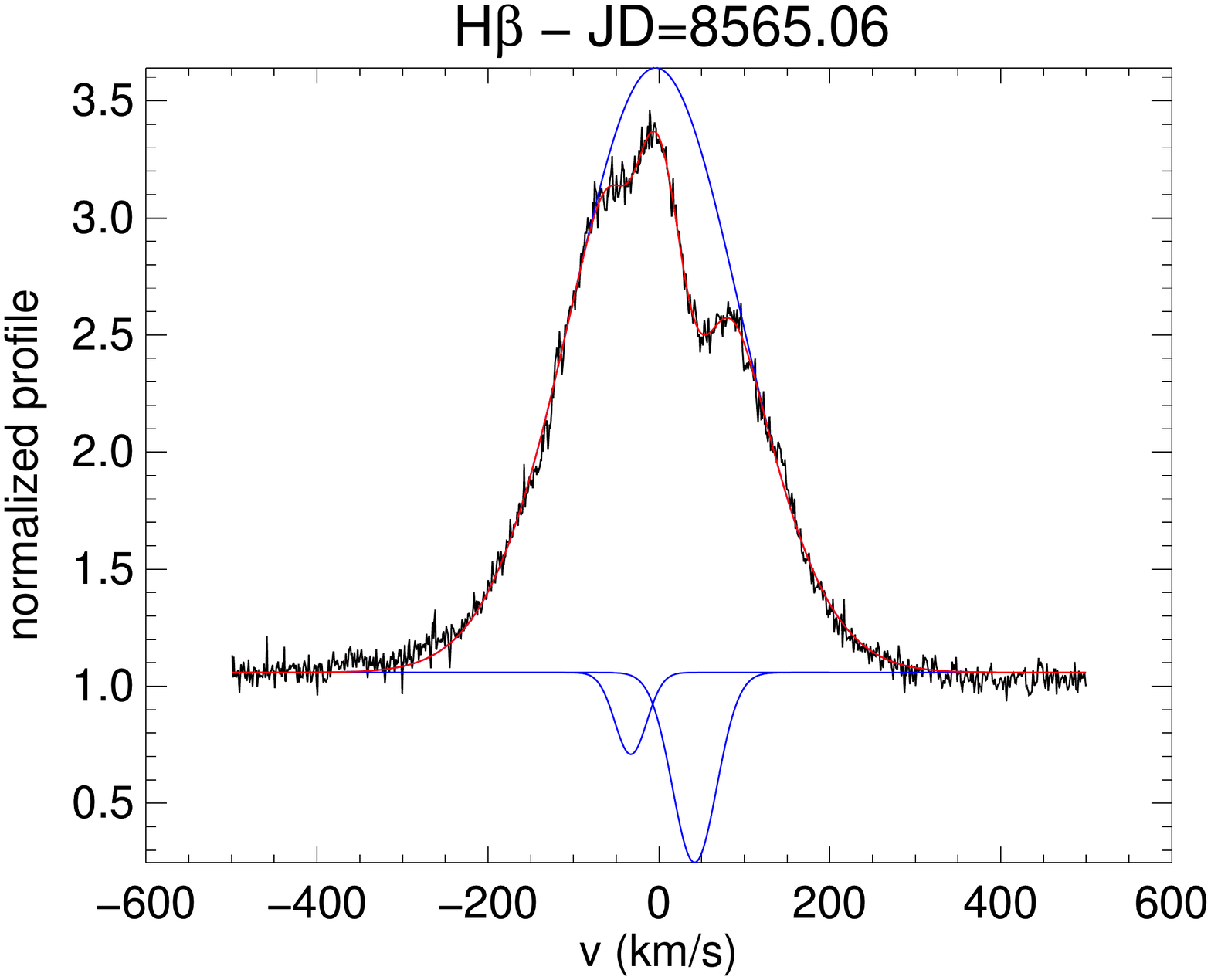}
   \includegraphics[width=0.45\hsize]{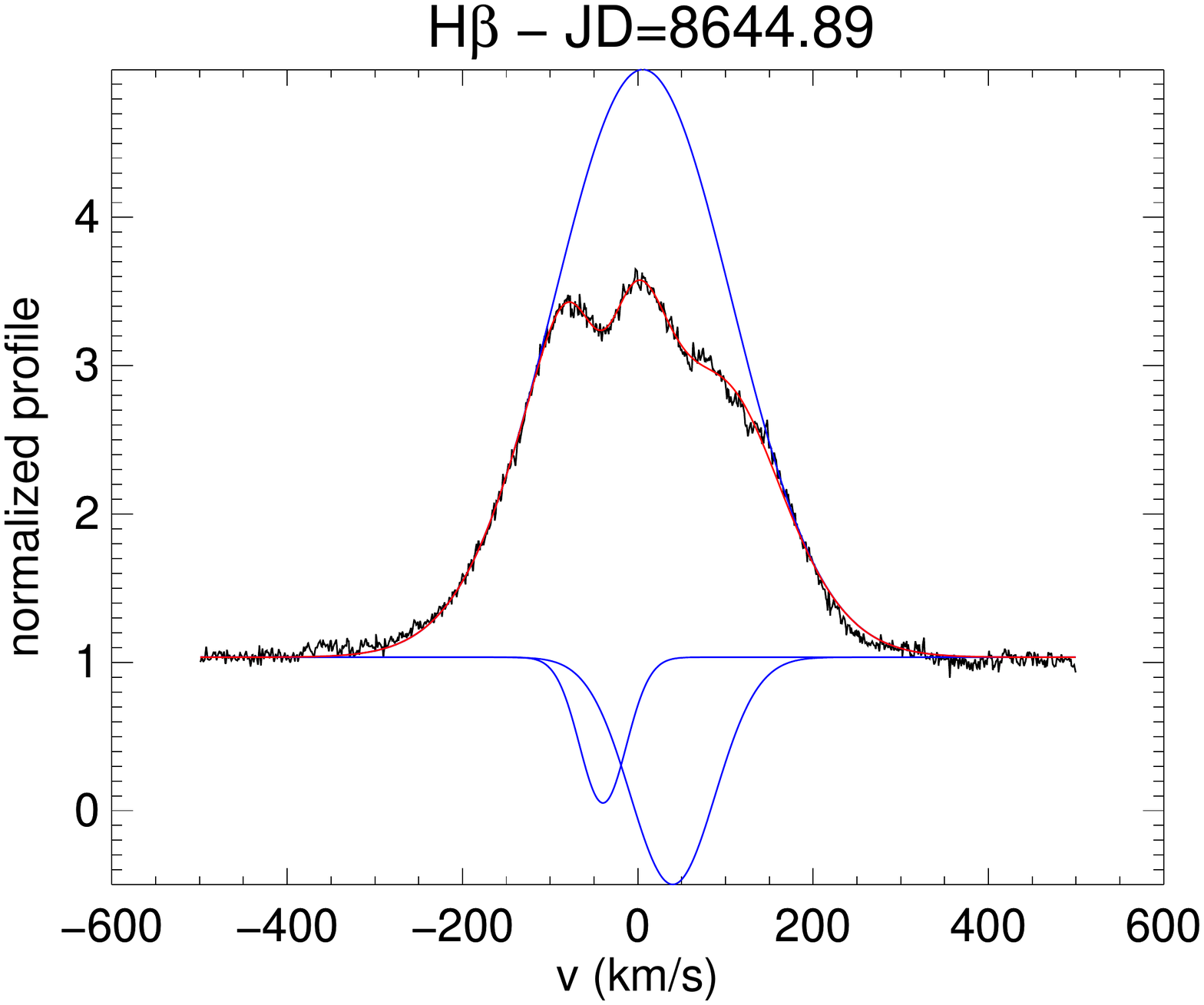}
   \includegraphics[width=0.45\hsize]{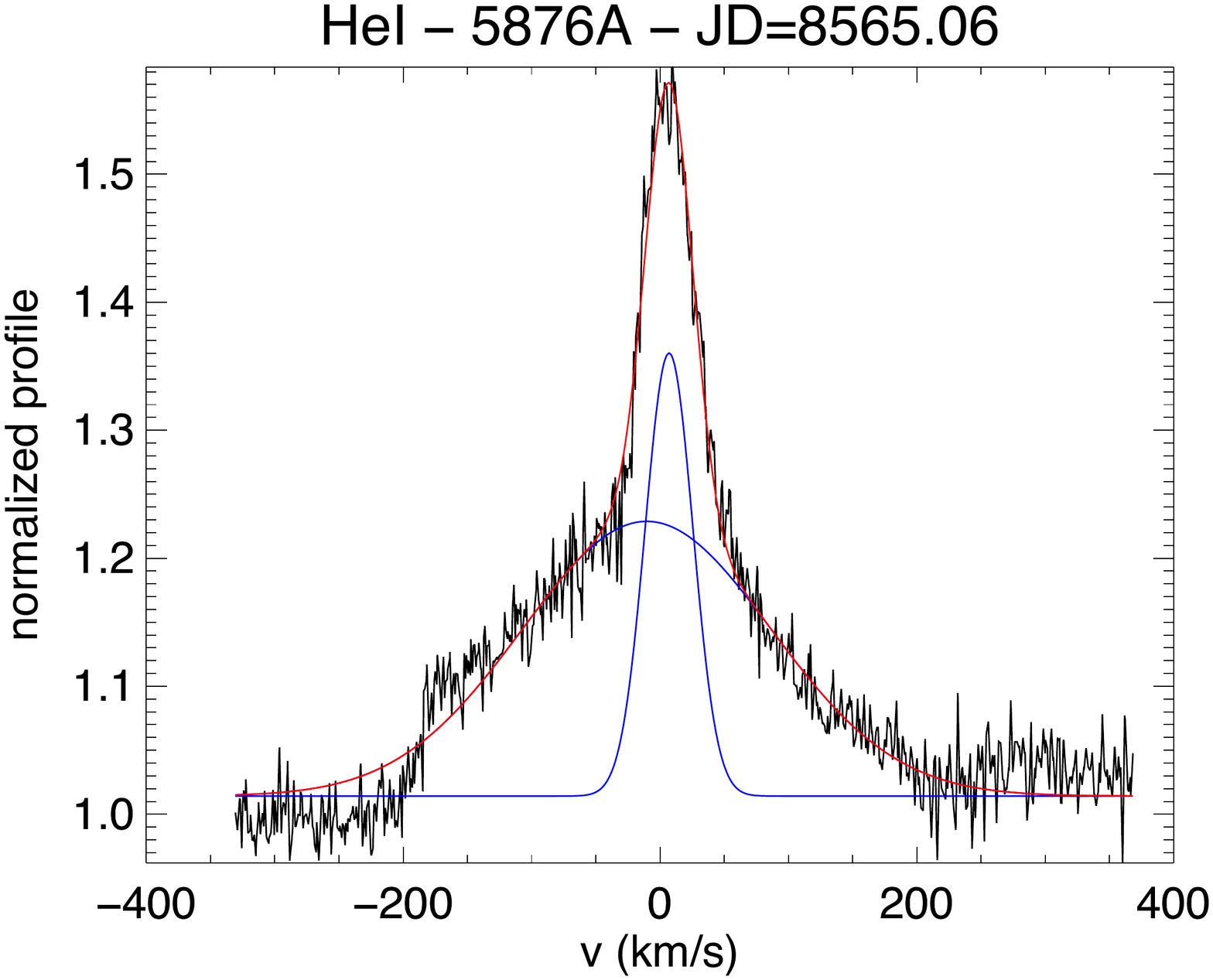}
   \includegraphics[width=0.45\hsize]{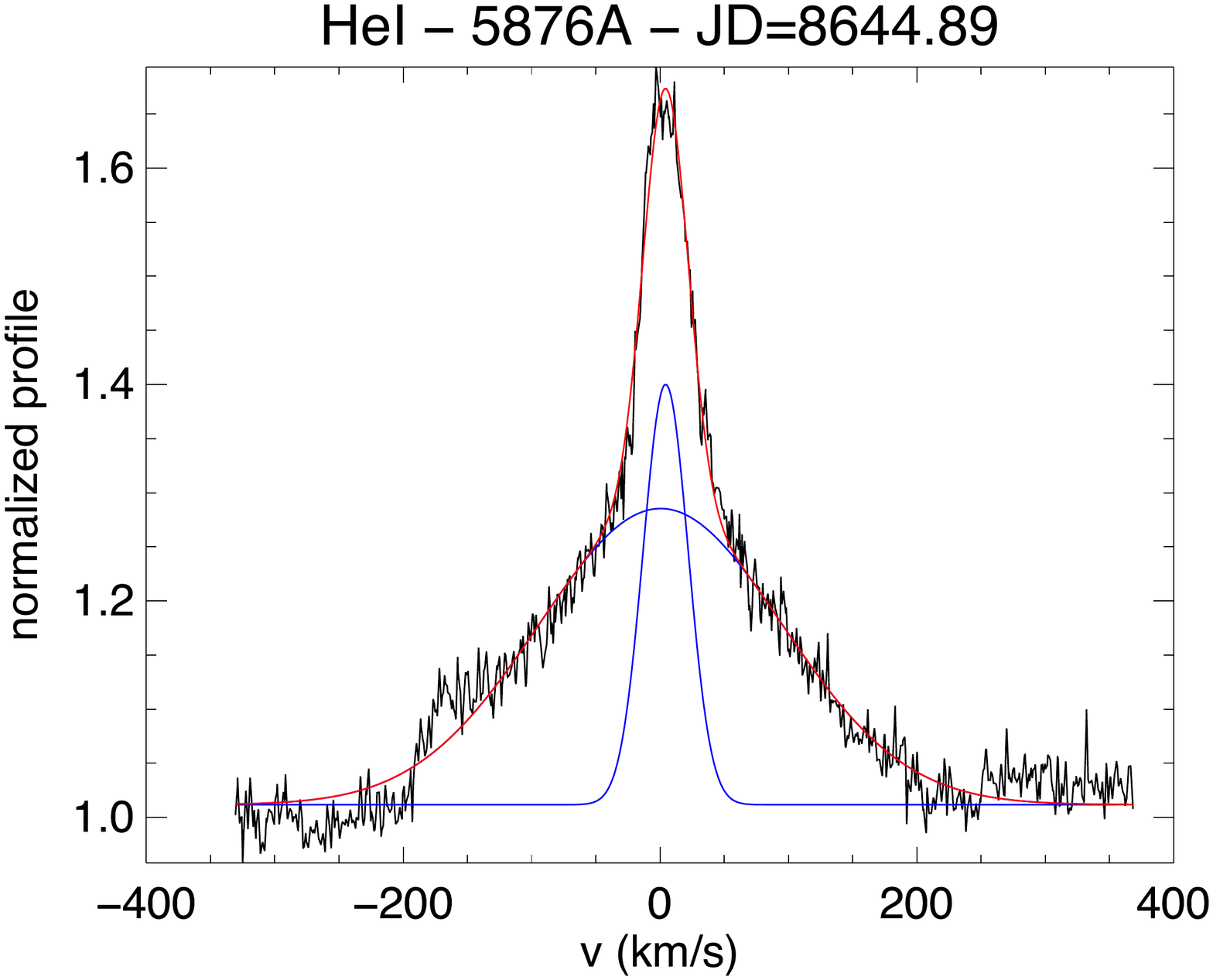}
   \includegraphics[width=0.45\hsize]{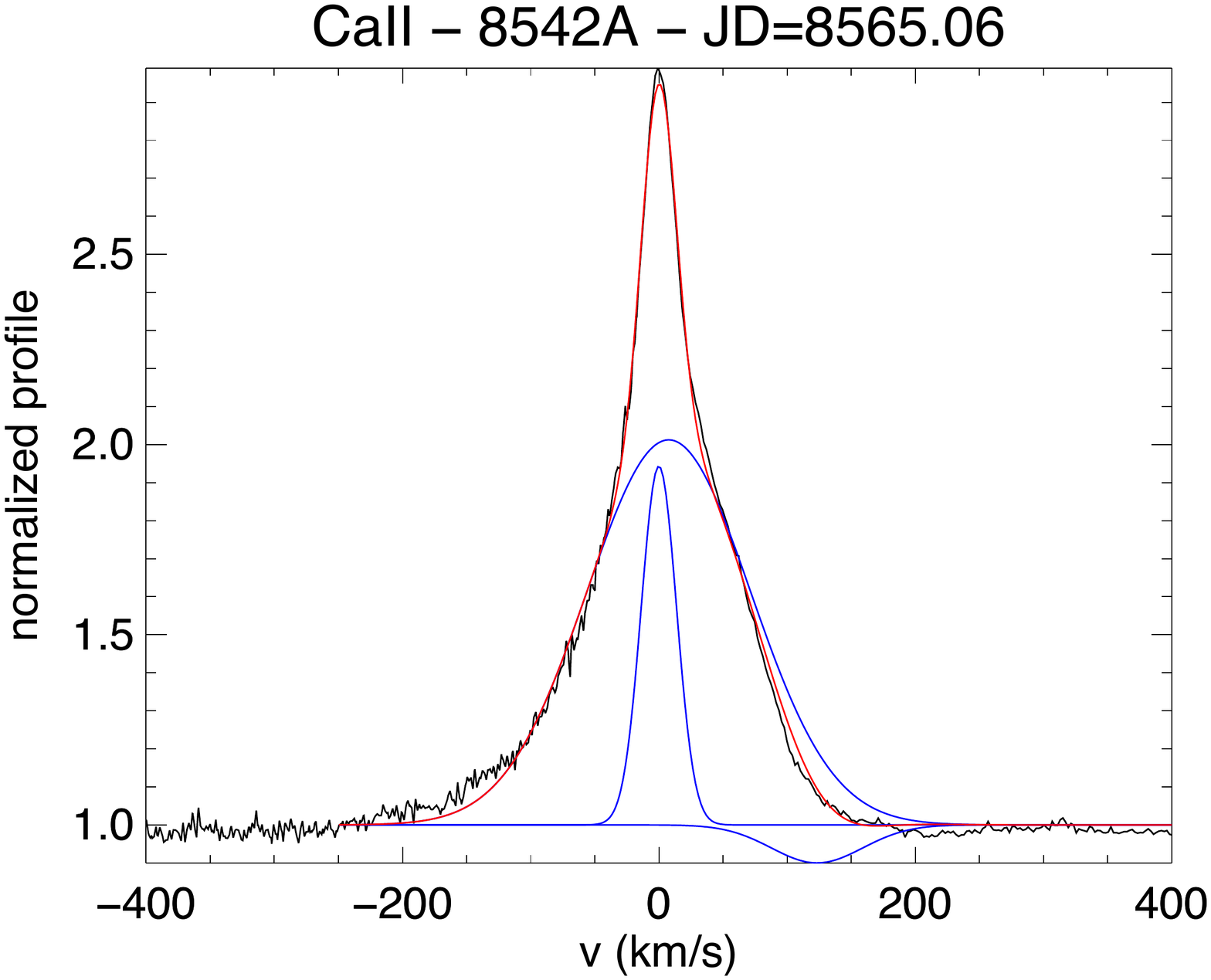}
   \includegraphics[width=0.45\hsize]{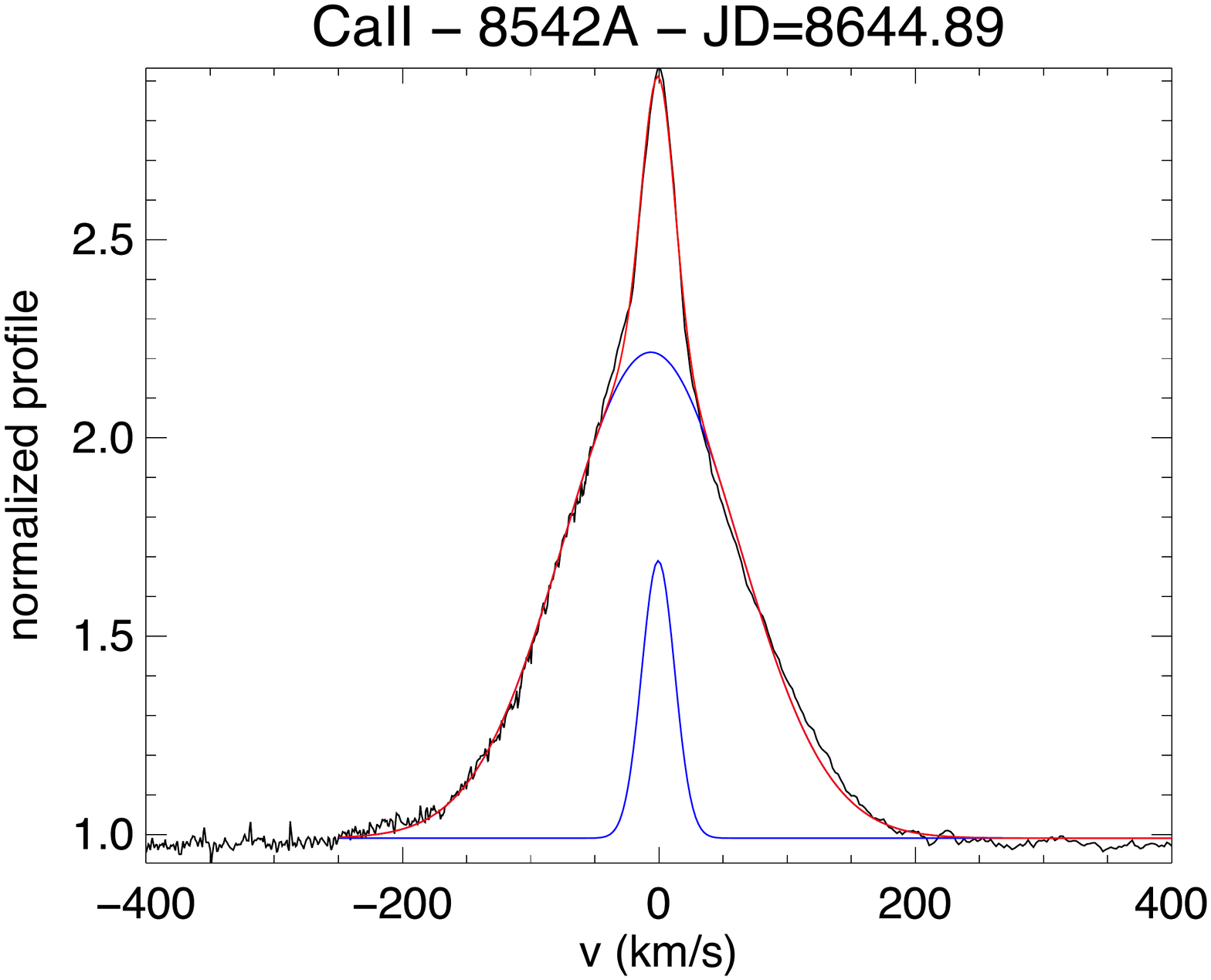}
   \caption{Line profile decomposition in Gaussian components. The Gaussian components are shown as blue curves, and the resulting modeled profile in red, while the observed profile is in black. {\it From top to bottom:} \ha, \hb, \hei, and Ca II 854.2 nm. }
              \label{decomp}%
    \end{figure}

\begin{table*}
\caption{Gaussian components of optical emission line profiles. The table lists the Julian date, the equivalent width, the peak intensity, radial velocity, and FWHM of the Gaussian fit for each component. }             
\label{decomptab}      
\centering                          
\begin{tabular}{l l l l l l l l l l l l l}        
\hline\hline                 
            HJD     & EW$_1$ & EW$_2$ &  EW$_3$  & Int$_1$ & \vrad$_1$ & FWHM$_1$ & Int$_2$ & \vrad$_2$ & FWHM$_2$ & Int$_3$ & \vrad$_3$ & FWHM$_3$ \\
            (-2,450,000) & (\AA) & (\AA) & (\AA) &  & \kms & \AA &  & \kms & \AA  &  & \kms & \AA  \\ 
\hline                        
\hline     
\multicolumn{13}{c}{\ha} \\
\hline
8558.10852 & 59.7 & -1 & -16.4 & 10 & 13.1 & 255.7 & -0.8 & -39.7 & 53.1 & -4.9 & 50.3 & 142.4 \\
8559.15082 & 50.3 & -5.8 & -5.5 & 9 & 11.9 & 239.2 & -2.7 & -7.6 & 92.8 & -3.4 & 47 & 67.7 \\
8561.13665 & 46 & -1.1 & -12.2 & 8.7 & 6.9 & 226.3 & -0.9 & -42.8 & 51.7 & -4.8 & 41.1 & 107.6 \\
8562.12015 & 60.9 & -8.7 & -5.4 & 10.7 & 5 & 242.3 & -3.1 & -4.2 & 119.1 & -2.7 & 50.4 & 84.1 \\
8563.09667 & 52.2 & -5.9 & -3.8 & 9.5 & -0.6 & 235.5 & -2.9 & -25.1 & 87.7 & -2.3 & 41.3 & 69.3 \\
8564.12225 & 61.2 & -2.6 & -14 & 10.5 & 8.5 & 249.8 & -2 & -41.9 & 55.9 & -5.2 & 39.1 & 114.9 \\
8565.06818 & 64 & -4.1 & -14.5 & 12.3 & 4.3 & 222.1 & -2.7 & -39.5 & 63.5 & -6.5 & 29 & 94.7 \\
8642.8923 & 86.3 & -10.7 & -13.8 & 15 & 4.9 & 246.5 & -5.4 & -34.5 & 85.1 & -6.9 & 40.2 & 85.3 \\
8643.89558 & 60.7 & -1.4 & -14.6 & 10.5 & 14.1 & 247.9 & -1.1 & -49.3 & 54.3 & -5 & 30.6 & 125.3 \\
8644.89415 & 69 & -2.6 & -12.1 & 11.3 & 6.5 & 261.6 & -1.9 & -41.2 & 59.2 & -4.4 & 28.4 & 116.1 \\
8645.93308 & 60.6 & -7.1 & -9.9 & 11.2 & 16.6 & 230.3 & -3.5 & -18.1 & 86.0 & -5.1 & 49.8 & 83.4 \\
8646.92779 & 58.2 & -2.1 & -16.7 & 11 & 12.7 & 226.5 & -1.5 & -33.7 & 57.3 & -5.9 & 33.2 & 120.8 \\
\\
\hline
\multicolumn{13}{c}{\hb} \\
\hline
8558.10852 & 13.9 & -0.48 & -4.4 & 3.3 & 18.9 & 241.3 & -0.5 & -41.8 & 55.2 & -2.1 & 56.4 & 119.1 \\
8559.15082 & 7.9 & -0.27 & -0.9 & 2 & 3.6 & 225.8 & -0.31 & -41 & 49.8 & -0.8 & 45 & 58.3 \\
8561.13665 & 7.9 & -0.13 & -1.6 & 2 & 1.5 & 223.3 & -0.24 & -37.5 & 31.0 & -1.1 & 48.9 & 88.1 \\
8562.12015 & 11.4 & -0.33 & -1.3 & 2.9 & 3.5 & 227.2 & -0.41 & -32.9 & 45.6 & -1 & 41.2 & 69.6 \\
8563.09667 & 8.9 & -0.55 & -0.6 & 2.4 & -5 & 214.8 & -0.59 & -34.7 & 53.8 & -0.5 & 42.5 & 66.3 \\
8564.12225 & 15 & -0.77 & -5.5 & 3.9 & 15.5 & 222.3 & -0.78 & -39.4 & 57.3 & -2.6 & 45.8 & 117.7 \\
8565.06818 & 10.2 & -0.26 & -0.8 & 2.5 & -3.9 & 229.4 & -0.34 & -32.8 & 43.5 & -0.8 & 41.7 & 60.6 \\
8642.8923 & 18.6 & -1.94 & -2.3 & 4.5 & 4.8 & 239.0 & -1.68 & -33.5 & 67.0 & -1.8 & 39.9 & 74.0 \\
8643.89558 & 13.4 & -0.45 & -4.2 & 3.3 & 19.1 & 231.9 & -0.57 & -43.7 & 45.4 & -1.8 & 41.1 & 128.5 \\
8644.89415 & 16.3 & -1.06 & -3 & 3.9 & 5.6 & 238.8 & -0.98 & -39.6 & 62.5 & -1.5 & 39.9 & 113.7 \\
8645.93308 & 12.9 & -0.92 & -2.1 & 3.2 & 14.5 & 232.2 & -0.9 & -34.7 & 58.8 & -1.5 & 44.2 & 79.9 \\
8646.92779 & 13.9 & -0.5 & -5.2 & 3.8 & 18.2 & 212.4 & -0.62 & -33.7 & 47.2 & -2.4 & 39.5 & 122.2 \\
\hline
\multicolumn{13}{c}{\hei\ 587.6 nm} \\
\hline
8558.10852 & 1.63 & 0.21 & & 0.31 & 17.5 & 248.6 & 0.22 & 5.8 & 47.0 \\
8559.15082 & 0.81 & 0.28 & & 0.16 & -4.1 & 238.8 & 0.31 & 7.1 & 43.0 \\
8561.13665 & 0.99 & 0.17 & & 0.18 & 11.1 & 259.0 & 0.2 & 5.6 & 41.4 \\
8562.12015 & 1.12 & 0.19 & & 0.25 & -10.4 & 215.5 & 0.21 & 5.9 & 43.0 \\
8563.09667 & 0.76 & 0.17 & & 0.16 & -8.5 & 229.1 & 0.21 & 10.9 & 38.5 \\
8564.12225 & 1.15 & 0.23 & & 0.21 & 5.3 & 256.6 & 0.24 & 5.6 & 46.8 \\
8565.06818 & 1.02 & 0.31 & & 0.21 & -10.3 & 228.0 & 0.34 & 7 & 43.5 \\
8642.8923 & 1.3 & 0.32 & & 0.28 & -0 & 221.6 & 0.4 & 11.4 & 38.8 \\
8643.89558 & 1.1 & 0.20 & & 0.2 & 14.1 & 264.1 & 0.23 & 6.1 & 41.8 \\
8644.89415 & 1.27 & 0.33 & & 0.27 & 0.4 & 222.8 & 0.38 & 4.4 & 40.9 \\
8645.93308 & 1.3 & 0.28 & & 0.27 & 4.7 & 228.9 & 0.35 & 11.2 & 38.3 \\
8646.92779 & 1.12 & 0.24 & & 0.22 & 12.7 & 236.4 & 0.25 & 6 & 46.3 \\
\hline
\multicolumn{13}{c}{\caii\ 854.2 nm} \\
\hline
8558.10852 & 6.9 & 0.62 &   & 1.26 & 19.7 & 180.5 & 0.74 & 0.2 & 27.3 &   &   & \\ 
8559.15082 & 4.0 & 0.75 & -0.25 & 0.87 & 16.7 & 149.9 & 0.83 & 2.1 & 29.6 & -0.1 & 122.4 & 84.4 \\
8561.13665 & 3.3 & 0.68 &   & 0.73 & 11.7 & 147.1 & 0.72 & 1 & 31.0 &   &   & \\ 
8562.12015 & 5.1 & 0.81 &   & 1.21 & 10.3 & 138.9 & 0.8 & 1.4 & 33.4 &   &   & \\ 
8563.09667 & 4.2 & 0.62 &   & 1 & 6.6 & 136.8 & 0.68 & 1.8 & 29.6 &   &   & \\ 
8564.12225 & 5.8 & 0.69 &   & 1.21 & 14.1 & 157.0 & 0.68 & -0.2 & 33.6 &   &   & \\ 
8565.06818 & 4.6 & 0.92 & -0.25 & 1.01 & 7.7 & 149.7 & 0.94 & -0 & 32.2 & -0.1 & 123.3 & 83.4 \\
8642.8923 & 7.4 & 0.69 &   & 1.5 & 1.6 & 162.2 & 0.79 & 1.7 & 28.7 &   &   & \\ 
8643.89558 & 4.6 & 0.60 &   & 0.99 & 13.4 & 152.0 & 0.64 & 1.9 & 30.8 &   &   & \\ 
8644.89415 & 6.0 & 0.64 &   & 1.22 & -6.3 & 161.2 & 0.69 & -0.5 & 30.1 &   &   & \\ 
8645.93308 & 5.3 & 0.77 &   & 1.1 & 13.7 & 157.5 & 0.8 & 2.7 & 31.3 &   &   & \\ 
8646.92779 & 6.0 & 0.65 &   & 1.28 & 12.7 & 151.8 & 0.65 & -0.7 & 32.9 &   &   & \\ 
\hline\end{tabular}
\end{table*}

\end{appendix}

\end{document}